\pdfoutput=1
\documentclass[a4paper,12pt]{article}
\usepackage{amssymb,lscape,psfrag}
\usepackage{epsfig,color}
\usepackage{array}
\usepackage{enumerate}
\usepackage{amsmath}
\numberwithin{equation}{section}
\usepackage{graphics}
\usepackage{graphicx}
\usepackage{epsfig}
\usepackage{verbatim}
\usepackage{cancel}
\usepackage{longtable}
\usepackage{multirow}
\usepackage{float}
\usepackage{ulem} 
\usepackage[toc,page]{appendix}
\usepackage{rotating}
\usepackage{amsmath}
\usepackage{amsthm}
\usepackage{url}

\usepackage{natbib}
\usepackage[colorlinks,citecolor=blue,urlcolor=blue,filecolor=blue,backref=page]{hyperref}

\numberwithin{equation}{section}
\theoremstyle{plain}

\newtheorem{definition}{Definition}[section]
\newtheorem{prop}{Proposition}[section]

\renewcommand{\&}{and}

\newcommand {\vg}{\dn{\gamma}}
\newcommand {\vgjo}{\dn{\gamma}_{j_1}}
\newcommand {\vgjz}{\dn{\gamma}_{j_0}}

\newcommand {\dn}[1] { {\boldsymbol#1} }

\newcommand {\mt}[1] {\mathrm #1}
\newcommand {\inispace}{\renewcommand{\baselinestretch}{1.}\normalsize}

\newcommand {\be} {\begin {equation} }
\newcommand {\ee} {\end {equation} }
\newcommand {\bmath} {\begin {displaymath} }
\newcommand {\emath} {\end {displaymath} }

\newcommand {\bg} {\boldsymbol{\gamma}}
\newcommand {\by} {\mathbf{y}}

\newcommand {\nstar} { n^* }

\newcommand {\bx} { \mt{X} }

\newcommand {\diag} { \mbox{diag} }

\newcommand{\bb}{\boldsymbol{\beta}}
\newcommand{\bbgmlestar}{\widehat{\boldsymbol{\beta}}{}_\vg^*}

\newcommand{\betab}{\boldsymbol{\beta}_\vg}
\newcommand{\betabg}{\boldsymbol{\beta}_{\setminus\vg}}

\newcommand{\invd}{1/\delta}
\newcommand{\PEP}{\mathrm{PEP}}

\newcommand{\I}{\mathbf{I}_n}
\newcommand{\X}{\mathbf{X}}
\newcommand{\XG}{\mathbf{X}_{\vg}}
\newcommand{\XT}{\mathbf{X}_{\vg}^{T}}

\newcommand{\HL}{\mathbf{H}_{\gamma}}

\newcommand{\LamdaOCR}{\mathbf{\Lambda}_{0}^{({\mathrm{CR}})}}

\newcommand{\LamdaO}{\mathbf{\Lambda}_{0}}
\newcommand{\IotaN}{\mathbf{1}_n}
\newcommand{\V}{\mathbf{V}_{\betab}}
\newcommand{\W}{\mathbf{W}_{\vg}}

\newcommand{\N}{\mathrm{N}}
\newcommand{\Z}{\mathrm{Z}}
\newcommand{\CR}{\mathrm{CR-PEP}}
\newcommand{\DR}{\mathrm{DR-PEP}}

\newcommand{\betabt}{\betab^{(t)}}
\newcommand{\betabprop}{\betab^{'}}
\newcommand{\betabtt}{\betab^{(t-1)}}
\newcommand{\betabgt}{\betabg^{(t)}}

\begin{document}
\normalem

\inispace
\title{Power-Expected-Posterior Priors for Generalized Linear Models}
\date{}
\author{
Dimitris Fouskakis$^1$, Ioannis Ntzoufras$^2$ and Konstantinos Perrakis$^{3,2}$\\\\   
{\it \small $^1$Department of Mathematics, National Technical University of Athens}\\
{\it \small $^2$Department of Statistics, Athens University of Economics and Business} \\
{\it \small $^3$German Center for Neurodegenerative Diseases (DZNE), Bonn}
}


\maketitle
\pagenumbering{arabic}
\begin{abstract}

The power-expected-posterior (PEP) prior 
provides an objective, automatic, consistent and parsimonious model selection procedure. 
At the same time it resolves the conceptual and computational problems due to the use of imaginary data. Namely, (i) it dispenses with the need to select and average across all possible minimal imaginary samples, and (ii) it diminishes the effect that the imaginary data have upon the posterior distribution. 
These attributes allow for large sample approximations, when needed, 
in order to reduce the computational burden under more complex models. In this work we generalize the applicability of the PEP methodology, focusing on the framework of generalized linear models (GLMs), by introducing two new PEP definitions which are in effect applicable to any general model setting. 
Hyper-prior extensions for the power parameter that regulates the contribution of the imaginary data are introduced.
We further study the validity of the predictive matching and of the model selection consistency, 
providing analytical proofs for the former and empirical evidence supporting the latter.
For estimation of posterior model and inclusion probabilities  we introduce a tuning-free Gibbs-based variable selection sampler.
Several simulation scenarios and one real life example are considered in order to evaluate the performance of the proposed methods compared to other commonly used approaches based on mixtures of $g$-priors. 
Results indicate that the GLM-PEP priors are more effective 
in the identification of sparse and parsimonious model formulations.

\vspace{0.2cm} \noindent \textit{Keywords: expected-posterior prior, $g$/hyper-$g$ priors, generalized linear models, imaginary data, objective Bayesian model selection, power-prior} 

\end{abstract}

\section{Introduction}

\subsection{Motivation}

In  this  article,  the  variable selection  problem  in  generalized linear models (GLMs) is analyzed from an objective and fully automatic 
Bayesian model choice perspective. The desire for an automatic Bayesian
procedure is motivated by the appealing property of creating a method that can be easily implemented 
in complex models without the need of 
specification of tuning parameters. Regarding the justification for the necessity of an objective model choice 
approach we can argue that in variable selection problems we are  rarely confident  about
any given set of regressors as explanatory variables, which translates to little prior
information  about the  regression  coefficients. Therefore, we would like 
to consider default prior distributions, which in many cases are improper, thus leading to undetermined Bayes factors.

Intrinsic priors \citep{berger_pericchi_96a,berger_pericchi_96b} and expected-posterior (EP) priors \citep{perez_berger_2002} can be considered as fully automatic, objective Bayesian methods for model comparison in regression models. They are developed through the utilization
of the device of ``training" or ``imaginary" samples, respectively, of ``minimal" size and therefore the resulting priors have a further advantage of being compatible across models; see \cite{consonni_veronese_2008}. Intrinsic and EP priors have been proposed in many articles for variable selection in Gaussian linear models  \citep[see for example][]{casella_moreno_2006}; however, to the best of our knowledge, there is only one study that proposes this methodology for GLMs, which is restricted to the case of the probit model \citep{Leon_Novelo_etal_2012}. We believe that this is due to the fact that derivation of such priors can be a very challenging task, especially under complex models, leading to computationally intensive solutions. Furthermore, by using minimal training samples, large sample approximations can not be applied in many cases. 

Our contribution with this article is two-fold. First, we develop an automatic, objective Bayesian variable selection procedure for GLMs based on the EP prior methodology. In particular we consider the power-expected-posterior (PEP) prior of \cite{fouskakis_etal_2015}, that diminishes the effect that the imaginary data have upon the posterior distribution and therefore the need of using minimal training samples. Through this approach we can consider imaginary samples of sufficiently large size and therefore be able to apply, when needed,  large sample approximations. Secondly, we introduce a simple tuning-free Gibbs-based variable selection sampler
for estimating posterior model and variable inclusion probabilities.

\subsection{Bayesian variable selection for generalized linear models}

Despite the importance and popularity of GLMs, Bayesian variable selection 
techniques for non-Gaussian models 
are scarce in relation to the abundance of methods that are available for the normal linear model. 
This is mainly due to the analytical intractability which arises outside the context of the normal model. 
The relatively limited studies that focus on non-Gaussian models, mainly aim to overcome analytical intractability through the use of Laplace approximations and/or stochastic model search algorithms. 

\cite{Chen_Ibrahim_2003} introduced a class of conjugate priors based on an initial prior prediction of the data (similar to the concept of imaginary data) associated with a scalar precision parameter. This approach essentially leads to a GLM analogue of the $g$ prior \citep{Zellner_siow_80, zellner_86} where the precision parameter has the role of $g$. However, the prior of \cite{Chen_Ibrahim_2003} is not analytically available for non-Gaussian GLMs and, therefore,  
\cite{Chen_etal_2008} proposed a Markov chain Monte Carlo (MCMC) based solution for this class of models. 
\cite{ntzoufras_etal_2003} used a unit-information $g$-prior \citep{kass_wasserman_95} for variable selection and link determination in binomial models through reversible-jump MCMC sampling. \cite{bove_held_2011} consider the asymptotic distribution of the prior of \cite{Chen_Ibrahim_2003}, which results in the same $g$-prior form used in 
\cite{ntzoufras_etal_2003}, and further consider mixtures of $g$-priors along the lines of \cite{liang_etal_2008}. 
Computation of the marginal likelihood in \cite{bove_held_2011} is handled through an 
integrated Laplace approximation, based on Gauss-Hermite quadrature, which allows variable selection through full enumeration for small/moderate model spaces or through MCMC model composition (MC$^3$) algorithms \citep{madigan_york_95} for spaces of large dimensionality. 
Other GLM variations of $g$-prior mixtures have an empirical Bayes (EB) flavor, using the observed or expected information matrix evaluated at the ML estimates as the prior variance-covariance matrix \citep{Hansen_Yu_2003,Wang_George_2007,Li_Clyde_2015}. A computational benefit of the EB approach is that the integrated Laplace approximation can be expressed in closed form as a set of functions of the ML estimates. For large model spaces, where full enumeration is infeasible, \cite{Li_Clyde_2015} recommend using the Bayesian adaptive sampling algorithm \citep{clyde_etal_2011}.  
A relevant prior specification is the information-matrix prior of \cite{gupta_ibrahim_09} which combines ideas from the $g$-prior and Jeffreys prior for GLMs \citep{ibrahim_laud_91};
under a Gaussian likelihood the information-matrix prior becomes the standard $g$-prior, while for $g \rightarrow \infty$ it reduces to Jeffreys prior which is proper only for the case of the binomial model. However, in applications \cite{gupta_ibrahim_09} do not directly consider the problem of stochastic search over the entire model space. 
Finally, one application of Bayesian intrinsic variable selection for probit models via MCMC is presented in \cite{Leon_Novelo_etal_2012}.

In this work we present an automatic, objective Bayesian variable selection procedure for GLMs based on the PEP methodology. The structure of the remainder of the paper is as follows. 
In Section \ref{pep_glms} we provide an overview of the PEP prior formulation and discuss the applicability problems that arise in the case of non-Gaussian models. We proceed with two alternative definitions, which generalize the applicability of the PEP prior for GLMs. 
In Section \ref{pep_post_inf} we introduce a Gibbs-based sampler suitable for variable selection and for single-model posterior inference.  
Section \ref{PEP_hyper_delta} 
presents an hierarchical extension of the methodology which involves assigning a hyper-prior to the power parameter that controls the contribution of the imaginary data.
In Section \ref{Desiderata} we examine the validity of certain desiderata proposed by \cite{Bayarri_etal_2012} and we proceed by presenting a general framework in Section \ref{Framework} for all PEP priors under consideration.
Illustrative examples and comparisons with other methods 
using both simulated and real life example are presented in Section \ref{examples}. We conclude with a summary and a discussion of future research directions in Section \ref{discussion}.

\section{PEP Priors for Generalized Linear Models}
\label{pep_glms}

\subsection{Model setting}
\label{Likelihood}
We consider $n$ realizations of a response variable $Y$ accompanied by a set of  potential predictors $X_1,X_2,...,X_p$ which may characterize the response. 
To fix notation, let $\vg\in\{0,1\}^p$ index all $2^p$ subsets of predictors  serving as a model indicator, 
where each element $\gamma_j$, for $j=1,\dots,p$, is an indicator of the inclusion of $X_j$ in the  structure of model $M_{\vg}$. 
Moreover, let $p_{\vg}=\sum_{j=1}^p \gamma_j$ denote the number of active covariates in model $M_{\vg}$.
Within the GLM framework, the response $Y$ follows a distribution which is a member of the exponential family. 
The sampling distribution of the response vector $\by=(y_1, \dots, y_n)^T$ under model $M_{\vg}$ is given by 
\begin{equation}
\label{loglike}
f_{\vg}( \by | \betab, \phi_{\vg} ) = 
\exp\left( \sum_{i=1}^n\frac{y_i \vartheta_{\vg (i)}  - b(\vartheta_{\vg (i)}) }{ a_{i}(\phi_{\vg})} + \sum_{i=1}^nc(y_i, \phi_{\vg}) \right).
\end{equation}
The functions  $a(\cdot)$, $b(\cdot)$ and $c(\cdot)$ determine the particular distribution of the exponential family. 
The parameter $\vartheta_{\vg(i)}$ is the canonical parameter which regulates the location of the distribution through the relationship $\vartheta_{\vg (i)} = \vartheta\big( \eta_{\vg (i)}\big)\equiv g\circ b'^{-1}\big( \eta_{\vg (i)}\big)$, where $g(\cdot)$ is the link function connecting the mean of the response $Y_i$ with the linear predictor $\eta_{\vg (i)}=\X_{\vg(i)} \betab$ and
$g\circ b'^{-1}(\eta_{\vg (i)})$ is the inverse function of $g\circ b'(\vartheta_{\vg (i)})\equiv g\big( b'(\vartheta_{\vg (i)}) \big)$. Commonly, a canonical $\vartheta$ function is used, so that $\vartheta_{\vg (i)} = \eta_{\vg (i)}$.
We assume that an intercept term is included in all $2^p$ models under consideration, so $\betab$ is the $d_{\vg}\times 1$ vector of regression coefficients, where $d_{\vg}=p_{\vg}+1$, and $\X_{\vg(i)}$ is the $i$--th row of the $n \times d_{\vg}$ design matrix $\X_{\vg}$  
with a vector of 1's in the first column and the $\vg$--th subset of the $\boldsymbol{X}_j$'s
in the remaining $p_{\vg}$ columns.  The parameter $\phi_{\vg }$ controls the dispersion and the function $\alpha(\cdot)$ is typically of the form $\alpha_i(\phi_{\vg })=\phi_{\vg}/w_i$, where $w_i$ is a known fixed weight that may either vary or remain constant per observation. In addition, the nuisance parameter $\phi_{\vg}$ is commonly considered as a common parameter across models, therefore we assume throughout that $\phi_{\vg}\equiv\phi$ without loss of generality. Given the above formulation, we have that 
$\mathrm{E}(Y_i)=b'(\vartheta_{\vg (i)})$ and 
$\mathrm{Var}(Y_i)=b''(\vartheta_{\vg (i)})\alpha_i(\phi)$. 

The GLM parameters $\dn{\theta}_\vg = ( \betab, \phi )$ are divided 
into the predictor effects $\betab$ and the parameter $\phi$ which affects dispersion. 
In the following we work along the lines of \cite{Fouskakis_Ntzoufras_2016_jcgs} considering the conditional PEP prior; i.e. we construct the PEP prior of $\betab$ conditional on $\phi$.

\subsection{An overview of the PEP prior}
\label{section_PEP}

The PEP prior, initially formulated in \cite{fouskakis_etal_2015} for the case of the normal linear model, creatively fuses ideas from the power prior \citep{ibrahim_chen_2000} and
the EP prior \citep{perez_berger_2002}. 
Let us first describe the EP prior approach. Consider that we have imaginary data   
$\by^* = (y^*_1,\dots,y^*_{n^{*} \color{black}})^T$  
coming from the prior-predictive distribution  $m^* ( \by^* )$ of a ``suitable" \textit{reference} model $M^*$.
Then, given $\by^*$, for any 
model $M_{\vg}$ with sampling distribution $f_\vg(\by^*| \betab,\phi)$ as defined in \eqref{loglike}
and a default \textit{baseline prior} of the form $\pi^{\mathrm{N}}_\vg( \betab,\phi)=\pi^{\mathrm{N}}_\vg( \betab|\phi)\pi^{\N}_{\vg}(\phi)$, 
we have a corresponding \textit{baseline posterior} distribution given by
\begin{equation}
\pi_\vg^{\N}( \betab,\phi |\by^*) = \frac{f_\vg(\by^*| \betab,\phi)\pi^{\mathrm{N}}_\vg( \betab|\phi)\pi_{\vg}^{\N}(\phi)}{m_{\vg}^{\N}(\by^*)},
\label{baseline posterior}
\end{equation} 
where $m_{\vg}^{\N}(\by^*)$ is the normalizing constant of the baseline posterior distribution under model $M_\vg$. 
The EP prior for the parameters of model $M_{\vg}$  
is then defined as the posterior distribution in \eqref{baseline posterior}, averaged over all possible
imaginary samples, i.e.
\begin{equation} 
\label{epp} 
\pi^{\mathrm{EP}}_{\vg} ( \betab,\phi ) = \int
\pi_{\vg}^{\N} (\betab,\phi | \by^* ) \, m^* ( \by^* ) \mathrm{d}
\by^* \, .
\end{equation}
The reference model $M^*$ is commonly considered to be the simplest model, i.e. the (null) intercept model in the regression framework. This selection makes 
the EP approach essentially equivalent to the arithmetic intrinsic Bayes factor of \cite{berger_pericchi_96b}.

A key issue in the implementation of the EP prior
is the selection of the size $n^*$ of the imaginary sample. 
In order to minimize the effect of the prior on posterior inference, 
the reasonable solution is to choose the smallest possible $n^*$ for which the posterior \eqref{baseline posterior}  is proper. This leads to the concept of the so-called {\it minimal training sample}, which however requires calculating the arithmetic mean (or other appropriate measures of centrality) of Bayes factors over all possible minimal training samples. 
In addition, when it comes to regression the same problem arises with the design matrix as one has to choose appropriate covariate values for each minimal training sample, and this further depends upon the choice of the reference model. 
Therefore,  under the EP prior, computation of the Bayes factors require calculations over all possible configurations of the design matrix for each minimal training sample 
\citep{perez_98} or, at least, calculations over an efficiently large number of random sub-samples of all possible configurations \citep{fouskakis_ntzoufras_2013}. 
An alternative and simpler computational solution has been proposed by \cite{casella_moreno_2006} and \cite{moreno_giron_2008}, however, this solution is only applicable under the normal linear regression model.  Additionally,  under this approach, it is not clear whether the resulting Bayes factors retain their intrinsic nature. Furthermore, the effect of the EP prior can become influential when the sample size is not much larger than the total number of predictors; see \cite{fouskakis_etal_2015} for details.  Finally, when  $n^*$ is small and (\ref{epp}) is hard to derive, large sample approximations cannot be applied. 

The PEP prior  
resolves the problem of defining and averaging over minimal training samples and at the same time scales down the 
effect of the imaginary data on the posterior distribution. 
The core idea lies in substituting the likelihood function involved in the calculation of \eqref{epp} 
by a powered-version of it, i.e. raising it to the power of $1/\delta$, similar to the power prior approach of \cite{ibrahim_chen_2000}.  
Following \cite{Fouskakis_Ntzoufras_2016_jcgs}, the conditional PEP (PCEP) prior in the GLM setup, under the null-reference model $M_0$, is defined as follows
\begin{equation}
\pi^{\mathrm{PEP}}_{\vg} ( \betab, \phi| \delta ) 
= \pi^{\mathrm{PEP}}_{\vg} ( \betab | \phi, \delta) \pi^{\mathrm{N}}_{\vg}( \phi),
\label{PEP-all}
\end{equation}
where
\begin{eqnarray}
\pi^{\mathrm{PEP}}_{\vg} ( \betab | \phi, \delta) 
&=&\int \pi_{\vg}^{\N} ( \betab | \by^*, \phi, \delta) m_0^{\mathrm{N}}(\by^* |\phi,\delta ) \mathrm{d}\by^*  , \label{PEP} \\
\pi_{\vg}^{\N} ( \betab | \by^*, \phi,\delta)  
\label{pep_original}
&=& \frac{f_{\vg}( \by^* | \betab, \phi, \delta ) \pi_{\vg}^{\mathrm{N}}(\betab\,|\phi)}{m_{\vg}^{\mathrm{N}} (\by^*|\phi,\delta)}, 
\label{general_power_prior}\\
m_{\vg}^{\mathrm{N}}( \by^*|\phi,\delta) 
&=&\int f_{\vg}( \by^* | \betab, \phi, \delta ) \pi_{\vg}^{\mathrm{N}}(\betab\,|\phi)
\mathrm{d}\betab,
\label{priorgamma}\\
f_{\vg}( \by^* | \betab, \phi, \delta ) & = & \frac{ f_{\vg}( \by^* | \betab, \phi)^{1/\delta} }{k_{\vg} (\betab, \phi, \delta )},
\label{pl}\\ 
m_0^{\mathrm{N}}( \by^*|\phi,\delta)&=&\int f_0( \by^* | \beta_0, \phi, \delta ) \pi_0^{\mathrm{N}}(\beta_0\,|\phi)
\mathrm{d}\beta_0
\label{priornull},\\
f_0( \by^* | \beta_0, \phi, \delta ) & = &  \frac{ f_0( \by^* | \beta_0, \phi)^{1/\delta} }{k_0 (\beta_0, \phi, \delta )}
\label{pl0}.
\end{eqnarray} 
For the original PEP prior of \cite{fouskakis_etal_2015}, we consider the choice  \linebreak 
$k_{\vg} (\betab, \phi, \delta )= \int f_{\vg}( \by^* | \betab, \phi)^{1/\delta} d \by^*$ for all models $\vg \in {\cal M}$. 
Under this choice, the PEP prior of the intercept $\beta_0$ of the reference $M_0$ reduces to the baseline prior; i.e. 
\linebreak 
$\pi_0^{\PEP}(\beta_0|\phi,\delta)=\pi_0^{\N}(\beta_0|\phi)$.
The selection of $k_{\vg} (\betab, \phi, \delta )$ and $k_{0} (\beta_0, \phi, \delta )$ is further discussed in Section \ref{Sec3}. 

Here the  power parameter $\delta$  controls the weight that the imaginary data contribute to the ``final'' posterior distributions of $\betab$ and $\phi$.
As  noted in \cite{fouskakis_etal_2015}, the choice of  $\delta=n^*$ leads to a minimally-informative prior with a unit-information interpretation \citep{kass_wasserman_95} where the  contribution of the imaginary data is down-weighted to account  overall for one data point. 
Furthermore, by setting $n^*=n$ we avoid the complicated problem of sampling over numerous imaginary design sub-matrices, as in this case we have that $\X_\vg^*\equiv\X_\vg$. 
Under this framework, the unit-information property in combination with the empirical evidence presented in  \cite{fouskakis_etal_2015} 
suggest that the PEP prior is robust with respect to the specification of $n^*$ and 
it also remains relatively non-informative even when the model dimensionality is close to  
the sample size. 

Another advantage of setting $\nstar=n$, which becomes more obvious in the GLM framework, is that one can now utilize
large-sample approximations when needed.
For instance, consider the baseline posterior in \eqref{general_power_prior}, which can be expressed as
\begin{eqnarray}
\pi_{\vg}^{\N}( \betab | \by^*,\phi, \delta )  
&\propto& \exp\left( \sum_{i=1}^{n^*}\frac{y_i^* \vartheta_{{\vg} (i)}  - b(\vartheta_{{\vg} (i)}) }{ \delta a_i(\phi)}  \right)
\pi^{\mathrm{N}}_{\vg}( \betab| \phi).
\label{power_prior}
\end{eqnarray} 
This unnormalized distribution is recognized as the  power prior for GLMs \citep{chen_etal_2000}.
If we assume a flat baseline prior for $\betab$, i.e. $\pi^{\mathrm{N}}_{\vg}( \betab | \phi ) \propto 1$, then, based on standard Bayesian asymptotic theory \citep{bernardo_smith_2000}, for $\nstar\rightarrow\infty$ the distribution in \eqref{power_prior} converges to 
\begin{equation}
\widehat{\pi}_{\vg}^{\N}( \betab | \by^*, \phi, \delta )
\approx 
\mathrm{N}_{d_{\vg}} \big(\bbgmlestar, \delta \mathcal{\boldsymbol{J}}_{\vg}^*\big(\bbgmlestar\big)^{-1}\big),
\label{asympt}
\end{equation}  
where
$\bbgmlestar$ is the MLE of $\betab$ for data $\by^*$ and design matrix $\X_{\vg}^*$, and $\mathcal{\boldsymbol{J}}_{\vg}^*\big(\bbgmlestar\big)$ is the observed information evaluated at $\bbgmlestar$.
Specifically, 
\begin{equation}
\mathcal{\boldsymbol{J}}_{\vg}^*\big(\bbgmlestar\big)=\big( \X_{\vg}^{*T} \mathbf{W}_{\vg}^* \X_{\vg}^* \big)^{-1},
\label{fischer}
\end{equation}
where
$\mathbf{W_{\vg}^*}= \diag(w_{{\vg} (i)}^*)$ with $w_{{\vg} (i)}^{*} =  \big(  \frac{ \partial \mu_{{\vg} (i)}}{\partial \eta_{{\vg} (i)}} \big)^2 \big[a_i( \phi ) b''( \vartheta_{{\vg} (i)})\big]^{-1} $ and $\mu_{{\vg} (i)}=b'( \vartheta_{{\vg} (i)})$. 
It is straightforward to see that the asymptotic distribution in \eqref{asympt} has a $g$-prior form 
according to the definitions for GLMs presented in 
\cite{ntzoufras_etal_2003} and \cite{bove_held_2011}.
The familiar zero-mean representation in \eqref{asympt} arises when the covariates are centered around their corresponding arithmetic mean and the imaginary response data are all the same, i.e. $\by^* = g^{-1}(0)\mathbf{1}_{\nstar}$, 
where $\mathbf{1}_{\nstar}$ is a vector of ones of size $\nstar$ since in this case we have that $\bbgmlestar=\mathbf{0}_{d_{\vg}}$; for details see
\cite{ntzoufras_etal_2003}.

\subsection{PEP prior extensions for GLMs via unnormalized power likelihoods}
\label{Sec3}


The sampling distribution of the imaginary data involved in the PEP prior via \eqref{general_power_prior}, \eqref{priorgamma} and \eqref{priornull} is a power version of the likelihood function. 
In the normal linear regression case \cite{fouskakis_etal_2015} and \cite{Fouskakis_Ntzoufras_2016_jcgs} naturally considered \linebreak 
$k_{\vg} (\betab, \phi, \delta )=\int f_{\vg}( \by^* | \dn{\theta}_{\vg},\phi )^{1/\delta} \mathrm{d}\by^*$, 
i.e.  the density normalized power likelihood
\begin{equation}
f_{\vg}( \by^* | \dn{\theta}_{\vg}, \phi,\delta ) =
\frac{f_{\vg}( \by^* | \dn{\theta}_{\vg}, \phi )^{1/\delta}}{\int f_{\vg}( \by^* | \dn{\theta}_{\vg}, \phi )^{1/\delta} \mathrm{d}\by^*},
\label{density_normalized_power_lik}
\end{equation} 
which is also a normal distribution
with variance inflated by a factor of $\delta$.
Similar results can be derived for specific distributions of the exponential family 
such as the Bernoulli, the exponential and the beta distributions
where the normalized power likelihood is of the same distributional form. This property simplifies calculations when using the PEP methodology, especially for Gaussian models where the resulting posterior distribution and marginal likelihood are available in closed form. An application of the PEP prior using the normalized power likelihood for MCMC-based variable selection in binary logistic regression can be found in \cite{perrakis_baysm}.  

However, this property does not hold for all members of the exponential family. For instance, for the binomial and Poisson regression models, the normalized power likelihoods are composed by products of discrete distributions that have no standard form. Although it is feasible to perform likelihood evaluations for each observation, the additional computational burden renders the implementation of the PEP prior methodology time-consuming and inefficient. 
One possible computational solution to the problem would be to utilize an exchange-rate algorithm for doubly-intractable distributions \citep{Murray_etal_2006}.
However, this approach would further increase MCMC computational costs. 

Here we pursue a more generic approach for the implementation of PEP methodology in GLMs 
by redefining the prior itself. 
Namely, we consider two adaptations of the PEP prior which, in principle, can be applied to any statistical model and, consequently, are applicable to all members of the exponential family. 
For the remainder of this paper, without loss of generality we restrict the scale parameter $\phi$ to be known, which is the case for the  binomial, Poisson and normal with known error variance regression models. 
Moreover, in order to alleviate notation we remove $\phi$ from all conditional expressions in the following of the paper.

The core idea is to use the unnormalized power likelihood (\ref{pl}) and (\ref{pl0}), 
i.e. set \linebreak $k_{\vg} (\betab, \delta )=k_{0} (\beta_0, \delta )=1$, 
and normalize the baseline posterior density (\ref{power_prior}) resulting in 
\begin{eqnarray}
\pi_{\vg}^{\N}( \betab | \by^*,  \delta )  
= \frac{     f_{\vg}( \by^*|\betab  )^{\invd}\pi_{\vg}^{\N}(\betab)                 }
{\int f_{\vg}( \by^*|\betab  )^{\invd}\pi_{\vg}^{\N}(\betab)\mathrm{d}\betab }
\label{power_posterior}
\end{eqnarray} 
and accordingly for the reference model $M_0$. This is also the approach of \citet[Eq.4]{Friel_Pettitt_2008} in the definition of the power posteriors which were used for the computation of marginal likelihoods. Given this first step, we proceed by proposing two versions of the PEP prior which differentiate with respect to the definition of the prior predictive distribution used to  
average the baseline posterior in \eqref{power_posterior} across imaginary data sets.
This prior predictive distribution can be alternatively viewed as a hyper-prior assigned to $\by^*$ \citep{Fouskakis_Ntzoufras_2016_jcgs}. 
More specifically we define the two PEP variants as follows. 

\begin{definition} 
	\label{def1} 	
	The {\bf concentrated-reference PEP 
		prior} of model parameters $\betab$ 
	is defined as the power posterior of $\betab$ in \eqref{power_posterior} ``averaged'' over all imaginary data coming from the  
	prior predictive distribution of the reference model $M_0$ based on the actual likelihood, that is
	\begin{eqnarray}
	\hspace{-1.2em}\pi_{\vg}^{\CR}(\betab|\delta) = 
	\mathbb{E}^{m_0^{\N}}_{\by^*}\Big[ \pi_{\vg}^{\N}( \betab | \by^*,  \delta ) \Big] \hspace{-1.2em}&~=~&\hspace{-1.2em} \pi_{\vg}^{\N}(\betab) \int{\frac{m_0^{\N}(\by^*)}{m_{\vg}^{\N}(\by^*|\delta)}f_{\vg}(\by^*|\betab)^{1/\delta}\mathrm{d}\by^*} \label{predictiveCR} \\
	\mbox{\rm~with~} m_0^{\N}(\by^*)  &=&  \int f_0(\by^*|\beta_0)  \pi_0^{\N}(\beta_0) \mathrm{d}\beta_0 \label{m0_ystar} \\ 
	\mbox{\rm~and~} m_{\vg}^{\N}(\by^*|\delta)  &=&  \int f_{\vg}(\by^*|\betab)^{1/\delta}  \pi_{\vg}^{\N}(\betab) \mathrm{d}\betab. \nonumber
	\end{eqnarray}
\end{definition}
In order for the above prior to exist we need to consider for each model $M_{\vg}$ similar assumptions as in  \cite{perez_berger_2002}, i.e.
\begin{equation}
0<m_{\vg}^{\mathrm{N}}( \by^*|\delta)<\infty, ~~~ 0<\int{\frac{m_0^{\N}(\by^*)}{m_{\vg}^{\mathrm{N}}( \by^*|\delta)}f_{\vg}( \by^* | \betab)^{1/\delta}}\mathrm{d}\by^*<\infty.
\label{CRPEP_conditions}
\end{equation}
In equation (\ref{predictiveCR}),  $m_0^{\N}$ will not necessarily be proper, but still, by slightly abusing notation we define the 
concentrated-reference PEP prior as the expectation of  $\pi_{\vg}^{\N}( \betab | \by^*,  \delta )$ with respect to $m_0^{\N}$.
Furthermore, impropriety of the baseline priors in (\ref{predictiveCR}) causes no indeterminacy of the resulting Bayes factors, 
since $\pi_{\vg}^{\CR}(\betab|\delta)$ depends only on the normalizing constant of the baseline prior of the parameter of the null model. 
Finally, the concentrated-reference PEP prior for the parameter of the null model is no longer equal to the baseline prior $\pi_0^{\N}(\beta_0)$, since
\begin{equation}
\pi_{0}^{\CR}(\beta_0|\delta) = 
\pi_0^{\N}(\beta_0) \int{\frac{m_0^{\N}(\by^*)}{m_0^{\N}(\by^*|\delta)}f_0(\by^*|\beta_0)^{1/\delta}\mathrm{d}\by^*}.
\end{equation}

\begin{definition} 
	\label{def2} 
	The {\bf diffuse-reference PEP 
		prior} of model parameters $\betab$ 
	is defined as the power posterior of $\betab$ in \eqref{power_posterior} ``averaged'' over all imaginary data coming from the 
	``normalized" prior predictive distribution of the reference model $M_0$ based on the unnormalized power likelihood, that is
	\begin{eqnarray}
	\hspace{-1.5em}\pi_{\vg}^{\DR}(\betab|\delta) = 
	\mathbb{E}^{m_0^{\Z}}_{\by^*|\delta}\Big[ \pi_{\vg}^{\N}( \betab | \by^*,  \delta ) \Big] \hspace{-1.5em}&~=~&\hspace{-1.5em} \pi_{\vg}^{\N}(\betab) \int{\frac{m_0^{\Z}(\by^*|\delta)}{m_{\vg}^{\N}(\by^*|\delta)}f_{\vg}(\by^*|\betab)^{1/\delta}\mathrm{d}\by^*} \label{predictiveDR} \\
	\mbox{\rm~with~~} m_0^{\Z}(\by^*|\delta)  =  \frac
	{m_{0}^{\N}(\by^*|\delta)}
	{\int m_{0}^{\N}(\by^*|\delta) \mathrm{d}\by^*} &=&  \frac{\int f_{0}(\by^*|\beta_0)^{1/\delta}  \pi_{0}^{\N}(\beta_0) \mathrm{d}\beta_0}{\int \int f_{0}(\by^*|\beta_0)^{1/\delta}  \pi_{0}^{\N}(\beta_0) \mathrm{d}\beta_0 \mathrm{d}\by^*} \nonumber \\
	\mbox{\rm~and~~} m_{\vg}^{\N}(\by^*|\delta)  &=&  \int f_{\vg}(\by^*|\betab)^{1/\delta}  \pi_{\vg}^{\N}(\betab) \mathrm{d}\betab. \nonumber
	\end{eqnarray}
\end{definition}
The conditions for the existence of the diffuse-reference PEP prior, for each model $M_{\vg}$, are similar to (\ref{CRPEP_conditions}), i.e.
\begin{equation}
0<m_{\vg}^{\mathrm{N}}( \by^*|\delta)<\infty, ~~~ 0<\int{\frac{m_0^{\N}(\by^*|\delta)}{m_{\vg}^{\mathrm{N}}( \by^*|\delta)}f_{\vg}( \by^* | \betab)^{1/\delta}}\mathrm{d}\by^*<\infty.
\label{DRPEP_conditions}
\end{equation}
Again the definition of the diffuse-reference PEP prior as an expectation of  $\pi_{\vg}^{\N}( \betab | \by^*,  \delta )$ with respect to $m_0^{\Z}$ 
is slightly abusive under improper baseline prior setups. 
The normalization of $m_0^{\N}(\by^*|\delta)$ is adopted in order to retain the ``expected-posterior'' interpretation 
under proper baseline prior setups. 
The induced normalizing constant 
$$
{\cal C}_0 = \int m_0^{\N}(\by^*|\delta) \mathrm{d}\by^* 
= \int \left\{ \int f_{0}( \by^* | \beta_0)^{1/\delta}\mathrm{d}\by^* \right\}   \pi_0^N(\beta_0) \mathrm{d}\beta_0
$$ 
exists under any proper baseline prior setup and 
has no effect on the posterior variable selection measures  since it is common in all models under consideration. 
Additionally, impropriety of the baseline priors causes no indeterminacy of the resulting Bayes factors, 
since $\pi_{\vg}^{\DR}(\betab|\delta)$ depends only on ${\cal C}_0$ which is common across all models. 
Note that the corresponding normalization is not needed for the CR-PEP since it will be equal to the normalizing constant of the prior 
and therefore equal to one for proper prior distributions. 
Finally, the diffuse-reference PEP prior for the parameter of the null model is no longer equal to the baseline prior, since 
\begin{equation*}
\pi_0^{\DR} (\beta_0|\delta)= \pi_0^{\N}(\beta_0) \frac{\int f_0(\by^*|\beta_0)^{1/\delta}\mathrm{d}\by^*}{{\cal C}_0} =  \frac{\int f_0(\by^*|\beta_0)^{1/\delta} \pi_0^{\N}(\beta_0) \mathrm{d}\by^*}{\int \int f_0(\by^*|\beta_0)^{1/\delta} \pi_0^{\N}(\beta_0) \mathrm{d}\beta_0 \mathrm{d}\by^*}.
\label{DRPEP_for_M0}
\end{equation*}

Definition \ref{def1}  is a special case of Definition \ref{def2} since $m_{0}^{\N}(\by^*)$ is a special case of $m_{0}^{\N}(\by^*|\delta)$ 
with $\delta=1$.
Because the likelihood in \eqref{m0_ystar} is not scaled down, 
it provides more information from the imaginary data 
resulting in a more concentrated (in relation to the alternative approach) 
predictive distribution. For this reason, this version is 
named  \textit{concentrated-reference} PEP (CR-PEP). 
The CR-PEP prior \eqref{predictiveCR} is also given by
\begin{eqnarray}
\pi_{\vg}^{\CR}(\betab|\delta) 
&=& \pi_{\vg}^{\N}(\betab) \int \int \frac{f_{\vg}( \by^*|\betab  )^{\invd}f_0(\by^*|\beta_0)}
{m_{\vg}^{\N}(\by^*|\delta)}   \pi_0^{\N}(\beta_0)\mathrm{d}\by^*\mathrm{d}\beta_0.
\label{CR-PEP}  
\end{eqnarray}   
In Definition \ref{def2}  
the likelihood involved in $m_{0}^{\N}(\by^*|\delta)$ in \eqref{predictiveDR} is raised to the power of $\invd$ and, therefore, 
the information incorporated in the prior predictive distribution 
becomes equal to $n^*/\delta$ points leading to a 
distribution which becomes increasingly diffuse 
as $\delta$ grows. 
Thus, this prior is coined 
\textit{diffuse-reference} PEP (DR-PEP). 
Specifically, we have that 
\begin{eqnarray}
\hspace{-0.8em}\pi_{\vg}^{\DR}(\betab|\delta) 
\hspace{-0.8em}&=&\hspace{-0.8em} {\cal C}_0^{-1} \pi_{\vg}^{\N}(\betab)  \int  \int \frac{f_{\vg}( \by^*|\betab  )^{\invd}f_0(\by^*|\beta_0)^{\invd}}
{m_{\vg}^{\N}(\by^*|\delta)}    \pi_0^{\N}(\beta_0)\mathrm{d}\by^*\mathrm{d}\beta_0. 
\label{DR-PEP}
\end{eqnarray}

In the normal regression case, the DR-PEP prior proposed here 
coincides with the conditional prior formulation of \cite{Fouskakis_Ntzoufras_2016_jcgs}, namely the PCEP prior. 
Assuming a Zellner's $g$-prior as baseline prior for $\betab$ with dispersion parameter $g=g_0$ 
and a reference baseline prior for the variance parameter $\pi(\sigma^2)\propto \sigma^{-2}$, 
then the DR-PEP is given by 
\begin{eqnarray}
\pi_{\vg}^{\DR} \big(\betab| \, \sigma^2 \,\delta, \X_\ell \big) 
& =& f_{N_{d_{\vg}}}\big(\betab; \, \mathbf{0}, \, \V\sigma^2 \big) , 
\label{PEPnormal} \\ 
\V &=& \delta\left( \XT\left[ w^{-1}\I-(\delta\LamdaO+w\HL)^{-1} \right]\XG\right)^{-1}		
\label{PEPnormal_cov}
\end{eqnarray}
with $w=g_0/(g_0+\delta)$, $\HL = \XG(\XT\XG)^{-1}\XT \nonumber$ and 
$\LamdaO = \delta^{-1} \left(\I-\tfrac{w}{n}\IotaN\IotaN^T \right)$. 
The CR-PEP prior has the same form as the DR-PEP in \eqref{PEPnormal}  
and differs only with respect to the variance-covariance matrix with $\LamdaO$ in \eqref{PEPnormal_cov} substituted by $\LamdaOCR=\I-\frac{g_0}{g_0+1}{n}^{-1}\IotaN\IotaN^T$.
Both approaches lead to a consistent variable selection procedure for normal regression models; details are provided in \cite{Fouskakis_Ntzoufras_Perrakis_2016_regression}.

\paragraph{Example:} Let $\dn{y}=(y_1, \dots, y_n)^T$ be a random sample from the exponential distribution with mean $\lambda$ and variance $\lambda^2$. 
We would like to test the hypothesis $H_0: \lambda =\lambda_0$ versus $H_1: \lambda \neq \lambda_0$. 
The baseline (reference) prior under $H_1$ is given by 
$$
\pi_1^N (\lambda) \propto \frac{1}{\lambda} ~.
$$
Let $\dn{ y }^*=(y_1^*, \dots, y_{n^*})^T$ be a training (imaginary) sample of size $n^*$: $1 \leq n^* \leq n$. For the following analysis we use $M_0$ (the model under the null hypothesis) as the reference model.

Under the original PEP methodology, the marginal likelihood, under the null hypothesis, is given by
$$
m_0^N(\dn{y} | \delta) \propto \frac{1}{\lambda_0^{n/\delta}} \exp\left(-\frac{1}{\lambda_0} \frac{1}{\delta} \sum_{i=1}^{n} y_i \right). 
$$
For the CR-PEP we consider $m_0^N(\dn{y} | \delta=1)$, while for the DR-PEP we consider the density normalized version of $m_0^N(\dn{y} | \delta)$, 
denoted by $m_0^Z(\dn{y} | \delta)$; see Definition \ref{def2}.

The posterior distribution of $\lambda$, under the baseline prior of $H_1$, for the CR and DR-PEP methods is given by 
$$
\lambda | \by  \sim \mbox{Inverse-Gamma} \Big( \frac{n}{\delta}, ~\frac{1}{\delta} \sum_{i=1}^{n} y_i \Big), 
$$
while for the original PEP (with the density normalized power likelihood) is given by 
$$
\lambda | \by  \sim \mbox{Inverse-Gamma} \Big( n, ~\frac{1}{\delta} \sum_{i=1}^{n} y_i \Big). 
$$

From the above we can obtain the original PEP prior (with the density normalized power likelihood) and the CR/DR-PEP priors:
\begin{eqnarray*}
	&&\mbox{PEP:  }  \frac{\lambda}{\lambda_0} \sim B' (n^*, n^*) \\
	&&\mbox{CR-PEP:  }  \frac{\lambda}{\delta \lambda_0} \sim B' \left(n^*, \frac{n^*}{\delta}  \right) \\
	&&\mbox{DR-PEP:  }  \frac{\lambda}{\lambda_0} \sim B' \left(n^*, \frac{n^*}{\delta}  \right), 
\end{eqnarray*}
where $B'(a,b)$ denotes the beta prime distribution with parameters $a$ and $b$ and p.d.f. \linebreak
$f(x)= x^{a-1}(1+x)^{-a-b}/B(a,b)$. Under this scenario the EP prior coincides with the original PEP prior.

Table \ref{exp} presents the prior mean and variance, under the alternative hypothesis, for the different PEP formulations. 
For fixed values of $\delta$, the variance of $\lambda$ under the original PEP and the DR-PEP priors  shrinks to zero as $n^*$ grows. 
Therefore, for large $n^*$,  the prior distributions degenerate to the value of $\lambda_0$ with probability one.
On the other hand, both the mean and the variance of $\lambda$ under either the CR or the DR-PEP priors are not defined 
for the default choice of $\delta=n^*$. 
But even for finite prior variances, 
the DR-PEP prior is more dispersed than the CR-PEP prior for any $\delta>1$ since 
$\mbox{Var}(\lambda \, | \, \mbox{DR-PEP})=\delta^2 \, \mbox{Var}(\lambda \, | \, \mbox{CR-PEP})$. 
Finally, when $\delta=an^* < n^*/2$, the prior variance of the CR-PEP 
converges to $a(1-2a)^{-1}(1-a)^{-2}$ for large $n^*$, 
while the corresponding variance of the DR-PEP grows with the same rate as ${n^*}^2$.

\begin{table}[!ht]
	\begin{center}
		\begin{tabular}{l|c|c} 
			\hline 
			Prior & Mean ($n^* > \delta$) & Variance ($n^* > 2 \delta$)\\ 
			\hline 
			&&\\[-0.8em]
			EP \& PEP prior              & $\frac{n^* }{n^*-1}\lambda_0$   &  $\frac{n^* (2n^*-1)}{(n^*-1)^2 (n^*-2)} \lambda_0^2$
			\\[0.5em]         
			CR-PEP prior              & $\frac{n^* }{n^*-\delta} \lambda_0$ & $\frac{1}{n^*}\frac{1+\delta-\delta/n^*} 
			{(1-\delta/n^*)^2 (1-2\delta/n^*)} \lambda_0^2$ 
			\\[0.5em]
			DR-PEP prior              & $\frac{\delta n^* }{(n^*-\delta)}\lambda_0$  
			&   $\frac{1}{ n^*}\frac{1+\delta-\delta/n^*} 
			{(1-\delta/n^*)^2 (1-2\delta/n^*)} \delta^2 \lambda_0^2$ 
			\\[0.5em]
			\hline
		\end{tabular}
	\caption{Prior mean and variance, under the alternative hypothesis, for the exponential case for different PEP variations}
	\label{exp} 
	\end{center}
\end{table}

\subsection{Further prior specifications} 
\label{other_priors}

To complete the model formulation we need to specify a baseline prior for $\betab$, under each model $M_{\vg}$,  
and also a prior distribution on the model space ${\cal M} = \{ 0, 1\}^p$, as we are interested in variable selection rather than single-model inference conditionally on the specific choice of  $\vg \in \mathcal{M}$. In addition, we do not need to specify a prior for $\phi$, which is considered known in our setting.   
For models with random $\phi$, but common across all models, 
we propose working along the lines of \cite{Fouskakis_Ntzoufras_2016_jcgs} 
and use a flat prior on $\phi$; this will just add one additional step to the MCMC algorithm presented in Section \ref{pep_gvs}.


Common choices for the baseline prior of the regression vector $\betab$ are either the flat improper prior 
\begin{equation}
\pi_{\vg}^{\N}(\betab)\propto 1 
\label{baseline_prior1}
\end{equation} 
or Jeffreys prior for GLMs \citep{ibrahim_laud_91} which is of the form
\begin{equation}
\pi_{\vg}^{\N}(\betab)\propto|\X_{\vg}^T\W(\betab)\X_{\vg}|^{1/2}~.
\label{baseline_prior2}
\end{equation}
For non-Gaussian GLMs, Jeffreys prior will depend on $\betab$ through the matrix $\W(\cdot)$; see Section \ref{section_PEP} for details. 
Note that Jeffreys prior for the parameter of the null model simplifies to
$\pi_0^{\N}(\beta_0)\propto \mathrm{tr}\big(\mathbf{W}_0(\beta_0)\big)^{1/2}$. 


A usual prior choice for $\vg$ is to use a product Bernoulli distribution where the prior inclusion probability of each predictor is equal to $0.5$. This leads to a discrete uniform prior on the model space, i.e. 
\begin{equation}
\pi(\vg)= 2^{-p}. 
\label{uniform_model}
\end{equation}
An alternative choice, better suited for moderate to large $p$, is to use the hierarchical prior design 
$$
\vg|\tau\sim\mathrm{Bernoulli}(\tau) \mbox{~and~}  \tau\sim\mathrm{Beta(1,1)},
$$ 
in order to account for an appropriate multiplicity adjustment \citep{scott_berger_2010}. In this case the resulting prior is given by 
\begin{equation}
\pi(\vg)=\frac{1}{p+1}\binom{p}{p_{\vg}}^{-1}.
\label{betabin_model}
\end{equation}

\section{Posterior Inference}
\label{pep_post_inf}

\subsection{Posterior distribution under the PEP prior}
\label{posterior}

In normal linear regression models the conditional PEP prior   
is a conjugate normal-inverse gamma distribution which leads to fast and efficients computations 
\citep{Fouskakis_Ntzoufras_2016_jcgs}. 
For non-Gaussian GLMs there exist no convenient conjugate distributions and the integrals involved in the derivation of the CR/DR-PEP priors are intractable. 
However, one can work with the hierarchical model, i.e. without marginalizing over the imaginary data, 
and use an MCMC algorithm in order to sample from the joint posterior distribution of $\betab$
and $\by^*$.  

For ease of exposition, for the remainder of this section we use the indicator $\psi$ to distinguish between the CR-PEP prior ($\psi=1$) and the DR-PEP prior ($\psi=\delta$) and we simply use the general term ``PEP'' to denote the joint posterior. Specifically, from \eqref{power_posterior}, \eqref{predictiveCR} and \eqref{predictiveDR} we have the following hierarchical form 
\begin{eqnarray}
\pi^{\PEP}_\vg ( \betab, \by^*|\by, \delta ) 
\hspace{-0.5em} &~\propto~& \hspace{-0.5em}  f_\vg(\by|\betab)\pi_\vg^{\N} ( \betab |\by^*,\delta) m_0^{\mathrm{N}}(\by^* | \psi ) \nonumber \\ 
\hspace{-0.5em} &\propto& \hspace{-0.5em} f_\vg(\by|\betab)\frac{f_\vg( \by^* | \betab)^{\invd}
	\pi_\vg^{\mathrm{N}}(\betab)}{m_\vg^{\mathrm{N}} (\by^*|\delta)} 
m_0^{\mathrm{N}}(\by^* | \psi ),
\label{posterior1}
\end{eqnarray}
where $m_0^{\mathrm{N}}(\by^* | 1 ) \equiv m_0^{\mathrm{N}}(\by^*)$.
A further computational problem in \eqref{posterior1} relates to the prior predictive distributions $ m_\vg^{\mathrm{N}}(\by^* |\delta )$ and $m_0^{\mathrm{N}}(\by^* |\psi )$ which are not available in closed form. One solution is to use a Laplace approximation for both. 
Alternatively, a more accurate solution can be obtained by augmenting the parameter space further and include   $\beta_0$ of   $M_0$ in the joint posterior, thus avoiding to use an approximation of $m_0^{\mathrm{N}}(\by^*|\psi)$. Based on \eqref{CR-PEP} and \eqref{DR-PEP} the posterior in \eqref{posterior1} is expanded as  
\begin{equation}
\pi^{\PEP}_\vg ( \betab, \beta_0,\by^*|\by, \delta ) 
\propto  f_\vg(\by|\betab)\frac{f_\vg( \by^* | \betab) ^{\invd}
	\pi_\vg^{\mathrm{N}}(\betab)}{m_\vg^{\mathrm{N}} (\by^*|\delta)}   
f_0(\by^* | \beta_0)^{1/\psi} \pi_0^{\mathrm{N}}(\beta_0),
\label{posterior2}
\end{equation}
which leaves us with the need of using only one  Laplace approximation for $m_\vg^{\mathrm{N}} (\by^*|\delta)$.

Sampling from \eqref{posterior2} for a single model $M_\vg$  is feasible using standard Metropolis-within-Gibbs algorithms. Note that under flat baseline priors the
posterior in \eqref{posterior2} and the corresponding MCMC scheme are simplified. 
Moreover, under a flat baseline prior one may also consider using the normal approximation in \eqref{asympt} for the entire fraction appearing in \eqref{posterior2}, instead of using a Laplace approximation for the prior predictive $m_\vg^{\mathrm{N}} (\by^*|\delta)$.
For variable selection,
which is the topic of the next section, we further assign a prior on $\vg$, based on the options discussed in Section \ref{other_priors}, and utilize the algorithm of \cite{dellaportas_etal_02}; see Section \ref{pep_gvs}. 

\subsection{Gibbs variable selection under the PEP prior} 
\label{pep_gvs}

The Gibbs variable selection \citep[GVS;][]{dellaportas_etal_02} method is a stochastic search algorithm based on the vector of binary indicators $\vg \in\{0,1\}^p$ which represents which of the $p$ covariates are included in a model.     
To formulate GVS we need to partition the regression vector $\boldsymbol \beta$ into  $(\betab,\betabg)$, corresponding to those components of $\boldsymbol\beta$ that are included 
and excluded
from the model, i.e. $\beta_j \in \betab$ if $\gamma_j=1$ and   
$\beta_j \in \betabg$ if $\gamma_j=0$, for $j=1,\dots,p$.
As we assume that the intercept term is included in all models under consideration, $\betab$ and $\betabg$ are of dimensionality $d_{\vg}=p_{\vg}+1$ and $d_{\setminus\vg}=p-p_{\vg}$, respectively. 

Under the GVS setting the joint prior of $\boldsymbol \beta$ and $\vg$ is specified 
as follows
\begin{equation}
\pi (\boldsymbol\beta,\vg)=\pi^\N_\vg(\boldsymbol\beta)\pi(\vg)=\pi^\N_\vg(\betab)\pi^{\N}_\vg(\betabg)\pi(\vg),
\label{GVSprior}
\end{equation} 
where the actual baseline prior choice involves only $\betab$, since $\pi^{\N}_\vg(\betabg)$ is just a \textit{pseudo-prior} used for balancing the dimensions between model spaces; see \cite{dellaportas_etal_02}. 
Suitable choices for the priors of $\betab$ and $\vg$ 
have been discussed in Section \ref{other_priors}, thus, 
in order to complete the GVS setup, we only need to specify the pseudo-prior for  the inactive part of the regression vector $\boldsymbol \beta$. 
In particular, we use a multivariate normal distribution of dimensionality $d_{\setminus\vg}$,  
with parameters specified by the ML estimates; namely, 
\begin{equation}
\pi^\N_\vg(\betabg)=\N_{d_{\setminus\vg}}\left( \widehat{\boldsymbol\beta}_{\setminus\vg}, \mathbf{I}_{d_{\setminus\vg}} \widehat{\sigma}_{\betabg}^2\right),
\label{pseudo}
\end{equation}
where $\widehat{\boldsymbol\beta}_{\setminus\vg}$ and $\widehat{\sigma}_{\betabg}$ are the respective ML estimates and  corresponding standard errors of $\betabg$ from the full model using the actual data $\by$ and $\mathbf{I}_{d_{\setminus\vg}}$ is the $d_{\setminus\vg} \times d_{\setminus\vg}$ identity matrix.
Based on this formulation, the full augmented posterior is
\small
\begin{equation}
\pi( \betab, \betabg, \beta_0,\by^*, \vg|\by, \delta ) 
\propto
f_\vg(\by|\betab)  
\frac{f_\vg( \by^* | \betab) ^{1/\delta} f_0(\by^* | \beta_0  )^{1/\psi} }{m_\vg^{\mathrm{N}} (\by^*|\delta)}  
\pi_\vg^{\mathrm{N}}(\betab) \pi^\N_\vg(\betabg) \pi(\vg)
\pi_0^{\mathrm{N}}(\beta_0),
\label{augment}
\end{equation}
\normalsize
where, as a reminder, $\psi=1$ in the CR-PEP setting and $\psi=\delta$ in the DR-PEP setting.

The proposed PEP-GVS sampling scheme is the following: 
\begin{description}
	\item[~] Set starting values $\vg^{(0)}, \bb^{(0)}=(\betab^{(0)},\betabg^{(0)}), \beta_0^{(0)}$ and $\by^{*(0)}$.  
	\item[~] For iterations $t=1,2,...,N$:
	\begin{description}
		
		\item [Step 1:] Set current values equal to $\bb=\bb^{(t-1)}$, $\beta_0 = \beta_0^{(t-1)}$ $\vg=\vg^{(t-1)}$ and $\by^*=\by^{*(t-1)}$. 
		
		\item [Step 2:] For $j=1,2,...,p$, sample  
		$\gamma_j \sim \pi \big( \gamma_j \big|  \bb,  \vg_{\setminus j}, \by^{*}, \by, \delta\big)$ 
		for $\gamma_j \in \{0,1\}$.
		
		\item [Step 3:] Update $\bb=(\betab,\betabg)$ based on the current configuration of $\vg$. 
		
		\item [Step 4:] Sample the active effects 
		$\betab \sim \pi \big( \betab \big|  \vg,  \by^{*}, \by, \delta \big)$ using a Metropolis-Hastings step.
		
		\item[Step 5:] Sample the inactive effects $\betabg$ from the pseudo-prior in (\ref{pseudo}). 
		
		\item[Step 6:] Sample $\beta_0$ from 
		$\pi (\beta_0 | \by^*, \psi) \propto f_0(\by^* | \beta_0)^{1/\psi}\pi_0^{\mathrm{N}}(\beta_0)$ 
		using a  Metropolis-Hastings step. 
		
		\item[Step 7:] Sample $\by^*$ from 
		$$
		\pi( \by^* | \betab, \beta_0, \vg, \delta, \psi ) \propto
		\frac{f_\vg( \by^* | \betab)^{\invd} f_0(\by^* | \beta_0)^{1/\psi} }{m_\vg^{\mathrm{N}} (\by^*|\delta)}   
		$$
		using a Metropolis-Hastings step. 
		
		\item [Step 8:] Update the parameter values at iteration $t$ as  $\bb^{(t)}=\bb$, $\beta_0^{(t)}=\beta_0 $ $\vg^{(t)}=\vg$ 
		and $\by^{*(t)}=\by^*$. 
		
	\end{description} 
\end{description} 
Note that the generation of $\vg$ and $\betabg$ (Steps 2 and 5) is straightforward 
since the corresponding conditional distributions are of known form. 
For the rest of parameters, $\betab$, $\beta_0$ and $\by^*$, we use Metropolis-Hastings steps. Implementation details and an analytic description of the algorithm are provided in Appendix \ref{PEPGVS_algorithm}.

\section{Hyper-$\delta$ Extensions}
\label{Sec4}
\label{PEP_hyper_delta}

The PEP prior for the normal regression model 
can be interpreted as a mixture of $g$-priors 
where the power parameter $\delta$ is equivalent to $g$ and the mixing density is the prior predictive of the reference model \citep{fouskakis_etal_2015}. Thus, under the PEP approach we assign a hyper-prior on the imaginary data $\by^*$, 
rather than to the variance multiplier, i.e. the power parameter $\delta$. As discussed in Section \ref{section_PEP}, the same representation holds asymptotically in the GLM setting given a flat baseline prior.

A natural extension of the PEP methodology arises by introducing an extra hierarchical level to the model formulation 
via the assignment of a hyper-prior on $\delta$. 
Moving from a fixed (but reasonable) choice of $\delta$ to a stochastic version of this parameter is desirable 
since it simplifies prior specifications by letting the data to ``speak'' for $\delta$ leading, eventually,  to a fully objective procedure. 
Under this approach one can estimate the power parameter 
instead of a-priori set it equal to a fixed predefined value. 
It should be noted, however, that when $\delta$ is not fixed at $\nstar$, then PEP priors loose their unit-information interpretation. 

The hyper-$\delta$ CR/DR-PEP priors can be approximately expressed as  
\begin{equation}
\pi^{\mathrm{CR/DR-PEP}}_{\vg} ( \betab ) \approx\int\int f_{\N_{d_{\vg}}} \Big( \betab ; \bbgmlestar, \delta  \mathcal{\boldsymbol{J}}_{\vg}^*\big(\bbgmlestar\big)^{-1} \Big) m_0^{\Z}(\by^* | \psi )\pi(\delta) \mathrm{d}\by^*\mathrm{d}\delta 
\label{hyper_d}
\end{equation}
under the baseline prior $\pi_{\vg}^{\N}(\betab )\propto 1$, 
where $m_0^{\Z}(\by^* | \psi )$ is equal to $m_0^{\N}(\by^* )$ for $\psi=1$ (CR-PEP) and equal to $ m_0^{\Z}(\by^* | \delta)$ for $\psi=\delta$ (DR-PEP), $\bbgmlestar$ is the ML estimate given the imaginary data,  $\mathcal{\boldsymbol{J}}_{\vg}^*\big(\bbgmlestar\big)$ is given in \eqref{fischer} and $f_{\N_{d_{\vg}}}(\cdot)$ denotes the $d_\vg$--dimensional multivariate normal distribution.
Sensible options for $\pi(\delta)$ are the hyper-$g$ analogues proposed in \cite{liang_etal_2008}. Specifically, we consider the hyper-$\delta$ prior 
\begin{equation}
\pi(\delta)=\frac{a-2}{2}(1+\delta)^{-a/2},~~~for~a>2,~\delta>0,
\end{equation}
which corresponds to a Beta$\big(1,\frac{a}{2}-1\big)$ for the shrinkage factor $\frac{\delta}{1+\delta}$. Thinking in terms of shrinkage, \cite{liang_etal_2008} propose setting $a=3$ in order to place most of the probability mass near 1 or $a=4$ which leads to a uniform prior. An alternative option is the hyper-$\delta/n$ prior given by   
\begin{equation}
\pi(\delta)=\frac{a-2}{2n}\Bigg(1+\frac{\delta}{n}\Bigg)^{-a/2},~~~for~a>2,~\delta>0.
\end{equation}
In principle, any other prior from the related literature can be incorporated in the PEP design; for instance, the inverse-gamma hyper-prior of \cite{Zellner_siow_80} or the recent $g$-prior mixtures proposed by 
\cite{maruama_george_2011} and \cite{Bayarri_etal_2012}. 

Of course, when working outside the context of the normal linear model the integration in \eqref{hyper_d} with respect to $\delta$ will not be tractable. 
Therefore, in order to incorporate the stochastic nature of $\delta$  
we need to introduce one additional MCMC sampling step. 
In this case the augmented posterior is given by 
\begin{equation}
\pi( \betab, \betabg, \beta_0, \by^*, \vg,\delta|\by) 
\propto \pi( \betab, \betabg, \beta_0,\by^*, \vg|\by, \delta )  \pi(\delta),
\label{augment_full}
\end{equation}
where the first quantity in the right-hand side of \eqref{augment_full} is given in \eqref{augment}.
The corresponding full conditionals we need to sample from are
\begin{eqnarray}
\pi^{\mathrm{CR-PEP}} ( \delta | \betab, \beta_0, \phi, \vg, \by^*, \by) 
&\propto &
\frac{ f_\vg( \by^* | \betab )^{1/\delta} \pi(\delta)}{m_\vg^{\mathrm{N}} (\by^*|\delta)},
\label{post_delta_cr} \\
\pi^{\mathrm{DR-PEP}} ( \delta | \betab, \beta_0, \phi, \vg, \by^*, \by) 
&\propto &
\frac{   f_\vg( \by^* | \betab )^{1/\delta}  f_0(\by^* | \beta_0 )^{1/\delta} \pi(\delta) }{m_\vg^{\mathrm{N}} (\by^*|\delta)}. 
\label{post_delta_dr}
\end{eqnarray}
Looking at the above expressions, a subtle point is that $\delta$ is not directly linked to the actual data $\by$; however, it is linked 
indirectly via the posterior values of the parameters of models $M_{\vg}$ (for both approaches) and $M_0$ (for the DR-PEP prior). 
Sampling from \eqref{post_delta_cr} or \eqref{post_delta_dr} is achieved by adding one simple step (after Step 7) in the
PEP-GVS algorithm described in Section \ref{pep_gvs}. 
Specifically, we use a random walk M-H step 
where we propose a candidate value $\delta'$ from 
$$
q(\delta'|\delta)=\mathrm{Gamma} \left( \, \frac{\delta^2}{s^2_\delta} , ~ \frac{\delta}{s^2_\delta} \, \right),
$$ 
which has mean equal to the current value $\delta$ and variance $s^2_\delta$. 
The latter is a tuning parameter which can be specified appropriately in order to have an acceptance rate between $0.2$ and $0.5$, 
as recommended by \cite{Roberts_Rosenthal_2001}. The value of $s^2_\delta=\delta$  proved to be efficient in the examples 
presented in Section \ref{examples}. 
Given this proposal, the new candidate $\delta'$ is accepted with probability 
$\alpha_{\delta}= \min ( 1, A_\delta )$, with $A_\delta$ given by 
\begin{equation*}
A_{\delta}
= \left( \frac{\delta}{\delta'} \right)^{d_\vg/2}
\left[\frac{ f_\vg(\by^{*}|\betab) }{ f_\vg(\by^{*}|\bbgmlestar)} \right]^{ \left\{\frac{1}{\delta'}-\frac{1}{\delta} \right\}}
f_0\big(\by^{*} \big|\beta_0 \big)^{ \left\{\frac{1}{\psi'}-\frac{1}{\psi} \right\} }
\frac{ \pi(\delta') } { \pi(\delta) }
\frac{ q(\delta|\delta') }{ q(\delta'|\delta) },
\end{equation*}
where 
$\psi'=\psi=1$ for the CR-PEP prior and $\psi'=\delta'$, $\psi=\delta$ for the DR-PEP prior; 
$\bbgmlestar$ is the MLE for $\dn{\beta}_\vg$ using data $\by^*$. 
The analytic description of the PEP-GVS algorithm in Appendix  \ref{GVSanalytic} includes the additional sampling step discussed here. 

\section{Desiderata for PEP priors in GLMs}
\label{Desiderata}

\subsection{Model selection consistency} 

With respect to the criteria discussed in \cite{Bayarri_etal_2012} we provided analytical proofs for the null and dimensional predictive matching criteria for all PEP priors proposed in this work.
With respect to model selection consistency, analytical proofs for the normal linear model are provided in \cite{Fouskakis_Ntzoufras_Perrakis_2016_regression}; in this work, we present empirical evidence suggesting that this criterion is also valid by the PEP priors under non-Gaussian GLMs. 
For further details and results, we defer the reader to Section \ref{sec_sim1} where we illustrate, 
for several simulation scenarios with binomial and Poisson response models, 
that the posterior probability of the true model approaches one as the sample size increases.


\subsection{Information consistency}

The definition of information consistency is unclear under GLMs with known dispersion parameters.
According to \cite{li_clyde_2016}, for models with discrete responses and known variance (such as the Poisson and binomial models), 
information inconsistency, as defined by \cite{Bayarri_etal_2012}, 
is not an issue since the likelihood is bounded even for saturated models.




\subsection{Predictive matching}

Under reasonable baseline assumptions, both the CR and the DR-PEP priors are satisfying the criteria of null as well as dimension predictive matching as defined in \cite{Bayarri_etal_2012}. 
To illustrate this, we express the baseline prior of $\betab$ as a product of  the functions $\psi(\dn{\eta}_\vg)$ and $\Psi_\vg(\dn{\beta}_{\setminus 0, \bg})$, where  
$\dn{\eta}_\vg =( \eta_{\vg (i)}, \eta_{\vg (2)}, \dots, \eta_{\vg (n)} )^T$ is the linear predictor and  
$\dn{\beta}_{\setminus 0, \bg}$ is the vector of all elements of $\betab$ excluding the intercept $\beta_{0, \vg}$ of  model $M_\vg$. The statements on predictive matching proceed as follows. 

Although in the paper we have considered $\phi$ to be known and therefore we have removed it from all conditional prior expressions, 
in this section we reintroduce it in order to make our findings more general and cover cases where $\phi$ is also under estimation such as 
the normal regression model. 

\begin{prop}
	\label{prop_null_pred}	
	If we consider the choice of $\delta=n^*=n$ and a baseline prior of the form
	\begin{equation}
	\pi^{\N}_\vg( \betab|\phi) = \psi( \dn{\eta}_\vg ) \Psi_\vg( \dn{\beta}_{\setminus 0, \vg}  ),
	\label{req1_predictive_matching}
	\end{equation}
	then the fixed $\delta$ PEP priors satisfy the  null predictive matching criterion 
	for samples of size one. \\ 
\end{prop}

\vspace{-2em}
\noindent{Proof of Proposition \ref{prop_null_pred}} is provided in Appendix \ref{proof1}. {\tiny $\square$}  
\vspace{1em}

\begin{prop}
	\label{prop_null_pred_hyperDR}	
	The hyper-$\delta$ and the hyper-$\delta/n$ DR-PEP priors with $n^*=n$ and  baseline priors of the form  \eqref{req1_predictive_matching} 
	satisfy the  null predictive matching criterion for samples of size one. \\ 
\end{prop}

\vspace{-2em}
\noindent{Proof of Proposition \ref{prop_null_pred_hyperDR}} is provided in Appendix \ref{proof2}.  {\tiny $\square$}  
\vspace{1em}

\begin{prop}
	\label{prop_null_pred_hyperCR}	
	The hyper-$\delta$ and the hyper-$\delta/n$  CR-PEP priors with $n^*=n$ and baseline priors of the form  \eqref{req1_predictive_matching} 
	satisfy the  null predictive matching criterion for samples of size one. \\ 
\end{prop}

\vspace{-2em}
\noindent{Proof of Proposition \ref{prop_null_pred_hyperCR}} is provided in Appendix \ref{proof3}.  {\tiny $\square$}  
\vspace{1em}

\begin{prop}
	\label{prop_dim_pred}	
	If we consider the choice of $n^*=n$ and a baseline prior of the form
	\begin{equation}
	\pi^{\N}_\vg( \betab|\phi) = \psi( \dn{\eta}_\vg ) 
	\label{req2_dim_predictive_matching},
	\end{equation}
	then all versions of DR-PEP priors (the fixed $\delta=n$,  the hyper-$\delta$ and the hyper-$\delta/n$) satisfy the  dimension predictive matching criterion for samples of size $p_\vg+1$. \\ 
\end{prop}

\vspace{-2em}
\noindent {Proof of Proposition \ref{prop_dim_pred}} is provided in Appendix \ref{proof4}. {\tiny $\square$}  
\vspace{1em}

\begin{prop}
	\label{prop_dim_pred2}	
	If we consider the choice of $n^*=n$ and a baseline prior of the form as in \eqref{req2_dim_predictive_matching}, 
	then all versions of CR-PEP priors (the fixed $\delta=n$,  the hyper-$\delta$ and the hyper-$\delta/n$)  satisfy the  dimension predictive matching criterion for samples of size $p_\vg+1$. \\ 
\end{prop}

\vspace{-2em}
\noindent {Proof of Proposition \ref{prop_dim_pred2}} can be obtained by using similar arguments as in the proof of Proposition \ref{prop_dim_pred}. {\tiny $\square$}

Both of the baseline prior distributions proposed in this article satisfy the requirements \eqref{req1_predictive_matching}  and \eqref{req2_dim_predictive_matching} since $\Psi(\dn{\beta}_{\setminus 0, \vg}) = 1$.
Specifically,  for the flat improper prior \eqref{baseline_prior1} we have $\psi(\dn{\eta}_\vg)\propto 1$ and $\Psi(\dn{\beta}_{\setminus 0, \vg}) = 1$ while for the Jeffreys prior \eqref{baseline_prior2}  
we have $\Psi(\dn{\beta}_{\setminus 0, \vg})= 1$ and 
\begin{eqnarray*}
	\psi(\dn{\eta}_\vg) &\propto&|\X_{\vg}^T\W(\dn{\eta}_\vg)\X_{\vg}|^{1/2} \mbox{~with~}  
	\W(\dn{\eta}_\vg)= \diag\big(w_{{\vg} (i)}\big) \mbox{~and~}\\
	&& w_{{\vg} (i)} =  \left(  \frac{ \partial \mu \big( \eta_{\vg, (i)} \big) }{\partial \eta_{\vg, (i)} } \right)^2 \big[a_i( \phi ) b''\big( \vartheta(\eta_{\vg, (i)} \big) \big]^{-1},~ 
	\mu \big( \eta_{\vg, (i)} \big)=b'\big( \vartheta\big(\eta_{\vg, (i)}\big) \big).  
\end{eqnarray*}

\section{A General Framework}
\label{Framework}

In this section we present a synopsis for the various priors under consideration. This requires introducing a set of separate power parameters $\delta_0$ and $\delta_1$, which respectively relate to the marginal likelihood and the posterior distribution components. 
Under this setting we have the following general prior formulation 
$$
\pi^\mathrm{G}( \dn{\theta}_\vg, \dn{\omega}, \delta_0, \delta_1 ) = 
\pi^\mathrm{G}( \dn{\theta}_\vg | \dn{\omega}, \delta_0, \delta_1 ) \pi( \dn{\omega} ) \pi(\delta_0) \pi(\delta_1), 
$$
where $\mathrm{G} \in \cal{P}$ with $\cal{P}$ being the set of PEP prior configurations considered in this paper, also including the EP prior. Here, $\dn{\theta}_\vg$ correspond to the model specific parameters, while $\dn{\omega}$ is a common nuisance parameter across all models.
When $\dn{\omega}$ does not exist or is known, $\pi( \dn{\omega} )$ should be omitted. Similarly, when $\delta_0$ and/or $\delta_1$ are fixed, $\pi(\delta_0)$ and/or $\pi(\delta_1)$ are omitted.

All priors in the set $\cal{P}$ are derived as follows: 
\begin{equation}
\label{general} 
\pi^\mathrm{G}( \dn{\theta}_\vg | \dn{\omega}, \delta_0, \delta_1) 
=  
\frac{\pi_\vg^\N( \dn{\theta}_\vg | \dn{\omega} ) }{k_\vg( \dn{\theta}_\vg, \dn{\omega}, \delta_1 ) {\cal C}_0}
\int \frac{m_0^\N( \by^* | \dn{\omega}, \delta_0 ) }{m_\vg^\N( \by^* | \dn{\omega}, \delta_1 ) } 
f_\vg ( \by^* | \dn{\theta}_\vg, \dn{\omega} )^{1/\delta_1} d\by^*
\end{equation}
with 
\[
m_\vg^\N( \by^* | \dn{\omega}, \delta_1 ) = 
\int \frac{ f_\vg( \by^* | \dn{\theta}_\vg, \dn{\omega} )^{1/\delta_1} }{k_\vg( \dn{\theta}_\vg, \dn{\omega}, \delta_1 )}
\pi_\vg^\N( \dn{\theta}_\vg | \dn{\omega} ) d\dn{\theta}_\vg
\]
and
\[
m_0^\N( \by^* | \dn{\omega}, \delta_0 ) = 
\int \frac{ f_0( \by^* | \dn{\theta}_0, \dn{\omega} )^{1/\delta_0} }{k_0( \dn{\theta}_0, \dn{\omega}, \delta_0 )}
\pi_0^\N( \dn{\theta}_0 | \dn{\omega} ) d\dn{\theta}_0.
\]
Each prior in the set $\cal{P}$ can be obtained from (\ref{general}); details are provided in Table \ref{general_table}. In Table \ref{issues_solutions} we summarize issues and proposed solutions for all priors under consideration.

\begin{table}[h!b!t!]
	\centering{}%
	\footnotesize
	\vspace{0.5em} 
	\begin{tabular}{l|c|c|ccc|c|c|c}
		\hline 
		Prior (G)	              & $\dn{\theta}_\vg$ & $\dn{\omega}$ & $\delta_0$ & $\delta_1$ & Hyper-prior $\pi(\delta)$  & $k_0( \dn{\theta}_0, \dn{\omega}, \delta_0 )$ & $k_\vg( \dn{\theta}_\vg, \dn{\omega}, \delta_1 )$ &  ${\cal C}_0$ \\[0.5em] 	 \hline 
		EP	                      & $\betab, \phi_\vg$& $\emptyset$   &$1$         & $1$        &  & 1 & 1 & 1 \\[0.5em] 	
		PEP                       & $\betab, \phi_\vg$& $\emptyset$   & $n^*$      & $n^*$      &  & $\kappa_0$& $\kappa_1$ & 1 \\[0.5em] 
		PCEP                      & $\betab$          & $\phi$        & $n^*$      & $n^*$      &  & $\kappa_0$& $\kappa_1$ & 1 \\[0.5em] 
		CR-PEP	                  & $\betab$          & $\phi$        & $1$        & $\nstar$   &  & 1 & 1 & 1 \\[0.5em] 
		DR-PEP	                  & $\betab$          & $\phi$        & $n^*$      & $n^*$      &  & 1 & 1 & $c_0$ \\[0.5em] 
		CR-PEP hyper-$\delta$	  & $\betab$          & $\phi$        & $1$        & $\delta$  & $\frac{a-2}{2}(1+\delta)^{-a/2}$ & 1 & 1 & 1 \\[0.5em] 
		DR-PEP hyper-$\delta$	  & $\betab$          & $\phi$        & $\delta$   & $\delta$  & $  \frac{a-2}{2}(1+\delta)^{-a/2}$ & 1 & 1 & $c_0$ \\[0.5em] 
		CR-PEP hyper-$\delta/n$	  & $\betab$          & $\phi$        & $1$        & $\delta$  & $ \frac{a-2}{2n}(1+\frac{\delta}{n})^{-a/2}$  & 1 & 1 & 1 \\[0.5em] 
		DR-PEP hyper-$\delta/n$	  & $\betab$          & $\phi$        & $\delta$ & $\delta$ & $\frac{a-2}{2n}(1+ \frac{\delta}{n})^{-a/2}$ & 1 & 1 & $c_0$\\ \hline  
		\multicolumn{9}{p{15cm}}{ }\\[-0.5em]
		\multicolumn{9}{p{15cm}}{ \scriptsize  $\kappa_0=\int f_0(\by^* | \dn{\theta_0}, \dn{\omega} )^{1/\delta_0} d\by^*$;~~
			$\kappa_1=\int f_\vg(\by^* | \dn{\theta_\vg}, \dn{\omega} )^{1/\delta_1} d\by^*$; ~~
			$c_0=\int\int f_0(\by^* | \dn{\theta_0}, \dn{\omega} )^{1/\delta_0} \pi_0^N( \dn{\theta}_0 | \dn{\omega} ) d \dn{\theta}_0d\by^*$.}\\[0.5em]
	\end{tabular} 
	\normalsize 
	\caption{Schematic presentation of all priors in $\cal{P}$}
	\label{general_table}
\end{table}

\begin{table}[h!b!t!]
	\centering{}%
	\footnotesize  
	\begin{tabular}{p{3.6cm}|p{5.5cm}|p{5.5cm}}
		\hline 
		\textbf{Prior}	& \textbf{Issues} & \textbf{Solutions} \\  \hline 
		\multirow{5}{5cm}{EP}	            &   -- Selection of imaginary sample size $n^*$  \newline 
		-- Sub-sampling of $\dn{X}^*_\vg$ \newline  
		-- Informative when using minimal training  sample and $p$ is close to $n$
		&  \multirow{5}{5cm}{-- Issues are solved using $\PEP$ with $\delta=n^*=n$ and $\dn{X}^*_\vg=\dn{X}_\vg$}\\ 	\hline 
		\multirow{6}{4cm}{PEP}	&  -- Cumbersome normalized power likelihood in GLMs  \newline 
		-- Monte Carlo is needed for the computation of the marginal likelihood even in the normal linear model 
		&  -- Use of unnormalized power likelihoods that lead to the CR/DR-PEP priors   \newline 
		-- Use PCEP that leads to a conjugate setup in the normal linear model  \\ \hline 
		\multirow{2}{3cm}{PCEP} 	&  \multirow{2}{6cm}{-- Not information consistent} & -- Use PEP which is information \newline consistent   \\ \hline 
		\multirow{4}{4cm}{CR-PEP}        &  -- There is no clear definition of $m_0^\N$ under the unnormalized power likelihood\newline  
		-- Selection of $\delta$ 
		& --   Use the original likelihood in $m_0^\N$ \newline 
		--   Set $\delta=n^*$ to have unit information \quad interpretation or consider random $\delta$ \\ \hline 
		\multirow{5}{3cm}{DR-PEP}       & -- There is no clear definition of $m_0^\N$ under the unnormalized power likelihood\newline  
		\multirow{1}{3cm}{-- Selection of $\delta$} 
		& --   Use the density normalized $m_0^\mathrm{Z}$ \quad under the unnormalized power likelihood  \newline 
		--   Set $\delta=n^*$ to have unit information\quad interpretation or consider random $\delta$ \\ \hline 
		\multirow{3}{4cm}{CR/DR-PEP hyper-$\delta$  }        & -- Demanding computation \newline 
		-- Prior of $\delta$ is not centered to unit-information  
		& -- Use fixed-$\delta$  CR/DR-PEP versions \newline 
		\multirow{2}{6cm}{-- Use the hyper-$\delta/n$ prior}  \\ 	\hline                                 
		\multirow{2}{5cm}{CR/DR-PEP hyper-$\delta/n$}     & \multirow{2}{5cm}{-- Demanding computation} \newline   &	\multirow{2}{6cm}{-- Use fixed-$\delta$ CR/DR-PEP versions} \\ \hline 
	\end{tabular} 
	\normalsize
	\caption{Issues and solutions of all priors in $\cal{P}$}
	\label{issues_solutions}
\end{table}

\section{Illustrative Examples} 
\label{examples}

\subsection{Methods}
\label{methods}
In this section we firstly present a simulation study for logistic and Poisson regression taking into account independent and correlated predictors as well as different levels of sparsity for the true model. 
In Section \ref{sec_sim2}, we proceed with a simulation study for logistic models where the number of predictors is larger and the correlation structure is more complicated. Section \ref{methods} concludes with a real life example with binary responses. 

In all illustrations we consider the CR-PEP and DR-PEP priors (introduced in Section \ref{Sec3}) and 
their hyper-$\delta$ and hyper-$\delta/n$ extensions (presented in Section \ref{Sec4}) with parameter $a=3$, 
which is one of values proposed in \cite{liang_etal_2008}. 
For all PEP prior configurations we consider $\nstar=n$ and $\X_{\vg}^*=\X_{\vg}$, where the columns of the design matrix are centered around their sample means. 
For fixed $\delta$, we consider the default unit-information approach, that is $\delta=\nstar$. 
The Jeffreys prior, given in \eqref{baseline_prior2}, is used as baseline prior for $\betab$.

We compare the PEP variants with standard $g$-prior methods, using the GLM version of \cite{bove_held_2011} for the parameters of the predictor variables and a flat improper prior for the intercept term. In particular, we consider the unit-information $g$-prior (with $g=n$) and three mixtures of $g$-priors; namely, the hyper-$g$ and hyper-$g/n$ priors with $a=3$ \citep{liang_etal_2008}, and the beta hyper-prior proposed by \cite{maruama_george_2011}. Henceforth, the latter will be referred to as MG hyper-$g$. 
Stochastic model search under these approaches is also implemented via GVS sampling.

\subsection{Simulation study 1}
\label{sec_sim1}

In this first example we consider two simulation scenarios for logistic and Poisson regression, presented in \cite{Hansen_Yu_2003} and \cite{Chen_etal_2008}, respectively. Both of these scenarios are also considered by \cite{Li_Clyde_2015}.
The number of predictors is $p=5$ in the logistic model and $p=3$ in the Poisson model, where 
each predictor is drawn from a standard normal distribution with pairwise correlations given by
\begin{equation*}
\mathrm{corr}(X_i,X_j)=r^{|i-j|}, \, \, \, 1\le i < j \le p.
\end{equation*}
Concerning the correlations between covariates we examine two cases: 
(i) independent predictors ($r=0$) and (ii) correlated predictors ($r=0.75$).  
In addition, four sparsity scenarios are assumed; the true data-generating models are summarized in Table \ref{Simulation1}. 
For the logistic case we use the same sample size as in \cite{Hansen_Yu_2003}, namely $n=100$, 
but with lower effects in order to reflect more realistic values of odds ratios.
Given the coefficients in Table \ref{Simulation1}, the odds ratios are approximately equal to 2, 2.5 and 3.5 for the sparse, medium and full models, respectively. 
For the Poisson simulation we use the same regression coefficients as in \cite{Chen_etal_2008}, 
but with sample size equal to $n=100$. Each simulation is repeated 100 times.  
Since the number of predictors 
in both regression models is small, we assign a uniform prior on model space as given in \eqref{uniform_model}. 
\begin{table}[h!b!t!]
	\centering{}%
	\def~{\hphantom{0}}
	{ %
		\begin{tabular}{l|cccccc|cccc}
			\hline
			\multirow{2}{*}{\textbf{Scenario}} & \multicolumn{6}{c|}{\textbf{Logistic}\,\,$(n=100)$}  & \multicolumn{4}{c}{\textbf{Poisson}\, \, ($n=100$)} \tabularnewline
			& $\beta_{0}$ & $\beta_{1}$ & $\beta_{2}$ & $\beta_{3}$ & $\beta_{4}$ & $\beta_{5}$ & $\beta_{0}$ & $\beta_{1}$ & $\beta_{2}$ & $\beta_{3}$\tabularnewline
			\hline 
			null & 0.1 & 0 & 0 & 0 & 0 & 0 & -0.3 & 0 & 0 & 0\tabularnewline
			sparse & 0.1 & 0.7 & 0 & 0 & 0 & 0 & -0.3 & 0.3 & 0 & 0\tabularnewline
			medium & 0.1 & 1.6 & 0.8 & -1.5 & 0 & 0 & -0.3 & 0.3 & 0.2 & 0\tabularnewline
			full & 0.1 & 1.75 & 1.5 & -1.1 & -1.4 & 0.5 & -0.3 & 0.3 & 0.2 & -0.15\tabularnewline
			\hline
	\end{tabular}}
    \caption{Logistic and Poisson regression scenarios for Simulation Study 1 using independent ($r=0$) and correlated ($r= 0.75$) predictors.}
	\label{Simulation1}
\end{table}

\subsubsection*{Comparison between different methods}
Results based on the frequency of identifying the true data-generating model through the maximum a-posteriori (MAP) model for the logistic regression simulation are summarized in Table \ref{logistic_simulation}. 
The comparison between the PEP prior approaches versus the rest of the methods indicates the following:
\begin{itemize}
	\item[i)] Overall the PEP based variable selection    	procedures perform well, since in 5 out of the 8 simulation scenarios the ``best'' prior for identifying the true model is at least one of the PEP priors.
	\item[ii)] The PEP procedures outperform all methods 
	under the null and sparse simulation scenarios.
	\item[iii)] Under the medium model scenario the PEP priors perform equally well to the rest of the methods in the case of independent predictors and slightly worse in the case of correlated predictors.
	\item[iv)]  Under the full model scenario $g$-prior methods perform better than PEP priors. This is no surprise 
	as PEP priors 
	tend to support more parsimonious solutions in general. 
\end{itemize}

With respect to the comparison between the CR-PEP and DR-PEP priors we find no obvious differences between the two approaches for fixed $\delta=n$. Concerning the fixed $\delta$ approach versus the hyper-$\delta$ and $\delta/n$ extensions, we see that under the DR-PEP approach the results are more or less the same in terms of MAP model success patterns. However, this is not the case under the CR-PEP approach as the hyper-$\delta$ prior support more complex models than the fixed-$\delta$ prior, while the hyper-$\delta/n$ prior is somewhere in the middle. Interestingly, a similar pattern is observed among the $g$-prior and the hyper-$g$,  hyper-$g/n$ priors.   
Boxplots of posterior inclusion probabilities can be found in Appendix \ref{PIPs_logistic}; from these results, the DR-PEP based approach is quite robust with respect to the choice between fixed versus random $\delta$, while among the category of $g$-prior mixtures the MG hyper-$g$ prior seems to have the strongest shrinkage effect.

\begin{sidewaystable}
	\centering{}{\scriptsize{}}%
	\begin{tabular}{ll|cccccccccc}
		\hline 
		\multirow{3}{*}{\textbf{\small{}Scenario}} & \multirow{3}{*}{\textbf{\small{}r}} & \multicolumn{10}{c}{\textbf{\small{}Prior distributions}}\tabularnewline
		&  & \multirow{2}{*}{{\small{}$g$-prior}} & {\small{}hyper} & {\small{}hyper} & {\small{}MG hyper} & {\small{}CR} & {\small{}CR PEP } & {\small{}CR PEP } & {\small{}DR} & {\small{}DR PEP } & {\small{}DR PEP }\tabularnewline
		&  &  & {\small{}$g$-prior} & {\small{}$g/n$-prior} & {\small{}$g$-prior} & {\small{}PEP} & {\small{}hyper-$\delta$} & {\small{}hyper-$\delta/n$} & {\small{}PEP} & {\small{}hyper-$\delta$} & {\small{}hyper-$\delta/n$ }\tabularnewline
		\hline 
		\multirow{2}{*}{{\small{}null}} & {\small{}0.00} & {\small{}77} & {\small{}35} & {\small{}63} & 75 & 79 & 46 & 80 & 79 & 73 & \textbf{82}\tabularnewline
		& {\small{}0.75} & 91 & 52 & 81 & 88 & \textbf{94} & 60 & 82 & 93 & 91 & 92\tabularnewline
		\hline 
		\multirow{2}{*}{{\small{}sparse}} & {\small{}0.00} & {\small{}67} & {\small{}57} & {\small{}63} & 67 & \textbf{72} & 58 & 68 & \textbf{72} & \textbf{72} & \textbf{72}\tabularnewline
		& {\small{}0.75} & 74 & 60 & 67 & 72 & 72 & 60 & \textbf{76} & 74 & 73 & 73\tabularnewline
		\hline 
		\multirow{2}{*}{{\small{}medium}} & {\small{}0.00} & {\small{}83} & {\small{}82} & \textbf{84} & \textbf{84} & 83 & \textbf{84} & 81 & 83 & \textbf{84} & \textbf{84}\tabularnewline
		& {\small{}0.75} & 33 & \textbf{38} & 34 & 30 & 26 & 37 & 32 & 27 & 29 & 27\tabularnewline
		\hline 
		\multirow{2}{*}{{\small{}full}} & {\small{}0.00} & {\small{}41} & {\small{}41} & {\small{}42} & \textbf{43} & 28 & 38 & 29 & 26 & 32 & 31\tabularnewline
		& {\small{}0.75} & 14 & 15 & \textbf{17} & 14 & 8 & 12 & 10 & 8 & 10 & 8\tabularnewline
		\hline 
	\end{tabular}
	\protect
	\caption{Number of simulated samples (over $100$ replications) that the MAP model coincides with  the true model in the logistic case of Simulation Study 1 (row-wise largest value in bold).}
	\label{logistic_simulation}
\end{sidewaystable}

The MAP-model results from the Poisson simulations are presented in Table \ref{poisson_simulation}. 
Boxplots of posterior inclusion probabilities under each method and  simulation scenario are presented in Appendix \ref{PIPs_poisson}. 
Overall, conclusions are similar to the logistic case. Specifically, looking at the differences between the PEP priors and all versions of $g$-priors, we conclude to the following:

\begin{itemize}
	
	\item[i)] The PEP procedures perform overall satisfactory; 6 out of the 8 best MAP success patterns are spotted by one of the PEP based methods.
	
	\item[ii)] The PEP procedures perform overall well under sparse conditions, i.e. under the null or the sparse models.
	
	\item[iii)] For the medium complexity scenarios, the hyper-$g$ and hyper-$\delta$ CR-PEP priors yield the best results; however, under the correlated predictors scenario the true model is rarely traced by any method.
	
	\item[iv)]  For the full model with independent covariates, 
	the MAP success rates under all methods are  low; the hyper-$g$ has the highest rate 
	but with the hyper-$\delta$ CR-PEP prior being close and rather competitive. 
	For the full model with correlated covariates, all methods 
	fail; the hyper-$\delta$ CR-PEP prior has the highest success rate which is only 3\%. 
	
\end{itemize} 

With respect to the various PEP prior distributions, the comparison in the Poisson case 
leads to the same findings as in the logistic regression case. 
Again, the most interesting finding is that inference under the DR-PEP prior is not affected by the choice   
of fixed versus random $\delta$. 
On the contrary, this is not the case for the CR-PEP prior, where the hyper-$\delta$ extension systematically supports more complex models. To a lesser extend the same holds for the CR-PEP hyper-$\delta/n$ prior.



\begin{sidewaystable}
	\centering{}{\scriptsize{}}%
	\begin{tabular}{ll|cccccccccc}
		\hline 
		\multirow{3}{*}{\textbf{\small{}Scenario}} & \multirow{3}{*}{\textbf{\small{}r}} & \multicolumn{10}{c}{\textbf{\small{}Prior distributions}}\tabularnewline
		&  & \multirow{2}{*}{{\small{}$g$-prior}} & {\small{}hyper} & {\small{}hyper} & {\small{}MG hyper} & {\small{}CR} & {\small{}CR PEP } & {\small{}CR PEP } & {\small{}DR} & {\small{}DR PEP } & {\small{}DR PEP }\tabularnewline
		&  &  & {\small{}$g$-prior} & {\small{}$g/n$-prior} & {\small{}$g$-prior} & {\small{}PEP} & {\small{}hyper-$\delta$} & {\small{}hyper-$\delta/n$} & {\small{}PEP} & {\small{}hyper-$\delta$} & {\small{}hyper-$\delta/n$ }\tabularnewline
		\hline 
		\multirow{2}{*}{{\small{}null}} & {\small{}0.00} & {\small{}86} & {\small{}68} & {\small{}80} & {\small{}87} & {\small{}88} & {\small{}71} & {\small{}83} & {\small{}90} & {\small{}91} & \textbf{\small{}94}\tabularnewline
		& {\small{}0.75} & {\small{}91} & {\small{}68} & {\small{}90} & {\small{}94} & {\small{}95} & {\small{}75} & {\small{}91} & {\small{}95} & \textbf{\small{}97} & {\small{}95}\tabularnewline
		\hline 
		\multirow{2}{*}{{\small{}sparse}} & {\small{}0.00} & {\small{}75} & {\small{}74} & {\small{}74} & {\small{}75} & {\small{}76} & {\small{}68} & \textbf{\small{}80} & {\small{}73} & {\small{}68} & {\small{}69}\tabularnewline
		& {\small{}0.75} & {\small{}40} & {\small{}43} & {\small{}41} & {\small{}38} & {\small{}35} & \textbf{\small{}44} & {\small{}40} & {\small{}32} & {\small{}30} & {\small{}28}\tabularnewline
		\hline 
		\multirow{2}{*}{{\small{}medium}} & {\small{}0.00} & {\small{}29} & {\small{}43} & {\small{}37} & {\small{}36} & {\small{}27} & \textbf{\small{}44} & {\small{}30} & {\small{}28} & {\small{}25} & {\small{}20}\tabularnewline
		& {\small{}0.75} & {\small{}0} & \textbf{\small{}5} & {\small{}0} & {\small{}0} & {\small{}0} & {\small{}4} & {\small{}0} & {\small{}0} & {\small{}0} & {\small{}0}\tabularnewline
		\hline 
		\multirow{2}{*}{{\small{}full}} & {\small{}0.00} & {\small{}6} & \textbf{\small{}23} & {\small{}13} & {\small{}9} & {\small{}5} & {\small{}18} & {\small{}11} & {\small{}5} & {\small{}4} & {\small{}3}\tabularnewline
		& {\small{}0.75} & {\small{}0} & {\small{}0} & {\small{}1} & {\small{}0} & {\small{}0} & \textbf{\small{}3} & {\small{}0} & {\small{}0} & {\small{}0} & {\small{}0}\tabularnewline
		\hline 
	\end{tabular}
	\protect
	\caption{Number of simulated samples (over $100$ replications) that the MAP model coincides with  the true model for the Poisson case in Simulation Study 1 (row-wise largest value in bold).}
	\label{poisson_simulation}
\end{sidewaystable}

\subsubsection*{Evaluation of model selection consistency of PEP methods} 
\begin{figure}[p]
	\centering{}\includegraphics[scale=0.45]{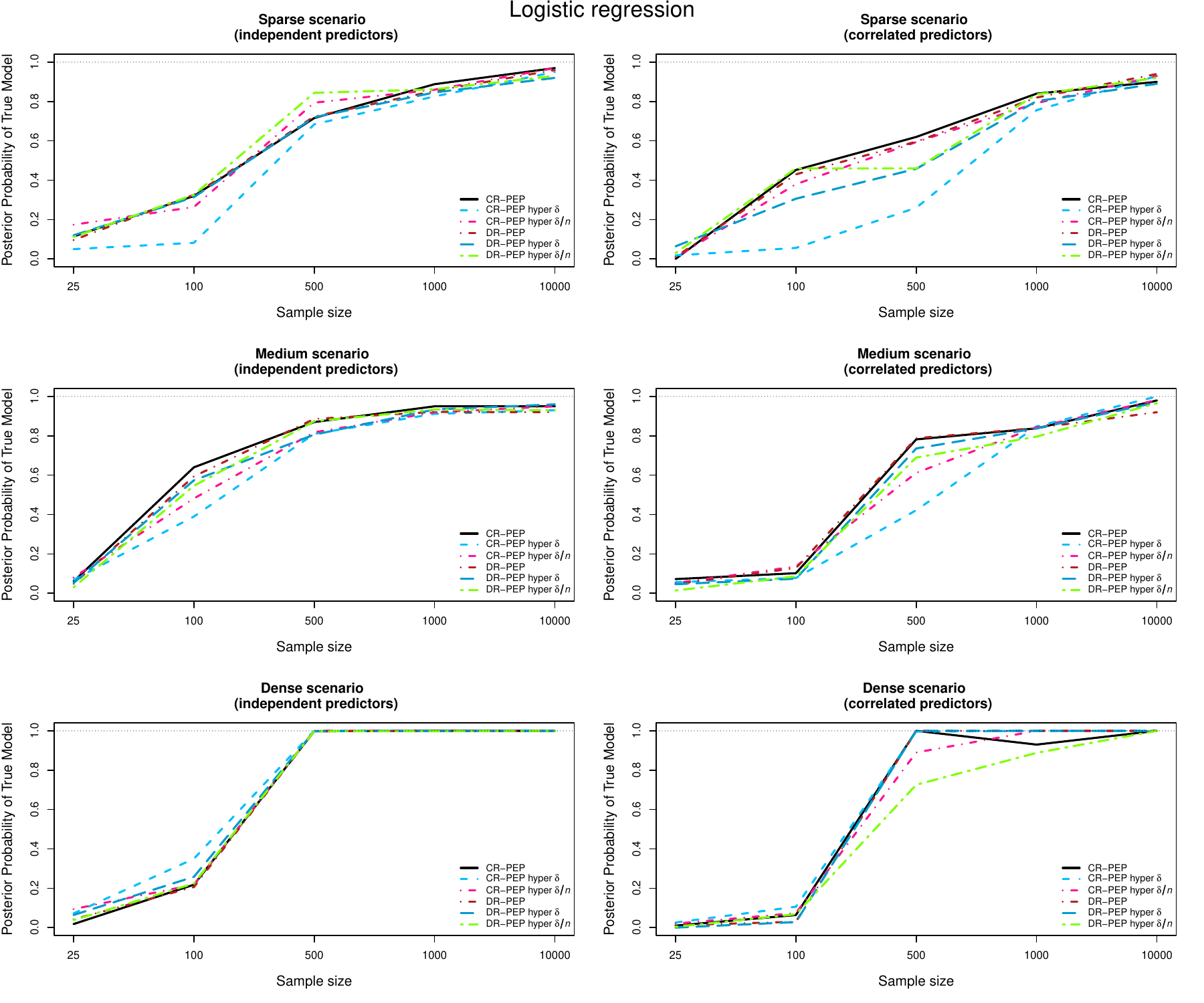}
	\caption{Posterior  probabilities of the true model vs. sample size for the sparse, medium and dense logistic  regression scenarios.}
	\label{plot_consistency1}
\end{figure}

\begin{figure}[p]
	\centering{}\includegraphics[scale=0.45]{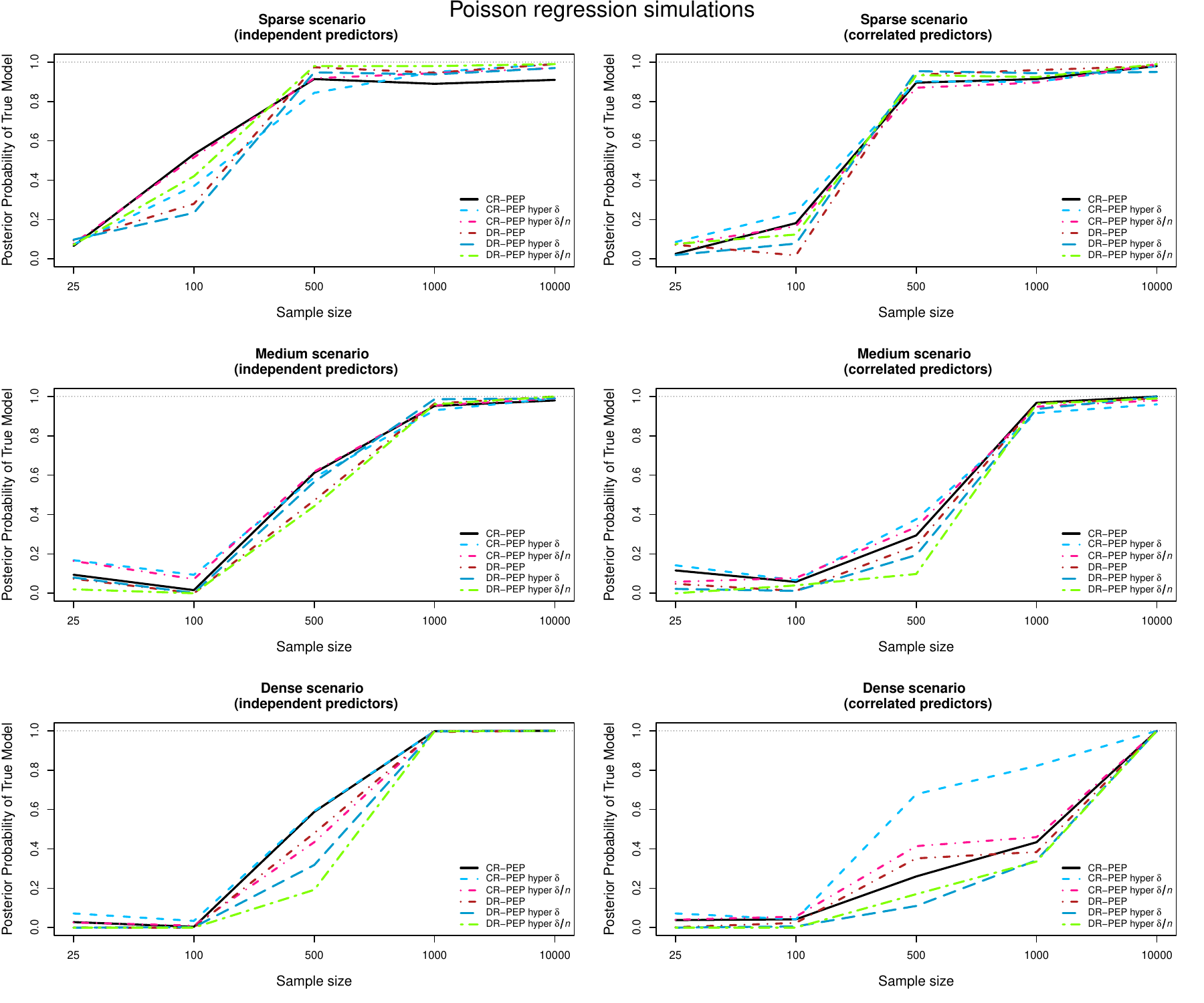}
	\caption{Posterior probabilities of the true model vs. sample size for the sparse, medium and dense Poisson regression scenarios.}
	\label{plot_consistency2}
\end{figure}

We conclude this illustration by studying the behaviour of PEP based methods for different sample sizes. 
Under the assumption of model selection consistency, we expect  that the posterior probability of the true model will approach the value of one as the sample size increases. 
Indeed, all PEP methods for all scenarios under study confirm the consistency criterion as it is evident in 
Figures \ref{plot_consistency1} and \ref{plot_consistency2}.

\subsection{Simulation study 2}
\label{sec_sim2}
In this illustration we consider a more sophisticated scenario with $p=10$ potential predictors ($1024$ models)
and a more intriguing correlation structure.  Similar to \cite{nott_kohn_05}, the first five covariates are generated from a standard normal distribution, while the remaining five covariates are generated from
\begin{equation*}
X_{ij}=N(0.3X_{i1}+0.5X_{i2}+0.7X_{i3}+0.9X_{i4}+1.1X_{i5},1),
\end{equation*} 
for $i=1,\dots,n$ and $j=6,\dots,10$. 
We assume that sample size $n$ is 200 and consider the three logistic regression data-generating models which are summarized in Table \ref{Simulation2}; the resulting odds ratios for the sparse and dense simulation models are approximately equal to 2 and 3, respectively. Each simulation is repeated 100 times.  

\begin{figure}[h!]
	\hspace{3em}\includegraphics[scale=0.37]{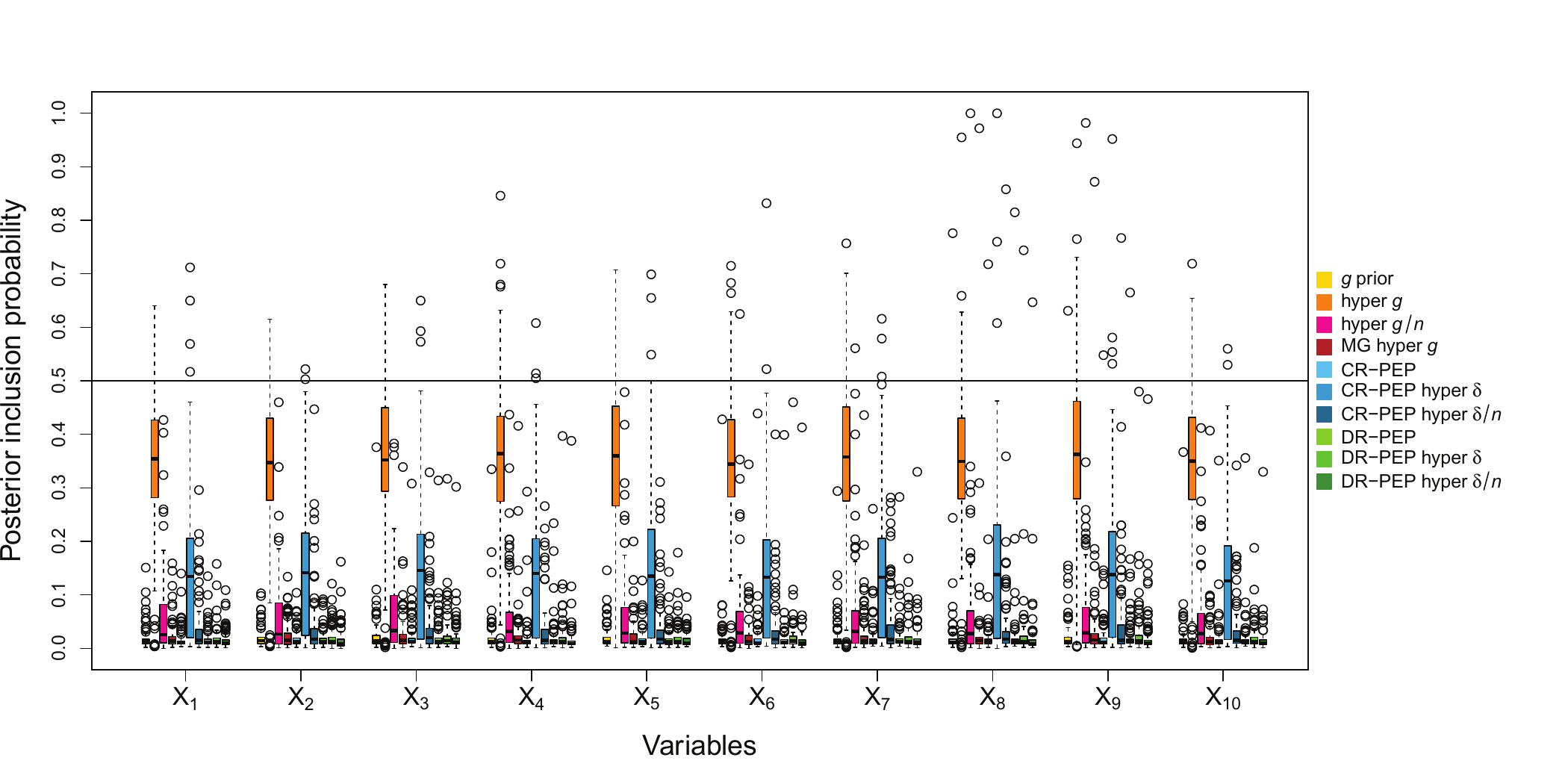}\caption{Posterior inclusion probabilities for Simulation Study 2 under the various priors from 100 repetitions of the null logistic simulation scenario.}
	\label{inclusion_sim0}
\end{figure}

\begin{figure}[h!t!]
	\hspace{3em}\includegraphics[scale=0.37]{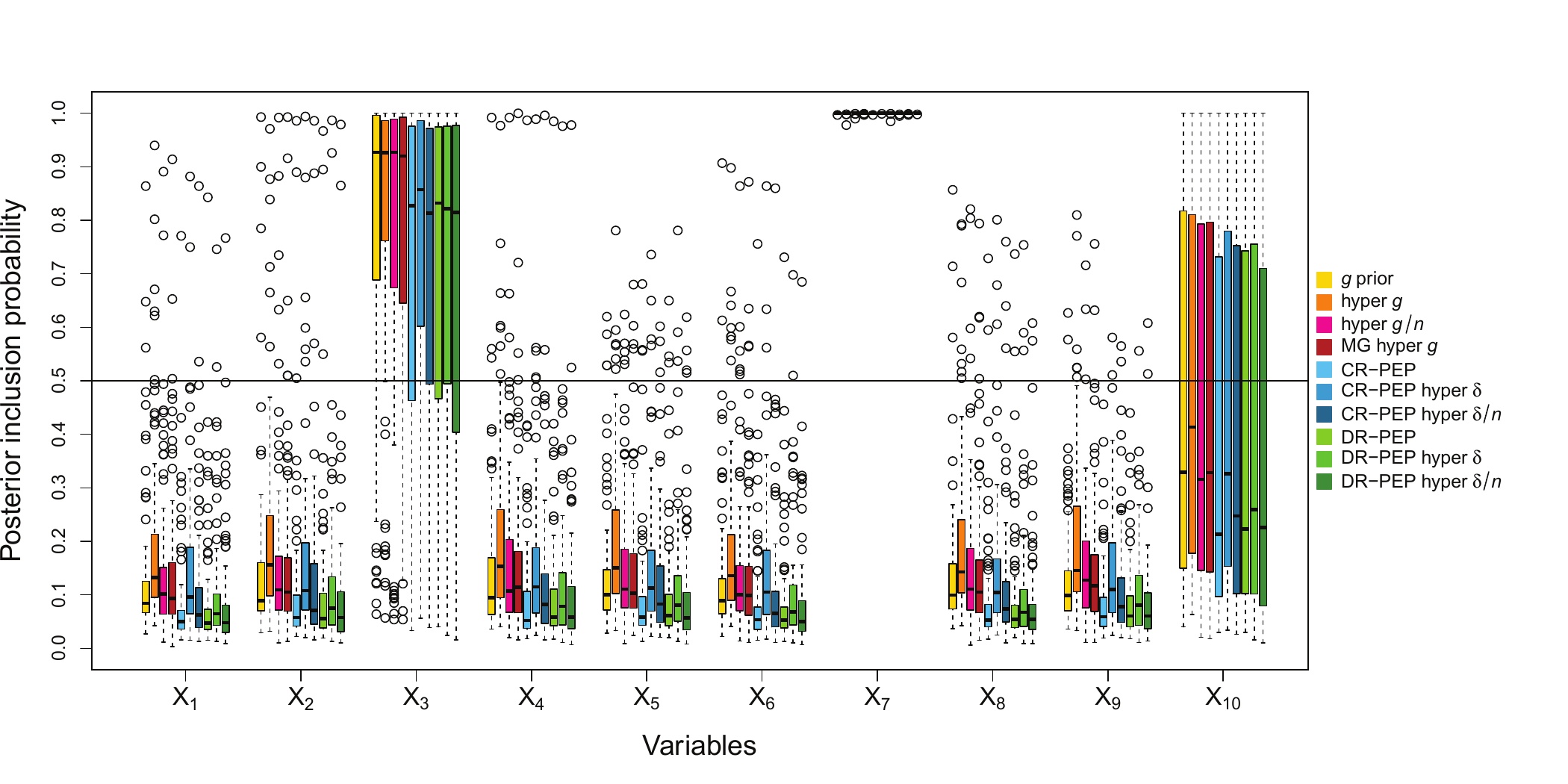}\caption{Posterior inclusion probabilities for Simulation Study 2 under the various priors from 100 repetitions of the sparse logistic simulation scenario where the true model is $X_3+X_7+X_{10}$.}
	\label{inclusion_sim3}
\end{figure}

\begin{figure}[h!t!]
	\hspace{3em} \includegraphics[scale=0.37]{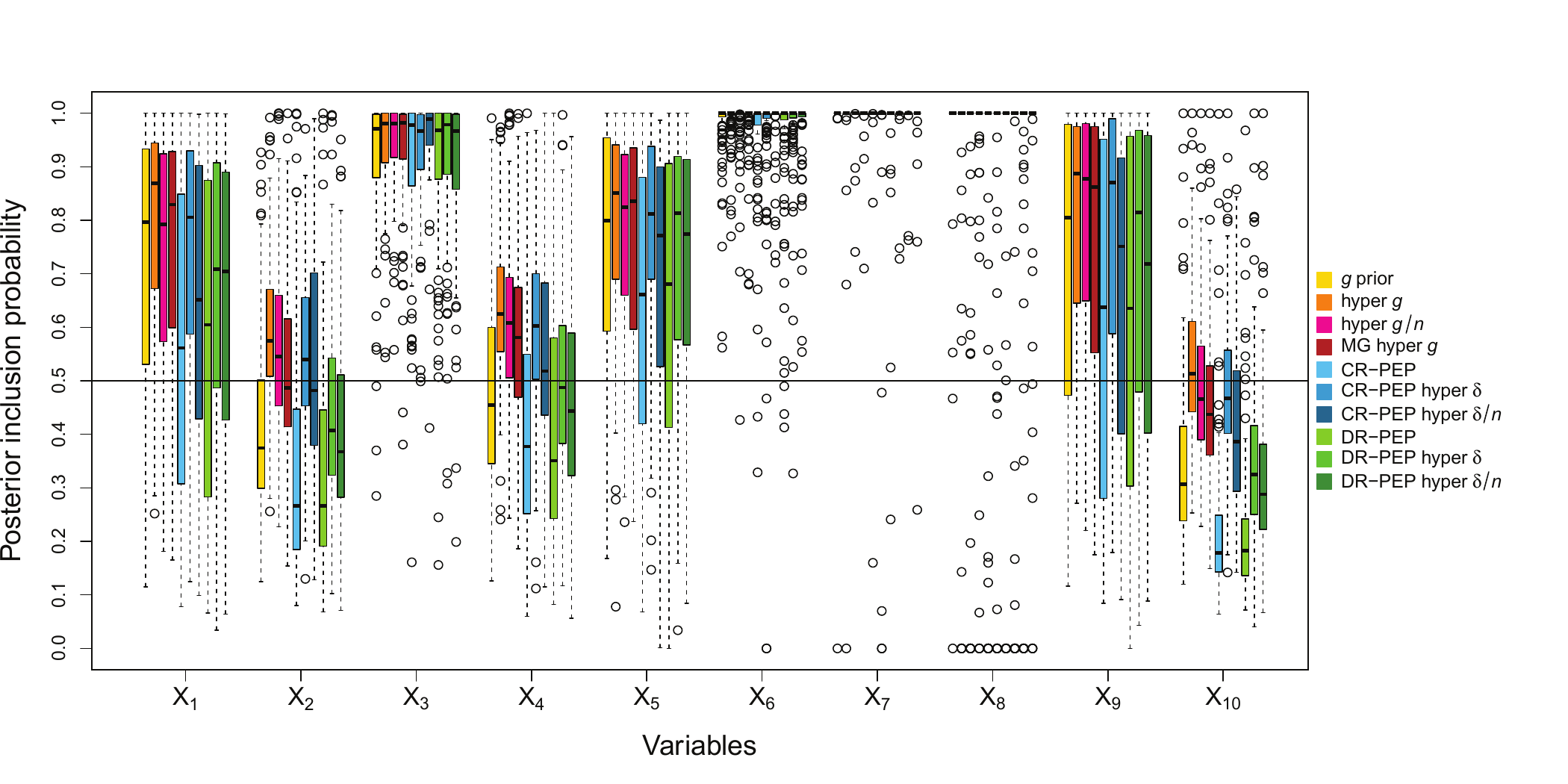}\caption{Posterior inclusion probabilities for Simulation Study 2 under the various priors from 100 repetitions of the dense logistic simulation scenario where the true model is $X_1+X_3+X_5+X_6+X_7+X_8+X_9$.}
	\label{inclusion_sim7}
\end{figure}

In this example we use the beta-binomial prior \eqref{betabin_model} with a Beta(1,1) mixing distribution; see \cite{scott_berger_2010}. 
The comparison that follows is based on the posterior inclusion probability of each covariate. Figures \ref{inclusion_sim0}, \ref{inclusion_sim3} and \ref{inclusion_sim7} present boxplots of the posterior inclusion probabilities from the 100 simulated data sets for the null, sparse and dense simulation scenarios, respectively.

\begin{table}[h!]
	\centering{}%
	{ %
		\begin{tabular}{l|ccccccccccc}
			\hline 
			\multirow{2}{*}{\textbf{Scenario}} & \multicolumn{10}{c}{\textbf{Logistic}\,\,$(n=200)$}  &\tabularnewline
			& $\beta_{0}$ & $\beta_{1}$ & $\beta_{2}$ & $\beta_{3}$ & $\beta_{4}$ & $\beta_{5}$ & $\beta_{6}$ & $\beta_{7}$ & $\beta_{8}$ & $\beta_{9}$ &$\beta_{10}$\tabularnewline
			\hline 
			null & 0.1 & 0 & 0 & 0 & 0 & 0 & 0 & 0 & 0 & 0&0\tabularnewline
			sparse & 0.1 & 0 & 0 & -0.9 & 0 & 0 & 0 & 1.2 & 0 & 0& 0.4\tabularnewline
			dense & 0.1 & 0.6 & 0 & -0.9 & 0 & 1 & 0.9 & 1.2 & -1.2 & -0.5&0\tabularnewline
			\hline
	\end{tabular}}
    \caption{Three logistic simulation scenarios for Simulation Study 2.}
	\label{Simulation2}
\end{table}

Under the null scenario, 
all priors, except the hyper-$g$ prior, exhibit strong shrinkage towards zero on the inclusion probabilities. Under the hyper-$g$ prior relatively large posterior inclusion probabilities with high variability across different samples are obtained. The hyper-$\delta$ CR-PEP prior also induces more variability, however, the resulting inclusion probabilities under this method are quite lower in comparison to those obtained under the hyper-$g$ prior. 

Under the sparse scenario (true model: $X_1+X_7+X_{10}$), there are no striking differences among all methods. All priors provide very strong support for the inclusion of $X_7$ and sufficient support for the inclusion of $X_3$, although the variability under PEP priors is larger for the latter variable. Moreover, all methods yield very wide posterior inclusion probability intervals for predictor $X_{10}$, implying high posterior uncertainty concerning the inclusion of this variable. For the non-important variables we observe that the fixed-$\delta$ CR-PEP and the DR-PEP priors yield the lowest posterior inclusion probabilities.

Finally, in the dense simulation scenario (Figure \ref{inclusion_sim7}), where the true model is $X_1+X_3+X_5+X_6+X_7+X_8+X_9$, the   fixed-$\delta$ PEP priors generally outperform other methods in terms of providing low posterior inclusion probabilities for the insignificant covariates $X_2$, $X_4$ and $X_{10}$. The $g$-prior and the hyper DR-PEP extensions yield similar posterior inclusion probabilities and generally perform well, however, they introduce some uncertainty concerning the inclusion of covariate $X_4$.  
The rest of the methods systematically support more complex models as they provide elevated support for the inclusion of variables $X_2$ and $X_4$.

\subsection{A real life example}

In our last example we consider the Pima Indians diabetes data set \citep{Ripley_1996}, which has been analyzed 
in several studies \citep[e.g.][]{Holmes_Held_2006, bove_held_2011}.
The data consist of $n=532$ complete records on diabetes presence (present=1, not present=0) according to the WHO criteria for signs of diabetes. The presence of diabetes is associated with $p=7$ potential covariates which are listed in Table \ref{Pima data}.

For each method we used 41000 iterations of the GVS  algorithm, discarding the first 1000 as burn-in period. We assigned a beta-binomial prior on model space (see Eq. \ref{betabin_model}) with both hyper-parameters 
equal to one. Table \ref{Pima_inclusion} shows the posterior inclusion probabilities of each covariate under the various methods. 
For comparison with the results presented in \cite{bove_held_2011}, 
we also include in Table \ref{Pima_inclusion}  the resulting posterior inclusion probabilities from  
the \cite{Zellner_siow_80} inverse gamma (ZS-IG) prior, the hyper-$g/n$  with $a=4$,
and a non-informative inverse gamma (NI-IG) hyper-$g$ prior with shape and scale equal to $10^{-3}$. As seen, the posterior inclusion probabilities that we obtain from the GVS algorithm are in agreement with the results presented in \cite{bove_held_2011}.

\begin{table}[h!t]
	\centering{}%
	{ %
		\begin{tabular}{l|l}
			\hline 
			\textbf{Covariate} & \textbf{Description}\tabularnewline
			\hline 
			$X_{1}$ & Number of pregnancies\tabularnewline
			$X_{2}$ & Plasma glucose concentration (mg/dl)\tabularnewline
			$X_{3}$ & Diastolic blood pressure (mm Hg)\tabularnewline
			$X_{4}$ & Triceps skin fold thickness (mm)\tabularnewline
			$X_{5}$ & Body mass index (kg/m$^{2}$)\tabularnewline
			$X_{6}$ & Diabetes pedigree function\tabularnewline
			$X_{7}$ & Age\tabularnewline
			\hline 
	\end{tabular}}
    \caption{Potential predictors in the Pima Indians diabetes data set.}
	\label{Pima data}
	\par
\end{table}

\begin{table}[b!]
	\centering{}%
	{ %
		\begin{tabular}{l|ccccccc}
			\hline 
			\multirow{2}{*}{\textbf{Method}} & \multicolumn{7}{c}{\textbf{Predictor}}\tabularnewline
			& $X_{1}$ & $X_{2}$ & $X_{3}$ & $X_{4}$ & $X_{5}$ & $X_{6}$ & $X_{7}$\tabularnewline
			\hline 
			ZS-IG hyper-$g$ & 0.961 & 1.000 & 0.252 & 0.250 & 0.998 & 0.994 & 0.530\tabularnewline
			NI-IG hyper-$g$  & 0.967 & 1.000 & 0.349 & 0.341 & 0.998 & 0.996 & 0.622\tabularnewline
			$g$-prior ($g=n$) & 0.952 & 1.000 & 0.136 & 0.139 & 0.998 & 0.992 & 0.382\tabularnewline
			hyper-$g$ ($a=3$) & 0.970 & 1.000 & 0.397 & 0.379 & 0.998 & 0.996 & 0.669\tabularnewline 
			hyper-$g/n$ ($a=3$) & 0.966 & 1.000 & 0.304 & 0.300 & 0.998 & 0.995 & 0.579\tabularnewline 
			hyper-$g/n$ $(a=4)$   & 0.965 & 1.000 & 0.307 & 0.299 & 0.997 & 0.995 & 0.582\tabularnewline
			MG hyper-$g$ & 0.958 & 1.000 & 0.262 & 0.259 & 0.998 & 0.994 & 0.548\tabularnewline 
			\hline 
			CR-PEP & 0.948 & 1.000 & 0.100 & 0.104 & 0.998 & 0.987 & 0.339\tabularnewline
			CR-PEP hyper-$\delta$ & 0.964 & 1.000 & 0.296 & 0.291 & 0.998 & 0.995 & 0.602\tabularnewline
			CR-PEP hyper-$\delta/n$ & 0.956 & 1.000 & 0.223 & 0.225 & 0.998 & 0.992 & 0.520\tabularnewline
			DR-PEP & 0.948 & 1.000 & 0.102 & 0.104 & 0.997 & 0.988 & 0.324\tabularnewline
			DR-PEP hyper-$\delta$ & 0.954 & 1.000 & 0.174 & 0.173 & 0.997 & 0.991 & 0.442\tabularnewline
			DR-PEP hyper-$\delta/n$ & 0.951 & 1.000 & 0.125 & 0.120 & 0.998 & 0.987 & 0.346\tabularnewline
			\hline
	\end{tabular}}
	\protect
	\caption{Posterior inclusion probabilities for the seven covariates of the Pima Indians data set.}
	\label{Pima_inclusion}
	\par
\end{table}

For the covariates $X_1,X_2,X_5$ and $X_6$, which seem to be highly influential, the results in Table \ref{Pima_inclusion} show no significant differences among methods.
On the contrary, the posterior inclusion probabilities for the ``uncertain'' covariates $X_3,X_4$ and $X_7$ vary  substantially; 
specifically, the inclusion probabilities from the fixed-$\delta$ CR/DR-PEP priors, the hyper-$\delta/n$ DR-PEP prior and the $g$-prior  are considerably lower than the inclusion probabilities resulting from the rest of the methods. 
In terms of the shrinkage factors $g/(g+1)$ and $\delta/(\delta+1)$,
results show that the shrinkage effect is stronger when  $g$ or $\delta$ is fixed, which  
leads to a drastic reduction in the effects (and the inclusion probabilities) of low-influential covariates.  
On the other hand, the priors with random $g$ or $\delta$ clearly result in higher posterior inclusion probabilities. Among this category of priors, the hyper-$\delta/n$ DR-PEP is evidently the most parsimonious, as it yields posterior inclusion probabilities which are actually quite close to those obtained from fixed $\delta$ PEP priors.

\begin{figure}[b!]
	\centering{}\includegraphics[scale=0.5]{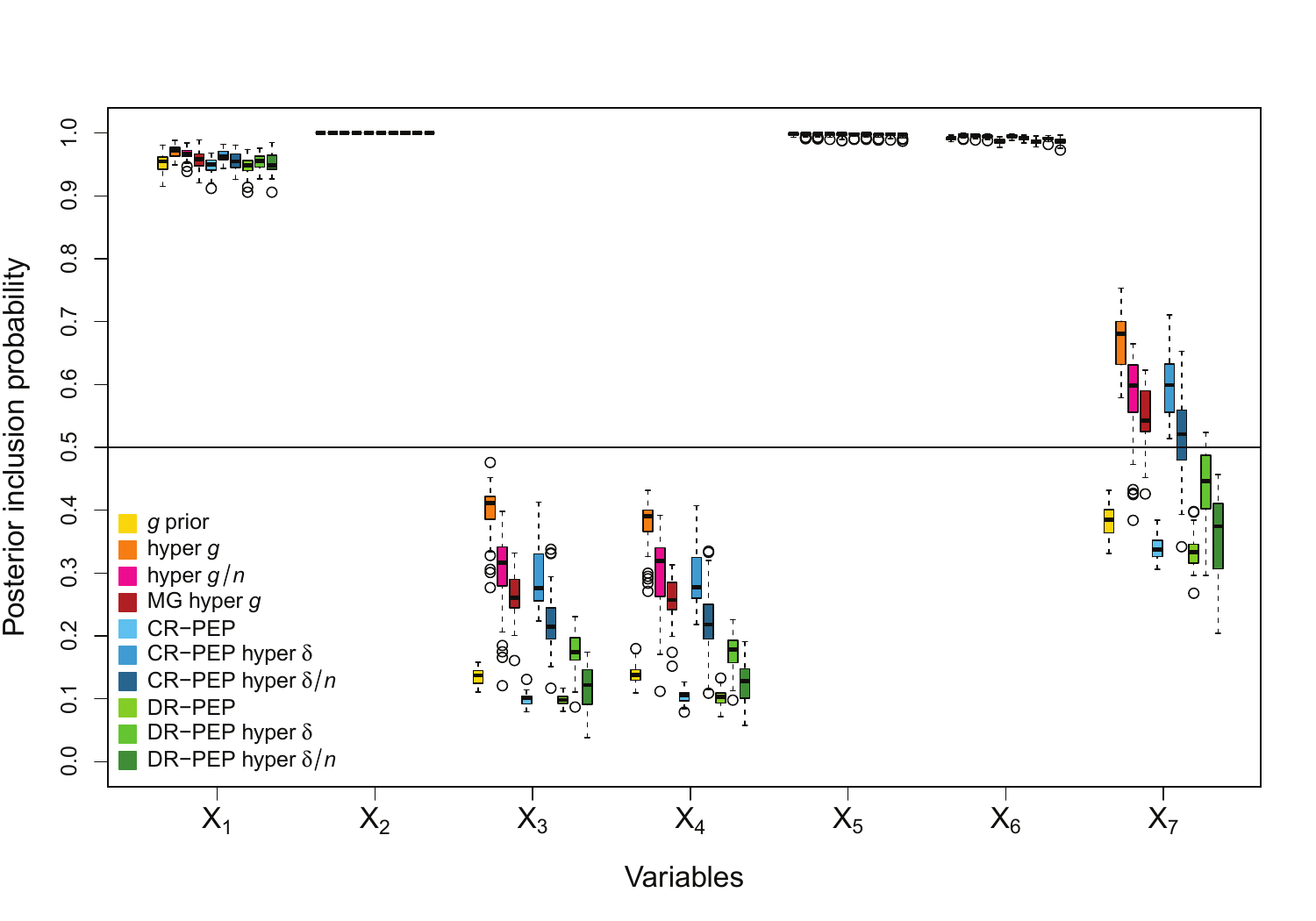}\caption{Boxplots of batched estimates of the posterior inclusion probabilities for the seven predictors in the Pima Indians data set based on 40 batches of size 1000.}
	\label{inclusion_pima}
\end{figure}

The uncertainty of the estimated posterior inclusion probabilities, for the standard methods considered in the previous examples, is depicted in Figure \ref{inclusion_pima}, where we present the corresponding boxplots produced by splitting the posterior samples into 40 batches of size 1000. 
As seen in Figure \ref{inclusion_pima}, stochasticity in $g$ and $\delta$ mainly affects the posterior inclusion probabilities of the ``uncertain'' covariates $X_3,X_4$ and $X_7$. 
For these variables the extra prior uncertainty induces higher posterior variability, as expected, and consequently larger Monte Carlo errors. 
Apart from this, we generally observe the same patterns of evidence leading to the conclusions discussed in Sections \ref{sec_sim1} and \ref{sec_sim2}. 

\begin{figure}[h!]
	\centering{}
	\begin{tabular}{cc} 
		\includegraphics[scale=0.3]{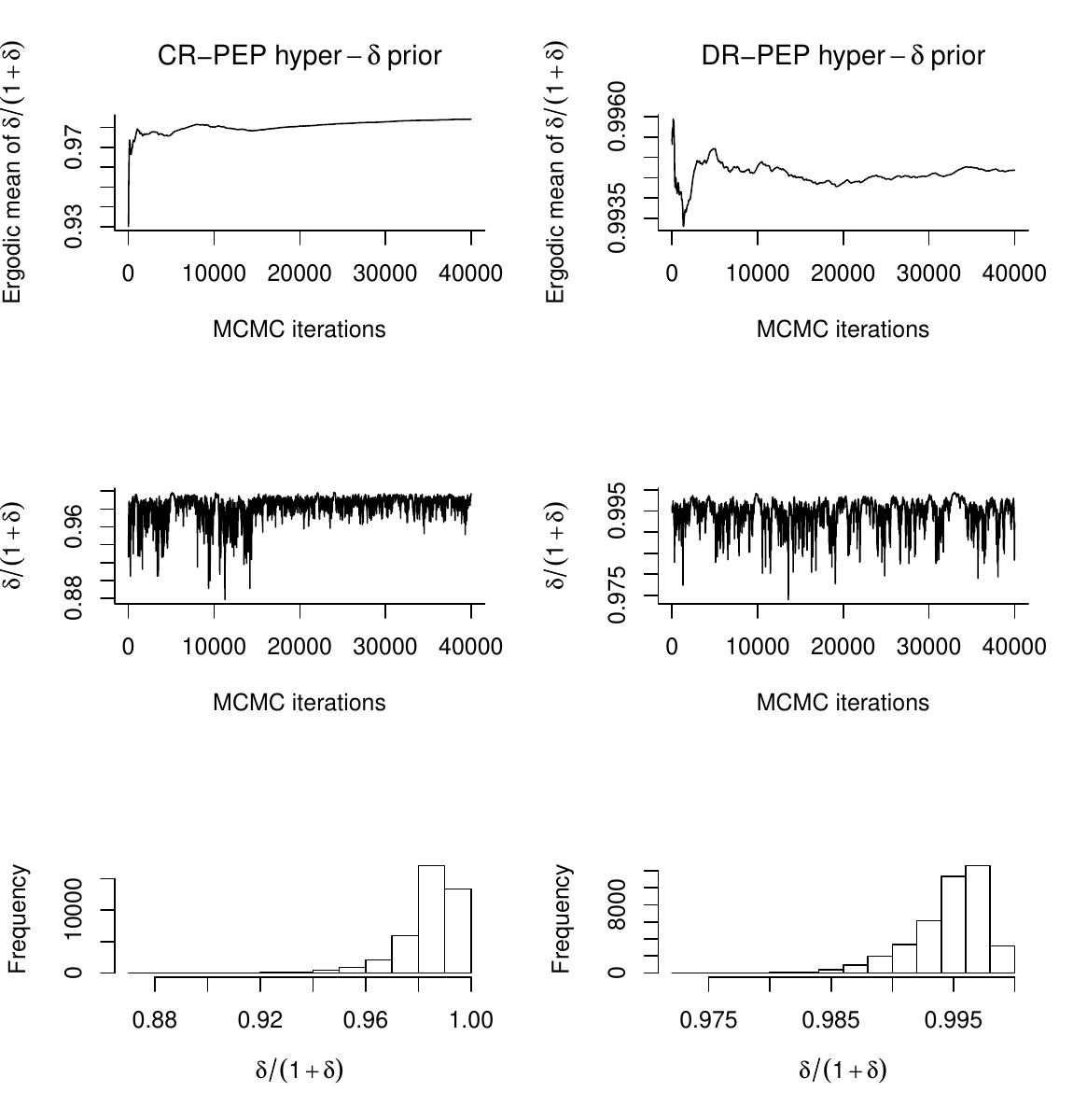} &  
		\includegraphics[scale=0.3]{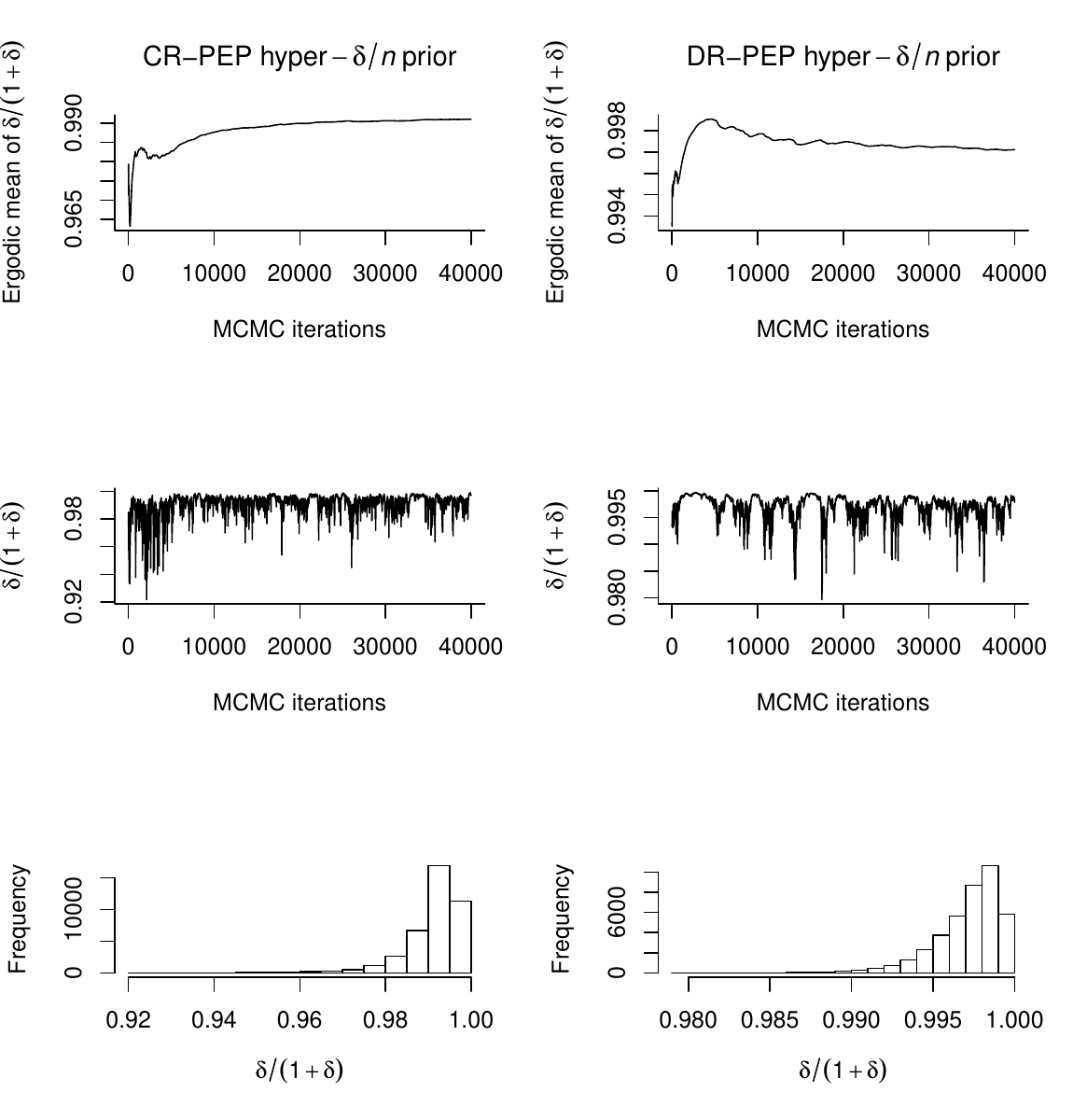} \\
		\footnotesize (a) Hyper-$\delta$ Priors &  \footnotesize  (b) Hyper-$\delta/n$ Priors
	\end{tabular} 
	\caption{Ergodic mean plots, time-series plots and histograms of the shrinkage factor $\delta/(1+\delta)$ for the hyper-$\delta$ and hyper-$\delta/n$ PEP  priors based on 40000 draws.}
	\label{shrinkage_d}
\end{figure}

Figure  \ref{shrinkage_d} depicts the convergence and the estimated posterior distribution of the shrinkage parameter $\delta/(1+\delta)$ under the four PEP hyper-prior approaches.
The posterior histograms are indicative of the behavior of the shrinkage parameter. 
Comparison between the hyper-$\delta$ (Figure \ref{shrinkage_d}a) and the hyper-$\delta/n$ (Figure \ref{shrinkage_d}b) approaches shows 
that the posterior distribution of the shrinkage parameter under the latter priors is more concentrated to values close to one, thus, resulting to a stronger shrinkage effect. 
Also, the histograms in Figures \ref{shrinkage_d}a and \ref{shrinkage_d}b indicate 
that the posterior distributions of the shrinkage parameter under DR-PEP are more concentrated to one in comparison to the corresponding posteriors under CR-PEP. 
Note that the shrinkage under the fixed-$\delta$ approaches is constant, equal to 0.998, which leads to considerably lower posterior inclusion probabilities as seen in Table \ref{Pima_inclusion} and Figure \ref{inclusion_pima}.

We conclude this example by examining the out-of-sample predictive accuracy of the various prior setups. We randomly split the data set in half in order to create a training sample and a test sample and identified  
the corresponding MAP and the median probability models, under all methods, from the training data set. Then, based on posterior samples from the predictive distribution we calculated the averages of false positive and false negative percentages for the test data set; the results are reported in Table \ref{Pima_prediction}. Overall, we cannot say that there is dominant method in terms of predictive accuracy as the predictions are more or less the same across the prior setups. We may note however that the most complex MAP model arises from the hyper-$g$ prior which also results in the highest false negative prediction rates. Also, the unit-information $g$-prior, the CR-PEP prior with fixed $\delta$, and the DR-PEP priors lead to the most parsimonious median probability model. This model is comparable in terms of predictive performance with the model that further includes $X_7$, which is indicated as the median probability model by the rest of the methods.

\begin{table}
	{ %
		\begin{tabular}{l|l c c |l c c }
			\hline 
			&   & \textbf{\small{False}} & \textbf{\small{False}}
			&   & \textbf{\small{False}} & \textbf{\small{False}}\tabularnewline
			&  & \textbf{\small{Neg.}} & \textbf{\small{Pos.}} 
			&  & \textbf{\small{Neg.}} & \textbf{\small{Pos.}} \tabularnewline
			\textbf{\small{Method}} & \textbf{\small{MAP }} & \textbf{\small{(\%) }} & \textbf{\small{(\%)}} 
			& \textbf{\small{MPM} } & \textbf{\small{(\%) }} & \textbf{\small{(\%)}} \tabularnewline
			\hline 
			{\small{$g$-prior ($g=n$)}}      & {\small{${\cal M}_A$}} & {\small{10.8}} & {\small{16.5}} 
			& {\small{${\cal M}_A$}} & {\small{10.8}} & {\small{16.5}} \tabularnewline
			{\small{hyper-$g$ ($a=3$)}}      & {\small{${\cal M}_A +X_{3}+X_{4}+X_{7}$}} & {\small{11.4}} & {\small{16.9}}
			& {\small{${\cal M}_A+X_{7}$}} & {\small{11.1}} & {\small{16.8}}\tabularnewline
			{\small{hyper-$g/n$ ($a=3$)}}    & {\small{${\cal M}_A$}}& {\small{11.0}} & {\small{16.6}} 
			& {\small{${\cal M}_A+X_{7}$}} & {\small{11.0}} & {\small{16.6}}\tabularnewline
			{\small{MG hyper-$g$}}           & {\small{${\cal M}_A$}} & {\small{10.9}} & {\small{16.6}}
			& {\small{${\cal M}_A+X_{7}$}} & {\small{10.9}} & {\small{16.6}}\tabularnewline
			{\small{CR-PEP }}                & {\small{${\cal M}_A$}} & {\small{10.9}} & {\small{16.9}}
			& {\small{${\cal M}_A$}} & {\small{10.9}} & {\small{16.9}} \tabularnewline
			{\small{CR-PEP hyper-$\delta$}}  & {\small{${\cal M}_A$}} & {\small{10.9}} & {\small{17.0}}
			& {\small{${\cal M}_A+X_{7}$}} & {\small{11.3}} & {\small{16.4}}\tabularnewline
			{\small{CR-PEP hyper-$\delta/n$}}& {\small{${\cal M}_A$}} & {\small{10.8}} & {\small{17.0}}
			& {\small{${\cal M}_A+X_{7}$}} & {\small{11.0}} & {\small{16.6}}\tabularnewline
			{\small{DR-PEP}}                 & {\small{${\cal M}_A$}} & {\small{10.9}} & {\small{16.8}}
			& {\small{${\cal M}_A$}} & {\small{10.9}} & {\small{16.8}} \tabularnewline
			{\small{DR-PEP hyper-$\delta$}}  & {\small{${\cal M}_A$}} & {\small{10.9}} & {\small{16.9}}
			& {\small{${\cal M}_A$}} & {\small{10.9}} & {\small{16.9}} \tabularnewline
			{\small{DR-PEP hyper-$\delta/n$}}& {\small{${\cal M}_A$}} & {\small{10.9}} & {\small{16.8}}
			& {\small{${\cal M}_A$}} & {\small{10.9}} & {\small{16.8}} \tabularnewline
			\hline 
			\multicolumn{7}{p{10cm}}{ \footnotesize ${\cal M}_A : X_{1}+X_{2}+X_{5}+X_{6}$}
	\end{tabular}}
    \caption{Percentages of false negative and false positive detections for the
    	Pima Indian data set under the MAP model and median probability model (MPM)
    	for the various priors.}
	\label{Pima_prediction}
	\par
\end{table}

\section{Discussion}
\label{discussion}

In this paper we presented an objective, automatic and compatible across competing models Bayesian procedure with applications to the variable selection problem in GLMs. Specifically we extended the PEP prior formulation through the use of unnormalized power likelihoods and defined two new PEP priors, called CR-PEP and DR-PEP, which differentiate with respect to the definition of the prior predictive distribution of the reference model.
Under the new definitions, the applicability of the PEP methodology is significantly enhanced. Although we focused on variable selection for GLMs, the CR/DR-PEP priors proposed here may in principle be used for any general model setting. 
At the same time the new approaches retain the desired features of the original PEP prior formulation; specifically, i) they resolve the problem  of selecting and averaging across minimal imaginary samples, thus, also allowing for large-sample approximations, and ii) they are minimally informative as they scale down the effect of the imaginary data on the posterior distribution.
We further studied the assignment of hyper-prior distributions to the power parameter $\delta$ that controls the contribution of the imaginary data. Following the hyper-$g$ and $g/n$ priors proposed in \cite{liang_etal_2008}, we effectively introduced the 
hyper-$\delta$ and $\delta/n$ analogues.

With respect to the criteria of \cite{Bayarri_etal_2012}, we provided analytical proofs for the null and dimensional predictive matching criteria for all PEP priors under consideration.
With respect to model selection consistency, analytical proofs for the normal linear model are provided in \cite{Fouskakis_Ntzoufras_Perrakis_2016_regression}; here, we illustrated through simulations that this criterion also seems to be valid within the PEP priors framework for specific GLMs scenarios. 

The empirical results presented in this paper suggest that the proposed PEP priors outperform mixtures of $g$-priors in terms of introducing larger shrinkage to the inclusion probabilities of non-influential or partially influential predictors, thus, leading to more parsimonious solutions with comparable predictive accuracy. 
When comparing PEP priors with fixed $\delta=n$ and random $\delta$
the results indicate that the former approach induces more stringent control in the inclusion of predictors. Therefore, fixed PEP priors support simpler models which is a desirable feature when the number of covariates is large. 
Concerning the choice between the CR and the DR prior setups, we conclude in favour to the use of the latter since it is rather robust with respect to the fixed vs. random specification of $\delta$. 

In future research we plan to extend the PEP methodology to high-dimensional problems, including the small $n$--large $p$ case, by incorporating shrinkage priors (e.g. ridge and LASSO procedures) into the PEP design. 
Another promising alternative is to embody the expectation-maximization variable selection approach of 
\cite{Rockova_George_2014} within the PEP prior.

\section*{Acknowledgement}
This research has been co-financed in part by the European Union (European Social Fund-ESF) and by Greek national funds through the Operational Program ``Education and Lifelong Learning" of the National Strategic Reference Framework (NSRF)-Research Funding Program: Aristeia II/PEP-BVS.


\bibliographystyle{agsm}
\bibliography{biblio2016}

\newpage
\setcounter{page}{1}
\pagenumbering{roman}

\appendix

\paragraph{\Large Appendix} 

\section{Proofs of Predictive Matching} 
\label{appendix_predictive}

\subsection{Proof of Proposition \ref{prop_null_pred}.} 
\label{proof1} 

Assuming known and common across all models $\phi_\vg=\phi$, then, for samples of size $n=1$ 
(and, therefore we have also $\delta=n^*=1$), both CR-PEP and DR-PEP priors coincide. 
Moreover, assuming that we have observed the response $y$ with covariate values $\dn{x}=(x_1, \dots, x_p)^T$ then we need to generate an imaginary data-point $y^*$ under the same set of covariates. 
Moreover, the linear predictor for any model $\vg$ is now given by 
\begin{equation} 
\eta_\vg= \beta_{0, \vg} + \dn{x}_\vg^T \dn{\beta}_{\setminus 0, \vg}
= \beta_{0, \vg} + \sum_{j=1}^p \gamma_j x_j \beta_j, 
\label{linearpred}
\end{equation} 
where $\dn{x}_\vg$ is the sub-vector of $\dn{x}$ with elements corresponding to covariates included in model 
$M_\vg$.

Under this formulation and \eqref{req1_predictive_matching},
the prior predictive density for model $M_0$  
under the baseline prior $\pi_0^N(\beta_{0} )$ is given by 
\begin{eqnarray}
m_0^{\N}(y^*|\delta=1) 
&=& \int \exp\left( \frac{y^* \vartheta(\beta_0)  - b\big(\vartheta(\beta_0)\big) }{  a(\phi)} 
+ c(y^*, \phi) \right) \psi( \beta_0 ) d\beta_0.  \nonumber 
\end{eqnarray}
By setting $\eta=\beta_0$ under model $M_0$, we obtain 
\begin{eqnarray}
m_0^{\N}(y^*|\delta=1) & =  & e^{c(y^*, \phi)} {\cal D}( y^*, 1, 1 ) \label{R1} \\ 
&& \mbox{where~}  {\cal D}( y, \omega, \delta ) = 
\int \exp\left( \frac{y^* \vartheta(\eta)  -  \omega b\big( \vartheta(\eta)\big) }{ \delta a(\phi)} \right) \psi( \eta ) d\eta.  \nonumber     
\end{eqnarray}
For model $M_\vg$, the prior predictive density under the baseline prior $\pi_\vg^N(\betab)$
is given by 
\begin{eqnarray}
m_\vg^{\N}(y^*|\delta=1) \hspace{-1em}
&=& \hspace{-1em} \int \exp\left( \frac{y^* \vartheta( \eta_\vg )  - b\big(\vartheta(\eta_\vg)\big) }{  a(\phi)} 
+ c(y^*, \phi) \right) \psi( \eta_\gamma ) \Psi_\vg( \dn{\beta}_{\setminus 0, \vg} ) d\betab. \hspace{2em}
\label{mgamma1}                   
\end{eqnarray}
By setting $\eta = \beta_{0, \vg}  + \sum_{j=1}^p \gamma_j x_j \beta_j= \eta_\vg$ (see Eq. \ref{linearpred}) and 
$\dn{b} = \dn{\beta}_{\setminus 0, \vg}$, then from \eqref{mgamma1} we obtain 
\begin{eqnarray}
m_\vg^{\N}(y^*|\delta=1) 
&=& e^{ c(y^*, \phi) } \int \left\{ \int \exp\left( \frac{y^* \vartheta( \eta )  - b\big(\vartheta(\eta)\big) }{  a(\phi)}  \right)  \psi( \eta ) d\eta \right\} \Psi_\vg( \dn{b} ) d\dn{b} \nonumber \\ 
&=& e^{ c(y^*, \phi)} {\cal D}( y^*, 1, 1 ) {\cal A}_\vg = m_0^N (y^* | \delta=1 ) {\cal A}_\vg,  \nonumber 
\label{mgamma2} 
\end{eqnarray}
\normalsize
where ${\cal A}_\vg = \int \Psi_\vg( \dn{b} ) d\dn{b}$.

The marginal likelihood of $M_0$ under the PEP prior is given by 
$$
m_0^{\PEP}( y | \delta=1 ) =  m_0^N( y | \delta=1),
$$ 
while for model $M_\vg$ is given by  
\begin{eqnarray}
m_\vg^{\PEP}( y | \delta=1) 
&=& \int \frac{m_0^{\N}(y^*| \delta=1)}{m_{\vg}^{\N}(y^*| \delta=1)} 
\left\{  \int{ f_{\vg}(y|\betab)f_{\vg}(y^*|\betab, \delta=1) \pi_{\vg}^{\N}(\betab) } \mathrm{d}\betab \right\} \mathrm{d}y^* \nonumber \\ 
&=& \int {\cal A}^{-1}_\vg 
\left\{  \int{ f_{\vg}(y|\betab)f_{\vg}(y^*|\betab, \delta=1) \pi_{\vg}^{\N}(\betab) } \mathrm{d}\betab \right\} \mathrm{d}y^* \nonumber \\ 
&=& \int {\cal A}^{-1}_\vg \left\{   f_{\vg}(y|\betab)
\left[ \int f_{\vg}(y^*|\betab, \delta=1)\mathrm{d}y^* \right] \pi_{\vg}^{\N}(\betab)  \right\} \mathrm{d}\betab \nonumber \\ 
&=& {\cal A}^{-1}_\vg  \int     f_{\vg}(y|\betab)\pi_{\vg}^{\N}(\betab)   \mathrm{d}\betab 
=  {\cal A}^{-1}_\vg  m_\vg^N(y| \delta=1)  \nonumber \\ 
&=& {\cal A}^{-1}_\vg m_0^N(y| \delta=1)  {\cal A}_\vg  = m_0^N(y| \delta=1) \nonumber
\end{eqnarray}
and hence, for known $\phi$, this concludes the proof. 

If $\phi$ is stochastic and common across models, then the two marginal likelihoods, obtained by integrating out $\phi$ over the common prior $\pi^N_0(\phi) = \pi^N_\vg(\phi)$ for all models $M_\vg$, still coincide.

\subsection{Proof of Proposition \ref{prop_null_pred_hyperDR}.}  
\label{proof2}
Following similar arguments as in Section \ref{proof1} we have that 
\begin{eqnarray}
m_0^{\N}(y^*|\delta) & =  & e^{c(y^*, \phi)/\delta} {\cal D}( y^*, 1, \delta ) \nonumber  \\
m_\vg^{\N}(y^*|\delta) 
&=&   e^{ c(y^*, \phi)/\delta} {\cal D}( y^*, 1, \delta ) {\cal A}_\vg = m_0^N (y^* | \delta ) {\cal A}_\vg.  \label{R2} 
\end{eqnarray}
The final marginal likelihood of $M_\vg$ under the DR-PEP prior, conditional on $\delta$, is given by 
\begin{eqnarray}
m_\vg^{\DR}( y | \delta) 
&=& {\cal C}_0^{-1} \int \frac{m_0^{\N}(y^*| \delta )}{m_{\vg}^{\N}(y^*| \delta )} 
\left\{  \int{ f_{\vg}(y|\betab)f_{\vg}(y^*|\betab, \delta) \pi_{\vg}^{\N}(\betab) } \mathrm{d}\betab \right\} \mathrm{d}y^* \nonumber \\ 
&=& {\cal C}_0^{-1} \int {\cal A}^{-1}_\vg 
\left\{  \int{ f_{\vg}(y|\betab)f_{\vg}(y^*|\betab, \delta) \pi_{\vg}^{\N}(\betab) } \mathrm{d}\betab \right\} \mathrm{d}y^* \nonumber \\ 
&=& {\cal C}_0^{-1} {\cal A}^{-1}_\vg\int\int     
\exp\left( \frac{y \vartheta( \eta_\vg )  - b\big(\vartheta(\eta_\vg)\big) }{  a(\phi)} 
+ c(y, \phi) \right)  \nonumber \\
&& \hspace{3.2em} \times 
\exp\left( \frac{y^* \vartheta( \eta_\vg )  - b\big(\vartheta(\eta_\vg)\big) }{  \delta a(\phi)} + \frac{c(y^*, \phi)}{\delta} \right) 
\pi_{\vg}^{\N}(\betab)   \mathrm{d}\betab\mathrm{d}y^*.   \nonumber
\end{eqnarray}
By using \eqref{req1_predictive_matching} and setting $\eta = \beta_{0, \vg}  + \sum_{j=1}^p \gamma_j x_j \beta_j= \eta_\vg$ (see Eq. \ref{linearpred}) and 
$\dn{b} = \dn{\beta}_{\setminus 0, \vg}$, we obtain
\small 
\begin{eqnarray}
&& \hspace{-2.7em} m_\vg^{\DR}( y | \delta) = {\cal C}_0^{-1} {\cal A}^{-1}_\vg   \times \nonumber \\ 
&&  \hspace{-0.5em} \times  \int e^{ c(y, \phi) +c(y^*, \phi)/\delta }  
\left\{   \int 
\exp\left( \frac{ (\delta y + y^*) \vartheta( \eta )  - (\delta+1) b\big(\vartheta(\eta)\big) }{  \delta a(\phi)} \right) \psi(\eta)  \mathrm{d}\eta  \int \Psi_\vg(\dn{b}) \mathrm{d}\dn{b} \right\} \mathrm{d}y^*.   \nonumber \\
&=& {\cal C}_0^{-1}   e^{ c(y, \phi) } \int e^{ c(y^*, \phi)/\delta }  {\cal D} ( \delta y + y^*, \delta+1, \delta ) \mathrm{d}y^*
\nonumber 
\end{eqnarray}
\normalsize 
which does not depend on the original model formulation $M_\vg$. 
Indeed, following similar logic we can prove that 
$$
m_0^\DR(y|\delta)= {\cal C}_0^{-1}  e^{ c(y, \phi)} \int e^{ c(y^*, \phi)/\delta }  {\cal D} ( \delta y + y^*, \delta+1, \delta )\mathrm{d}y^*,  
$$
and hence, for known $\phi$, it is obvious that after integrating out $\delta$ over the hyper-prior 
$\pi(\delta)$, $m_\vg^{\DR}( y ) = m_0^\DR(y)$ and this concludes the proof. 

If $\phi$ is stochastic and common across models, then the two marginal likelihoods, obtained by integrating out $\phi$ over the common prior $\pi^N_0(\phi) = \pi^N_\vg(\phi)$ for all models $M_\vg$, still coincide.

\subsection{Proof of Proposition \ref{prop_null_pred_hyperCR}.}  
\label{proof3}
For the CR-PEP prior, conditional on $\delta$, using 
\eqref{req1_predictive_matching}, \eqref{R1} and \eqref{R2} and following similar steps as in Appendix \ref{proof1} we have that 
\begin{eqnarray}
m_\vg^{\CR}( y | \delta) 
\hspace{-0.9em} &=& \hspace{-1em} \int \frac{m_0^{\N}(y^*| \delta=1 )}{m_{\vg}^{\N}(y^*| \delta )} 
\left\{  \int{ f_{\vg}(y|\betab)f_{\vg}(y^*|\betab, \delta) \pi_{\vg}^{\N}(\betab) } \mathrm{d}\betab \right\} \mathrm{d}y^* \nonumber \\ 
\hspace{-0.9em} &=& \hspace{-1em}  \int {\cal A}^{-1}_\vg e^{ \frac{\delta-1}{\delta} c(y^*, \phi) } \frac{ {\cal D}(y^*,1,1) }{ {\cal D}(y^*,1,\delta) }
\left\{  \int{ f_{\vg}(y|\betab)f_{\vg}(y^*|\betab, \delta) \pi_{\vg}^{\N}(\betab) } \mathrm{d}\betab \right\} \mathrm{d}y^* \nonumber \\ 
\hspace{-0.9em} &=& \hspace{-0.7em}     e^{ c(y, \phi) } \int e^{ c(y^*, \phi) }  \frac{ {\cal D}(y^*,1,1) }{ {\cal D}(y^*,1,\delta) }
{\cal D} ( \delta y + y^*, \delta+1, \delta ) \mathrm{d}y^*
\nonumber  
\end{eqnarray}
which (again) does not depend on the original model formulation $M_\vg$. 
Hence, for known $\phi$, it is obvious that after integrating out $\delta$ over the hyper-prior 
$\pi(\delta)$, $m_\vg^{\CR}( y ) = m_0^\CR(y)$ and this concludes the proof. 

If $\phi$ is stochastic and common across models, then the two marginal likelihoods, obtained by integrating out $\phi$ over the common prior $\pi^N_0(\phi) = \pi^N_\vg(\phi)$ for all models $M_\vg$, still coincide.

\subsection{Proof of Proposition \ref{prop_dim_pred}.} 
\label{proof4} 

Let us assume known and common across models $\phi_\vg=\phi$ and 
samples of size $n=p_\vg+1= \sum_{j=1}^p+1$.  
Then the linear predictor is given by $\dn{\eta}_\vg = \bx_\vg \betab $ 
where $\bx_\vg$ is a matrix of dimension $(p_\vg+1)\times(p_\vg+1)$. 
By considering $n^*=n=1+p_\vg$ and further assuming that is invertible, 
then the prior predictive density for model $M_0$  under the baseline prior $\pi_0^N(\beta_0)$ is given by 
\begin{eqnarray}
m_0^{\N}(\by^*|\delta) 
&=& \int \exp\left( \frac{ n^* \overline{y}^* \vartheta(\beta_0)  - n^* b\big(\vartheta(\beta_0)\big) }{  \delta a(\phi)} 
+ \sum_{i=1}^{n^*}\frac{ c(y_i^*, \phi)}{\delta} \right) \psi( \beta_0 ) d\beta_0  \nonumber 
\end{eqnarray}
By setting $\eta=\beta_0$, we obtain 
\begin{eqnarray}
m_0^{\N}(\by^*|\delta) & =  & \exp \left( \sum_{i=1}^{n^*} \frac{c(y_i^*, \phi)}{\delta} \right) {\cal D}( n^*\overline{y}^*, n^*, \delta ). \nonumber 
\end{eqnarray}
For model $M_\vg$, the prior predictive density under the baseline prior $\pi_\vg^N(\betab)$
is given by 
\begin{eqnarray}
m_\vg^{\N}(\by^*|\delta ) 
\hspace{-0.7em} &=& \hspace{-1em}  \int \exp\left( \frac{ 
	\sum_{i=1}^{n^*}y_i^* \vartheta( \eta_{\vg(i)} )  - \sum_{i=1}^{n^*}b\big(\vartheta(\eta_{\vg(i)})\big) }{ \delta a(\phi)} 
+ \sum_{i=1}^{n^*} \frac{c(y_i^*, \phi)}{\delta} \right) \psi( \dn{\eta}_\gamma )  d\betab ~.
\nonumber          
\end{eqnarray}
By setting $\dn{\eta} = \bx_\vg \betab = \dn{\eta}_\gamma$,  we obtain 
\begin{eqnarray}
m_\vg^{\N}(\by^*|\delta ) 
\hspace{-0.7em} &=& \hspace{-0.7em} \exp \left( \sum_{i=1}^{n^*} \frac{c(y_i^*, \phi)}{\delta} \right)
\int \exp\left( \frac{ 
	\sum_{i=1}^{n^*}y_i^* \vartheta( \eta_i )  - \sum_{i=1}^{n^*}b\big(\vartheta(\eta_i)\big) }{ \delta a(\phi)} \right)
\psi( \dn{\eta} ) |\bx_\vg^{-1}| d\dn{\eta} \nonumber \\ 
\hspace{-0.7em} &=& \hspace{-0.7em}  \exp \left( \sum_{i=1}^{n^*} \frac{c(y_i^*, \phi)}{\delta} \right)|\bx_\vg^{-1}| {\cal E}( \by^*, 1, \delta )  \nonumber \\ 
&& \mbox{where~} 
{\cal E}( \by^*, \omega, \delta ) = 
\int \exp\left( \frac{ 
	\sum_{i=1}^{n^*}y_i^* \vartheta( \eta_i )  - \omega \sum_{i=1}^{n^*}b\big(\vartheta(\eta_i)\big) }{ \delta a(\phi)} \right) \psi( \dn{\eta} ) d\dn{\eta}; \nonumber
\end{eqnarray}
\normalsize
note that ${\cal E}( y^*, \omega, \delta )={\cal D}( y^*, \omega, \delta )$ for $n^*=1$.

The marginal likelihood of $M_0$ under the DR-PEP prior, conditional on $\delta$, is given by 
$m_0^{\DR}( \by | \delta ) =  m_0^N( \by | \delta)$ 
while for model $M_\vg$ is given by  
\begin{eqnarray}
m_\vg^{\DR}( \by | \delta) 
\hspace{-0.7em} &=& \hspace{-0.7em} {\cal C}_0^{-1}\int |\bx_\vg| \frac{ {\cal D}( n^*\overline{y}^*, n^*, \delta ) }{ {\cal E}( \by^*, 1, \delta ) } 
\left\{  \int{ f_{\vg}(y|\betab)f_{\vg}(y^*|\betab, \delta) \pi_{\vg}^{\N}(\betab) } \mathrm{d}\betab \right\} \mathrm{d}\by^* \nonumber \\ 
\hspace{-0.7em} &=& \hspace{-0.7em}  {\cal C}_0^{-1} |\bx_\vg|  \int \frac{ {\cal D}( n^*\overline{y}^*, n^*, \delta ) }{ {\cal E}( \by^*, 1, \delta ) } 
\left\{  \int{ f_{\vg}(y|\betab)f_{\vg}(y^*|\betab, \delta) \pi_{\vg}^{\N}(\betab) } \mathrm{d}\betab \right\} \mathrm{d}\by^* \nonumber \\ 
\hspace{-0.7em} &=& \hspace{-0.7em}  {\cal C}_0^{-1} |\bx_\vg|  \int \frac{ {\cal D}( n^*\overline{y}^*, n^*, \delta )}{ {\cal E}( \by^*, 1, \delta ) } 
\exp \left( \sum_{i=1}^{n^*}  c(y_i, \phi)  + \sum_{i=1}^{n^*} \frac{c(y_i^*, \phi)}{\delta} \right) \nonumber \\ 
&& \hspace{15em} \times |\bx_\vg|^{-1}
{\cal E} ( \delta \by + \by^*, \delta + 1, \delta )   \mathrm{d}\by^* \nonumber \\ 
\hspace{-0.7em} &=& \hspace{-0.7em}  {\cal C}_0^{-1}   \int 
\prod_{i=1}^{n^*} e^{ c(y_i, \phi)  + \frac{c(y_i^*, \phi)}{\delta} }
{\cal D}( n^*\overline{y}^*, n^*, \delta ) 
\frac{ {\cal E} ( \delta \by + \by^*, \delta + 1, \delta ) }{ {\cal E}( \by^*, 1, \delta ) }  \mathrm{d}\by^*  \nonumber  
\end{eqnarray}
which coincides for any model of the same dimension with training samples of the same size.
Hence, for known $\phi$, this concludes the proof for the (fixed $\delta$) DR-PEP prior. It is obvious that for random $\delta$ 
integrating out the above marginal likelihoods over any hyper-prior 
$\pi(\delta)$ will result to $m_\vg^{\DR}( \by ) = m_0^\DR(\by)$ and this concludes the proof for hyper-$\delta$ DR-PEP. 

Finally if $\phi$ is stochastic and common across models, then the two marginal likelihoods (under fixed or random $\delta$), obtained by integrating out $\phi$ over the common prior $\pi^N_0(\phi) = \pi^N_\vg(\phi)$ for all models $M_\vg$, still coincide.

\section{The PEP-GVS algorithm} 
\label{PEPGVS_algorithm}

\subsection{Implementation details}
\label{DETAILS}

Concerning the binary inclusion indicators $\gamma_j$, 
the  conditional posterior distribution \linebreak 
$\pi\big( \gamma_j \big|  \bb,  \vg_{\setminus j},  \by^{*}, \by, \delta\big)$ is a Bernoulli distribution 
with success probability $O_j/(1+O_j)$ and 
\begin{eqnarray}
O_j & = &
\frac{ f_{\vgjo} \big(\by|\bb_{\vgjo} \big)  }{ f_{\vgjz} \big(\by|\bb_{\vgjz} \big) }  
\left[ \frac{ f_{\vgjo}(\by^*|\bb_{\vgjo})}{ f_{\vgjz}(\by^*|\bb_{\vgjz}) } \right]^{\invd}
\frac{ \pi^{\N}_{\vgjo}(\bb ) }{ \pi^{\N}_{\vgjz}(\bb) }    
\frac{ {m}^{\N}_{\vgjz}(\by^*| \delta) }{ {m}^{\N}_{\vgjo}(\by^*| \delta) }  
\frac{ \pi(\vgjo) } { \pi(\vgjz) }, 
\label{success_prob_gamma_j}
\end{eqnarray}
where 
$\vgjo = (\gamma_j=1, \vg_{\setminus j})$, 
$\vgjz = (\gamma_j=0, \vg_{\setminus j})$ and $\pi^\N_\vg(\boldsymbol\beta)=\pi^\N_\vg(\betab)\pi^{\N}_\vg(\betabg)$ for $\vg \in \{ \vgjo, \vgjz \}$.
All of the quantities involved in \eqref{success_prob_gamma_j} are available in closed form expressions 
except of the marginal likelihood $m_\vg^{\mathrm{N}}(\by^*|\delta)$. 
The latter is estimated through the following Laplace approximation
\begin{equation}
\widehat{m}_{\vg}^\N(\by^*|\delta)=
(2\pi \delta )^{d_{\vg}/2}|\XT \W(\bbgmlestar) \XG|^{-1/2}f_\vg\big(\by^*\big| \bbgmlestar \big)^{\invd}
\pi^{\mathrm{N}}_\vg \big( \bbgmlestar  \big),
\label{GVS-Laplace}
\end{equation}
where $\bbgmlestar$ is the MLE for data $\by^*$ given the configuration of $\vg$ and 
$\delta  \Big[ \XT \W(\bbgmlestar) \XG \Big]^{-1}$ is equal to minus the inverse Hessian matrix evaluated at $\bbgmlestar$.
Under a Jeffreys baseline prior for $\betab$, the Laplace approximation simplifies to 
$
\widehat{m}_\vg^\N(\by^*|\delta)=
(2\pi\delta)^{d_{\vg}/2}f_\vg(\by^*|\bbgmlestar)^{\invd}
$. A comparison with respect to numerical integration, in terms of the the marginal likelihood log-ratios, is provided in Appendix \ref{LAPLACE}.     

For the active effects $\betab$ of model $M_{\vg}$ and the intercept term $\beta_0$ of the reference model $M_0$, we use independence sampler M-H steps. Specifically, for $\betab$ we generate new candidate values as
$$ \betab^{'} \sim q(\betab^{'})\equiv \mathrm{N}_{d_{\vg}} \left(\widehat{\boldsymbol{\beta}}{}_\vg^{\mathrm{all}},\widehat{\Sigma}_{\betab^{\mathrm{all}}} \right), 
$$ 
where $\widehat{\boldsymbol{\beta}}{}_\vg^{\mathrm{all}}$ is the ML estimate from a weighted regression on $\by^{\mathrm{all}}=(\by,\by^*)^T$, using weights $\mathbf{w}^{\mathrm{all}}=(\mathbf{1}_n,\mathbf{1}_n\delta^{-1})^T$, and $\widehat{\Sigma}_{\betab^{\mathrm{all}}}$ is the estimated variance-covariance matrix of $\widehat{\boldsymbol{\beta}}{}_\vg^{\mathrm{all}}$.
The proposed move is accepted with probability 
\begin{equation*}
\alpha_{\betab}=
\min\left[1, \frac{f_\vg(\by|\betabprop)}{f_\vg(\by|\betab)}  
\left( \frac{ f_\vg(\by^*|\betabprop)}{ f_\vg(\by^*|\betab) } \right)^{\invd}
\frac{ \pi^{\N}_\vg(\betabprop)q(\betab)} {\pi^{\N}_\vg(\betab)q(\betabprop)} \right],\\
\end{equation*}
where $\betab$ denotes the current value of the chain. The proposal distribution of $\beta_0$  is 
$q(\beta_0)=\mathrm{N}(\widehat{\beta_0},\psi\widehat{\sigma}^2_{\beta_0})$ 
with $\widehat{\beta_0}$ and $\widehat{\sigma}_{\beta_0}$ 
being the respective ML estimate of $\beta_0$ and the standard error of $\widehat{\beta_0}$
from the null model with response data $\by^*$. 
The proposed move is accepted with the usual M-H transition probability 
where the likelihood of the reference model is raised to the power of $1/\psi$. 
Note that no specific fine tuning is required for the proposal distributions of
$\betab$ and $\beta_0$.

Finally, for the generation of the imaginary data 
we propose candidate values $\by^{*'}$ from a proposal distribution $q(\by^{*'})$ 
and accept the proposed move with probability 
\begin{equation*}
\alpha_{\by^*}= \min\left[ 1, 
\left( \frac{ f_\vg(\by^{*'}|\betab) }{f_\vg(\by^*|\betab) } \right)^{\invd}
\left( \frac{ f_0(\by^{*'}|\beta_0   ) }{f_0(\by^*|\beta_0) } \right) ^{1/\psi}
\frac{\widehat{m}^\N_{\vg}(\by^{*}|\delta) } {\widehat{m}^\N_{\vg}(\by^{*'}|\delta)} 
\frac{ q(\by^{*})} { q(\by^{*'})} \right],
\end{equation*}
where the marginal likelihood estimates are obtained through \eqref{GVS-Laplace} and $\by^{*}$ denotes the current value of the chain. 
The joint proposal density is formed by the product of independent distributions, i.e. $q(\by^*)=\prod_{i=1}^{n^*}q(y^*_i)$, 
where the proposal of each imaginary observation $y_i^*$ is constructed by 
combining the two likelihood components of the PEP prior. 
Hence, for the logistic regression model we use 
\begin{equation*}
q(y^*_i) \equiv \mbox{Binomial}( N_i, \pi_i^* )  \mbox{~with~}
\pi_i^*=\frac{\pi_0^{1/\psi}\pi_{\vg(i)}^{\invd}}{\pi_0^{1/\psi}\pi_{\vg(i)}^{\invd}+(1-\pi_0)^{1/\psi}(1-\pi_{\vg(i)})^{\invd}},
\end{equation*}
where  $\pi_0=(1+\exp(-\beta_0))^{-1}$, $\pi_{\vg(i)}=(1+\exp(-\X_{\vg(i)}\betab))^{-1}$ and $N_i$ denotes the number of trials of the observed data.
Equivalently, for Poisson regression models we consider 
$$
q(y^*_i) \equiv \mbox{Poisson}\left( \lambda_0\lambda_{\vg(i)}^{\invd} \right)
$$ 
for the CR-PEP prior; where $\lambda_0=\exp(\beta_0)$ and $\lambda_{\vg(i)}=\exp(\X_{\vg(i)}\betab)$. 
For the DR-PEP prior, the corresponding choice of a Poisson proposal with mean $(\lambda_0\lambda_{\vg})^{\invd}$ 
was not found to be efficient in practice. 
Therefore, we use instead a Poisson random-walk proposal  
with mean equal to the value of $y^*_i$ at the current iteration.     

\subsection{An analytic description}
\label{GVSanalytic}

Given the posterior distribution in Eq. \ref{augment}, with $\psi=1$ for the CR-PEP prior and $\psi=\delta$ for the DR-PEP prior, the PEP-GVS sampler proceeds as follows: 

\begin{itemize}
	\item[\textbf{A.}] Set starting values $\vg^{(0)}, \bb^{(0)}=(\betab^{(0)},\betabg^{(0)}), \beta_0^{(0)}$ and $\by^{*(0)}$. For fixed $\delta$ set $\delta=n$, for random $\delta$ set starting starting value $\delta^{(0)}.$
	\item[\textbf{B.}] For iterations $t=1,2,...,N$:
	\begin{itemize}
		
		\item[\textbf{Step 1:}] Sampling of $\gamma_j^{(t)}$, for $j=1,2,...,p$, given the current state of $\betab, \betabg, \vg_{\setminus j}, \by^*$ and $\delta$.
		\begin{itemize}
			
			\item[(a)] Calculate the MLEs under $\vgjo = (\gamma_j=1, \vg_{\setminus j})$, 
			$\vgjz = (\gamma_j=0, \vg_{\setminus j})$ and compute the Laplace approximations $\widehat{m}_{\vgjo}^{\N}(\by^*|\delta)$, 
			$\widehat{m}_{\vgjz}^{\N}(\by^*|\delta)$ through Eq. \ref{GVS-Laplace}.
			
			\item[(b)] Evaluate 
			the odds:
			\begin{eqnarray*}
				O_j & = &
				\frac{ f_{\vgjo} \big(\by|\bb_{\vgjo} \big)  }{ f_{\vgjz} \big(\by|\bb_{\vgjz} \big) } 
				\left[ \frac{ f_{\vgjo}(\by^*|\bb_{\vgjo})}{ f_{\vgjz}(\by^*|\bb_{\vgjz}) } \right]^{\invd} 
				\frac{ \pi^{\N}_{\vgjo}(\bb_{\vgjo}) }{ \pi^{\N}_{\vgjz}(\bb_{\vgjz})} 
				\frac{ \pi^{\N}_{\vgjo}(\bb_{\setminus\vgjo}) }{ \pi^{\N}_{\vgjz}(\bb_{\setminus\vgjz})} \\         
				&  &   
				\times   
				\frac{ \widehat{m}^{\N}_{\vgjz}(\by^*| \delta) }{ \widehat{m}^{\N}_{\vgjo}(\by^*| \delta) } 
				\frac{ \pi(\vgjo) } { \pi(\vgjz) }
			\end{eqnarray*}
			
			\item[(c)] Sample $\gamma_j^{'}\sim\mathrm{Bernoulli}\left(\frac{O_j}{1+O_j}\right)$ and set $\gamma_j^{(t)}=\gamma_j^{'}$ with probability equal to 1.\\
		\end{itemize}

		\item[\textbf{Step 2:}] Update $\boldsymbol{\beta}^{(t-1)}=(\betab^{(t-1)},\betabg^{(t-1)})$ based on the current configuration of $\vg$.\\

		\item[\textbf{Step 3:}] Sampling of $\betabt$ given the current state of $\vg,\by^*$ and $\delta$.
		\begin{itemize}
			
			\item[(a)] Generate $\betab^{'}$ from the proposal distribution $q(\betab^{'})=\mathrm{N}_{d_{\vg}} (\widehat{\boldsymbol{\beta}}{}_\vg^{\mathrm{all}},\widehat{\Sigma}_{\betab^{\mathrm{all}}} )$, where $\widehat{\boldsymbol{\beta}}{}_\vg^{\mathrm{all}}$ is the ML estimate from a weighted regression on $\by^{\mathrm{all}}=(\by,\by^*)^T$, using weights $\mathbf{w}^{\mathrm{all}}=(\mathbf{1}_n,\mathbf{1}_n\delta^{-1})^T$, and $\widehat{\Sigma}_{\betab^{\mathrm{all}}}$ is the estimated variance-covariance matrix of $\widehat{\boldsymbol{\beta}}{}_\vg^{\mathrm{all}}$.
			
			\item[(b)] Calculate the probability of accepting the proposed move:
			\begin{equation*}
			\alpha_{\betab}=
			1 \land \left[\frac{f_\vg(\by|\betabprop)}{f_\vg(\by|\betab^{(t-1)})}  
			\left\{ \frac{ f_\vg(\by^*|\betabprop)}{ f_\vg\big(\by^*|\betab^{(t-1)} \big) } \right\}^{\invd}
			\frac{ \pi^{\N}_\vg(\betabprop)} {\pi^{\N}_\vg \big( \betab^{(t-1)} \big)} \frac{q(\betab^{(t-1)})}{q(\betabprop)}\right].
			\end{equation*}
			
			\item[(c)] Set 
			$\betabt=\begin{cases}
			\betab^{'} \;\;\;\;\; \text{with probability}\; \alpha_{\betab},\\
			\betabtt\; \text{with probability}\; 1-\alpha_{\betab}. 
			\end{cases}$\\[0.5cm]
		\end{itemize}

		\item[\textbf{Step 4:}] Sampling of $\betabgt$ given the current state of $\vg$.
		\begin{itemize}
			
			\item[(a)] Generate $\betabg^{'}$ from the pseudo-prior  $\pi^\N_\vg(\betabg^{'})=\N_{d_{\setminus\vg}}\big( \widehat{\boldsymbol\beta}_{\setminus\vg}, \mathbf{I}_{d_{\setminus\vg}} \widehat{\sigma}_{\betabg}^2\big)$,where $\widehat{\boldsymbol\beta}_{\setminus\vg}$ and $\widehat{\sigma}_{\betabg}$ are the respective MLEs and  corresponding standard errors of $\betabg$ from the full model given data $\by$.
			
			\item[(b)] Set $\betabgt=\betabg^{'}$ with probability equal to 1.\\
		\end{itemize}

		\item[\textbf{Step 5:}] Sampling of $\beta_0^{(t)}$ given the current state of $\by^*$ and $\delta$.
		\begin{itemize}
			\item[(a)]  Generate $\beta_0^{'}$ from the proposal distribution $q(\beta_0^{'})=\mathrm{N} \big(\widehat{\beta}_0,\psi\widehat{\sigma}^2_{\beta_0}\big)$, where $\widehat{\beta}_0$ and $\widehat{\sigma}_{\beta_0}$ are the respective MLE of $\beta_0$ and the standard error of $\widehat{\beta}_0$ from the null model given data $\by^*$. 
			
			\item[(b)] Calculate the probability of accepting the proposed move:
			\begin{equation*}
			\alpha_{\beta_0}=1\land\left[
			\left\{ \frac{f_0(\by^*|\beta_0^{'})}
			{f_0 \big( \by^*|\beta_0^{(t-1)} \big) } \right\}^{1/\psi}
			\frac{\pi_0^{\N}(\beta_0^{'})}
			{\pi_0^{\N} \big( \beta_0^{(t-1)} \big) }
			\frac{q \big( \beta_0^{(t-1)} \big) }
			{q(\beta_0^{'})} 
			\right].
			\end{equation*}
			
			\item[(c)] Set 
			$
			\beta_0^{(t)}=\begin{cases}
			\beta_0^{'} \;\;\;\;\;\;\; \text{with probability}\; \alpha_{\beta_0},\\
			\beta_0^{(t-1)}\;\; \text{with probability}\; 1-\alpha_{\beta_0}.
			\end{cases}
			$  \\
		\end{itemize}
		
		\item[\textbf{Step 6:}] Sampling of $\by^{*(t)}$ given the current state of $\betab$, $\beta_0$, $\vg$ and $\delta$.
		\begin{itemize}
			
			\item[(a)]Generate $\by^{*'}$ from a proposal distribution $q(\by^{*'})$; for details about this proposal, see the remarks in the subsection which immediately follows the description of the algorithm.
			
			\item[(b)] Calculate the MLEs given $\by^{*(t-1)}$ and $\by^{*'}$ and compute the Laplace approximations $\widehat{m}_{\vg}^\N(\by^{*(t-1)}|\delta)$ and $\widehat{m}_{\vg}^\N(\by^{*'}|\delta)$ through Eq. \ref{GVS-Laplace}.
			
			\item[(c)] Calculate the probability $\alpha_{\by^*}$ of accepting the proposed move equal to \\  
			\small   
			\begin{equation*}
			1\land\left[
			\left\{\frac{f_{\vg}(\by^{*'}|\betab) }
			{f_{\vg}(\by^{*(t-1)}|\betab)} \right\}^{\invd}
			\left\{\frac{f_0(\by^{*'}|\beta_0)}
			{f_0(\by^{*(t-1)}|\beta_0)}\right\}^{1/\psi}
			\frac{\widehat{m}_{\vg}^\N (\by^{*(t-1)}|\delta \big)}
			{\widehat{m}_{\vg}^\N(\by^{*'}|\delta)}
			\frac{q\big(\by^{*(t-1)}\big)}
			{q(\by^{*'})}
			\right].
			\end{equation*}
			\normalsize 
			
			\item[(d)] Set 
			$
			\by^{*(t)}=\begin{cases}
			\by^*{'} \;\;\;\;\;\;\; \text{with probability}\; \alpha_{\by^*},\\
			\by^{*(t-1)}\;\; \text{with probability}\; 1-\alpha_{\by^*}.
			\end{cases}
			$ \\
		\end{itemize}

		\item[\textbf{Step 7:}] Sampling of $\delta^{(t)}$ given the current state of $\betab$, $\beta_0$ and $\vg$.
		\begin{itemize}
			\item[(a)] If $\delta$ is fixed at $n$ go to Step 1, else implement (b)-(e) of Step 7.
			
			\item[(b)] Generate $\delta'$ from the proposal distribution $q(\delta'|\delta^{(t-1)})=\mathrm{Gamma}(a,b)$ with $a=\delta^{(t-1)^2}/s^2_{\delta}$ and $b=\delta^{(t-1)}/s^2_{\delta}$.
			
			
			\item[(d)] Calculate the probability of accepting the proposed move:
			\begin{eqnarray*}
				\alpha_{\delta} & = & 1\land \left\{  \left[
				\left( \frac{\delta^{(t-1)}}{\delta'} \right)^{d_\vg/2}
				\left(\frac{ f_\vg(\by^{*}|\betab) }{ f_\vg(\by^{*}|\bbgmlestar)} \right)^{ \left\{\frac{1}{\delta'}-\frac{1}{\delta^{(t-1)}} \right\}}  
				\right]\times \right.  \\
				&  & 
				\left. 
				\hspace{5em}
				\times\Bigg[ 
				f_0(\by^{*}|\beta_0)^{ \left\{\frac{1}{\psi'}-\frac{1}{\psi^{(t-1)}} \right\}}
				\frac{ \pi(\delta') } { \pi(\delta^{(t-1)}) }
				\frac{ q(\delta^{(t-1)}|\delta') }{ q(\delta'|\delta^{(t-1)}) }\Bigg] \right\},
			\end{eqnarray*}
			where $\psi'=\psi^{(t-1)}=1$ for the CR-PEP prior and $\psi'=\delta'$, $\psi^{(t-1)}=\delta^{(t-1)}$ for the DR-PEP prior.
			
			\item[(e)] Set 
			$
			\delta^{(t)}=\begin{cases}
			\delta^{'} \;\;\;\;\;\;\; \text{with probability}\; \alpha_{\delta},\\
			\delta^{(t-1)}\;\; \text{with probability}\; 1-\alpha_{\delta}.
			\end{cases}
			$\\
		\end{itemize}
		
	\end{itemize}
	\item[\textbf{C.}] Repeat the steps in B until convergence. \\
\end{itemize}

\paragraph{Suggested proposals for Step 6:}
For $\by^*$ we recommend the following proposals depending on the likelihood of the model and on the PEP prior that is used:

\begin{itemize}
	\item[i)] For logistic regression a product binomial proposal distribution given by 
	\begin{equation*}
	q(\by^*) = \prod\limits_{i=1}^{\nstar}\mbox{Binomial}(N_i, \pi_i^* )  \mbox{~with~}
	\pi_i^*=\frac{\pi_0^{1/\psi}\pi_{\vg(i)}^{\invd}}{\pi_0^{1/\psi}\pi_{\vg(i)}^{\invd}+(1-\pi_0)^{1/\psi}(1-\pi_{\vg(i)})^{\invd}},
	\end{equation*}
	where  $\pi_0=\big\{1+\exp(-\beta_0)\big\}^{-1}$, $\pi_{\vg(i)}=\big\{1+\exp(-\X_{\vg(i)}\betab)\big\}^{-1}$ and $N_i$ denotes the number of trials of the observed data.
	
	\item[ii)] For Poisson regression  a product Poisson proposal distribution given by
	$
	q(\by^*) = \prod\limits_{i=1}^{\nstar} \mbox{Pois}(\lambda_i^*).
	$ 
	For the CR-PEP prior $\lambda_i^*= \lambda_0\lambda_{\vg(i)}^{\invd}$, where $\lambda_0=\exp(\beta_0)$ and $\lambda_{\vg(i)}=\exp(\X_{\vg(i)}\betab)$. For the DR-PEP prior we utilize a random-walk proposal, i.e. $\lambda_i^*=y_i^{*(t-1)}$.     
\end{itemize}

\newpage
\section{Laplace Approximation vs. Numerical Integration}
\label{LAPLACE}

\begin{figure}[h]
	\centering{}\includegraphics[scale=0.57]{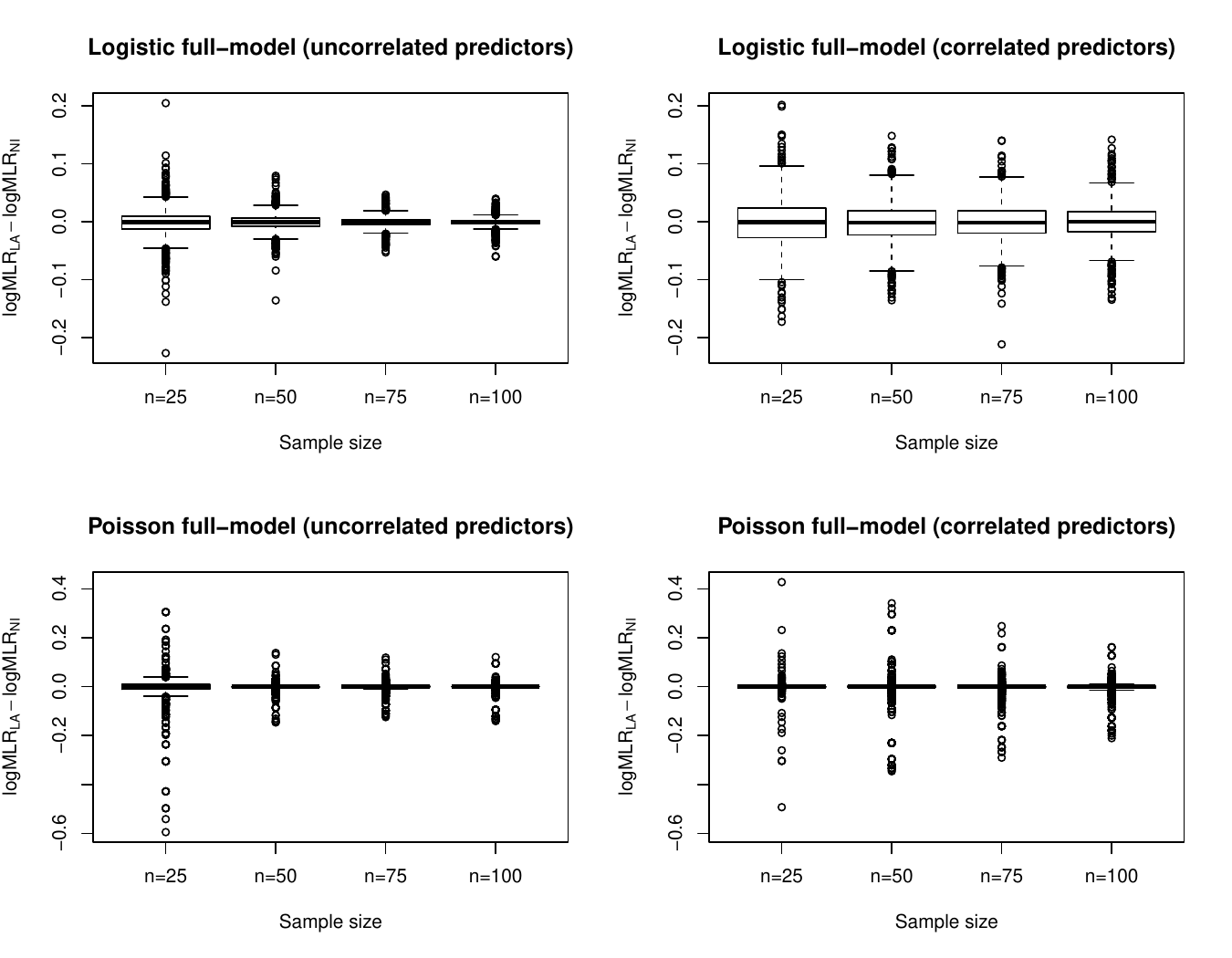}
	\caption{Log-differences of Laplace approximation MLRs and numerical integration MLRs from 1000 GVS-PEP iterations under different sample sizes.}
	\label{MLR}
\end{figure}

The Laplace approximation given in Eq. \eqref{GVS-Laplace} is used during the MCMC algorithm in order to calculate the marginal likelihood ratios (MLR) appearing in Step 1(b) and in Step 6(c) of the GVS-PEP algorithm of Appendix \ref{GVSanalytic}. For instance, in Step 6(c) at iteration $t$ we use the Laplace approximation to evaluate the quantity
\begin{equation*}
\mathrm{MLR}(\by^{*(t)}|\delta)=\frac{\widehat{m}_{\vg}^\N(\by^{*(t-1)}|\delta)}
{\widehat{m}_{\vg}^\N(\by^{*(t)}|\delta)}.
\end{equation*}
Here we consider comparing the Laplace approximation (LA) to numerical integration (NI), in terms of estimating the above quantity, for the corresponding full models (i.e. models including all covariates under consideration)
under the logistic and Poisson simulation scenarios  presented in Section \ref{sec_sim1}. In Figure \ref{MLR} we present the differences $\log\mathrm{MLR_{LA}}-\log\mathrm{MLR_{NI}}$ from 1000 MCMC iterations and sample sizes $n=25,50,75,100$. As seen the differences between the two approaches are negligible in all cases.    

\newpage
\section{Results from Simulation 1} 
\label{Boxplots}

\subsection{Posterior inclusion probabilities (logistic model)}
\label{PIPs_logistic}
\begin{figure}[h!]
	\centering{}\includegraphics[scale=0.37]{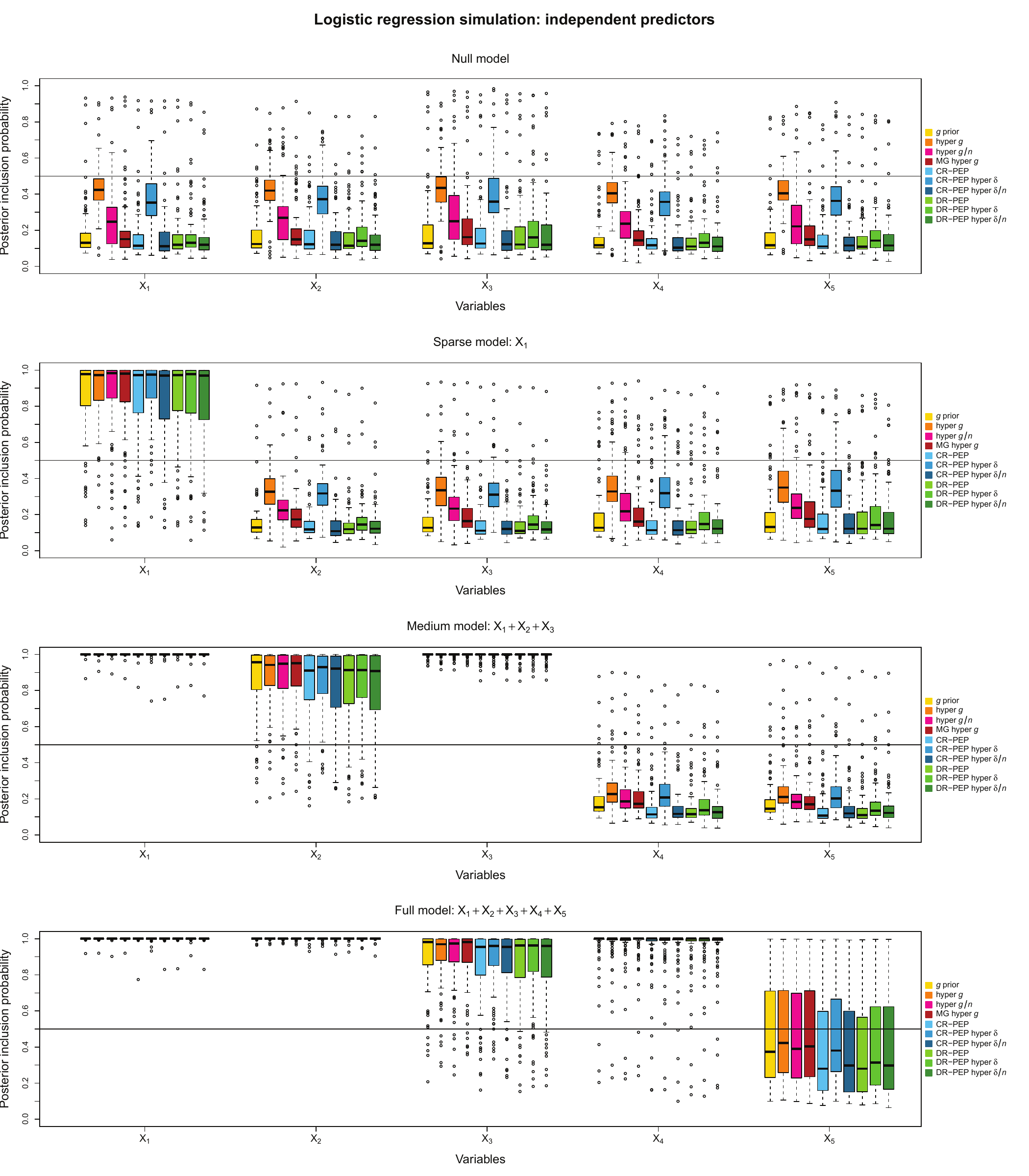}
	\caption{Posterior inclusion probabilities for Simulation Study 1 from 100 replicated samples of the null, sparse, medium and full logistic regression model scenarios with independent predictors ($r=0$).}
	\label{ex1_logistic_sim1}
\end{figure}
\begin{figure}[h!]
	\centering{}\includegraphics[scale=0.37]{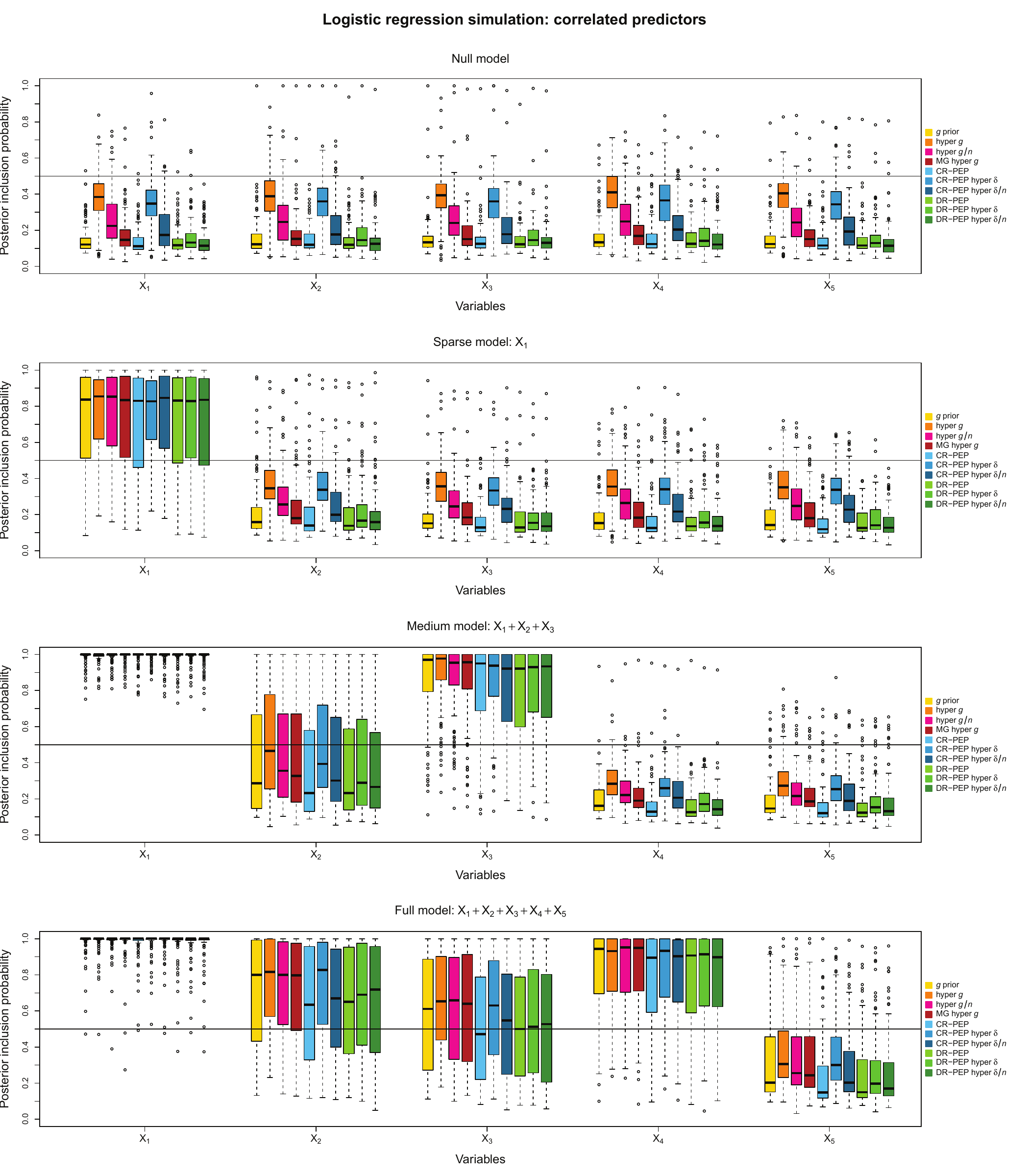}
	\caption{Posterior inclusion probabilities for Simulation Study 1 from 100 replicated samples of the null, sparse, medium and full logistic regression model scenarios with correlated predictors ($r=0.75$).}
	\label{ex1_logistic_sim2}
\end{figure}

\subsection{Posterior inclusion probabilities (Poisson model)}
\label{PIPs_poisson}

\begin{figure}[h!]
	\centering{}\includegraphics[scale=0.37]{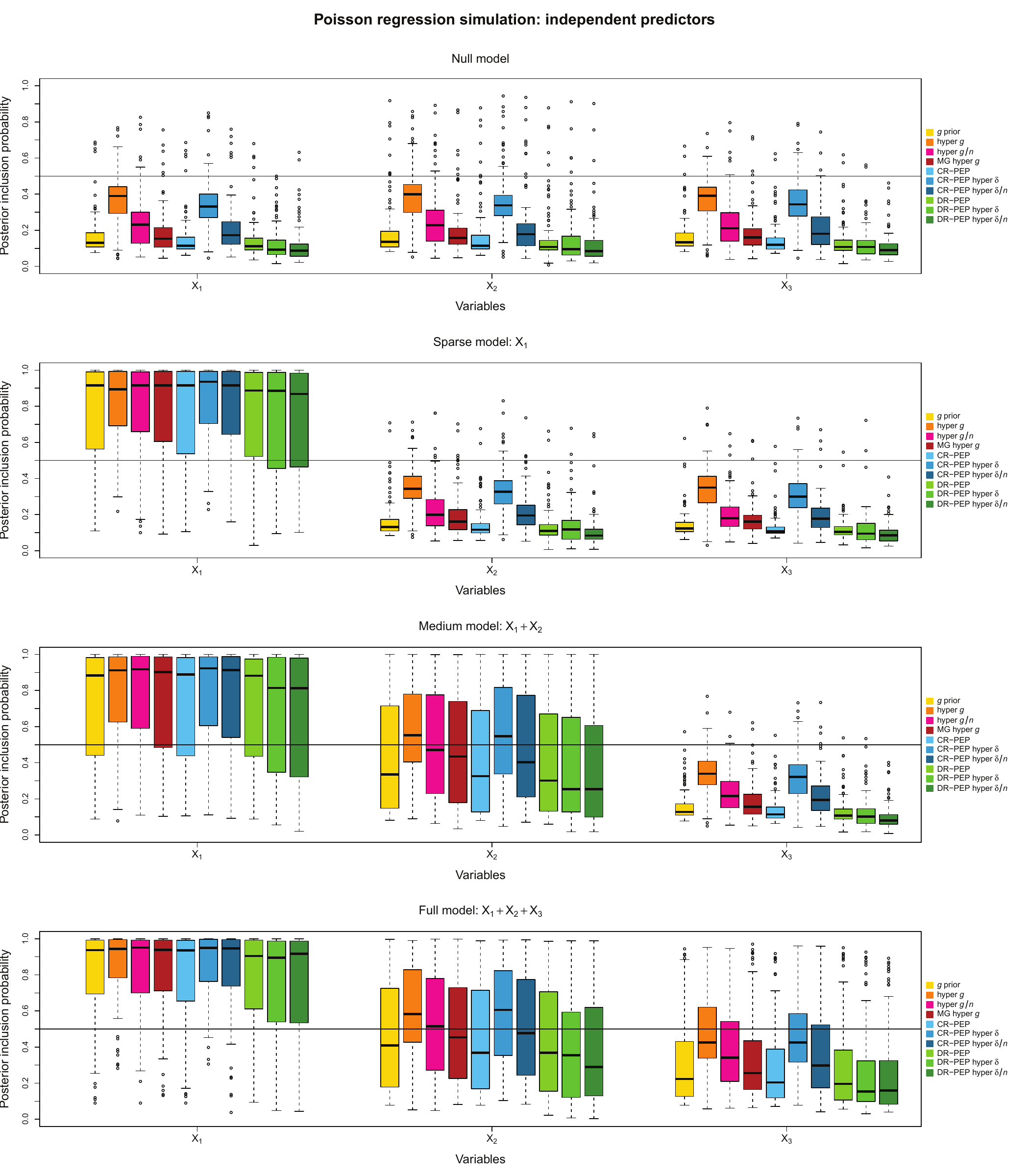}
	\caption{Posterior inclusion probabilities for Simulation Study 1 from 100 replicated samples of the null, sparse, medium and full Poisson model scenarios with independent predictors ($r=0$).}
	\label{ex1_poisson_sim1}
\end{figure}

\begin{figure}[h!]
	\centering{}\includegraphics[scale=0.37]{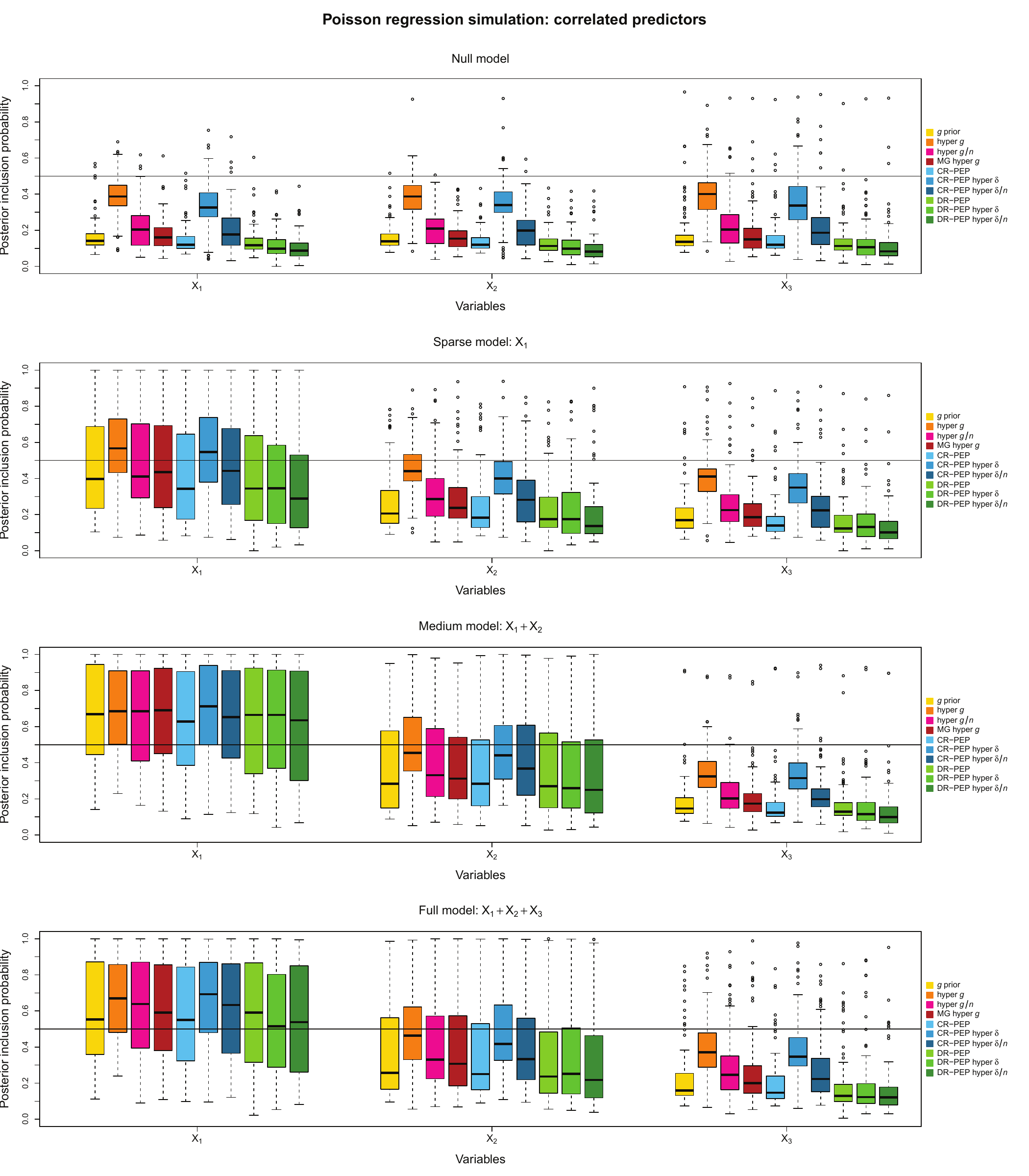}
	\caption{Posterior inclusion probabilities for Simulation Study 1 from 100 replicated samples of the null, sparse, medium and full Poisson model scenarios with correlated predictors ($r=0.75$).}
	\label{ex1_poisson_sim2}
\end{figure}

\end{document}